\documentclass[]{aa}

\RequirePackage{snapshot}

\usepackage{txfonts}
\usepackage[pdftex]{graphicx}
\usepackage{natbib}
\usepackage{rotating}
\usepackage{colortbl}
\usepackage{hyperref}
\usepackage{grffile}
\usepackage{rotating}
\usepackage{color,soul}
\usepackage{soul}

\bibpunct{(}{)}{;}{a}{}{,} % to follow the A&A style
\newcommand{\lineflux}{f_\mathrm{0}} %% flux of main line
\newcommand{\filter}[1]{T({#1})}
\newcommand{\spectrum}[1]{f_{#1}(#1)}

\newcommand{\lya}[0]{$\mathrm{Ly}\,\alpha$}
\newcommand{\ha}[0]{$\mathrm{H}\alpha$}
\newcommand{\fion}[2]{[\ion{#1}{#2}]}
\newcommand{\sext}[0]{\emph{SExtractor}}
\newcommand{\ultvis}[0]{UltraVISTA}
\newcommand{\minestbb}[0]{\delta m_\mathrm{BB}^\mathrm{est}}

\setstcolor{green}

\title{A Method to improve line flux and redshift measurements with narrowband filters}

	\institute{Dark Cosmology Centre, Niels Bohr Institute, University of Copenhagen, Juliane Maries Vej 30, 2100 Copenhagen {\O}, Denmark \email{johannes@dark-cosmology.dk}\label{inst1} \and European Southern Observatory, Karl-Schwarzschild-Stra\ss{}e 2, 85748 Garching bei M{\"u}nchen, Germany \label{inst2} \and Aix Marseille Universit\'e, CNRS, LAM (Laboratoire d'Astrophysique de Marseille) UMR 7326, 13388, Marseille, France \label{inst3}}

	\author{J.~Zabl \inst{\ref{inst1}}, W.~Freudling \inst{\ref{inst2}}, P.~M{\o}ller \inst{\ref{inst2}},  B.~Milvang-Jensen \inst{\ref{inst1}}, K.K.~Nilsson \inst{\ref{inst2}}, J.P.U.~Fynbo \inst{\ref{inst1}},  O.~Le F\`evre \inst{\ref{inst3}}, L.A.M.~Tasca \inst{\ref{inst3}}}
\authorrunning{J. Zabl}

\begin{document}

\abstract{High redshift star-forming galaxies are discovered routinely through
	the presence of a flux excess in narrowband filters caused by an
	emission line.  In most cases, the width of such filters is broad
	compared to typical line widths, and the throughput of the filters
	varies substantially within the bandpass. This leads to substantial
uncertainties in redshifts and fluxes that are derived from the observations
with one specific narrowband filter. }
{ The uncertainty in measured line parameters can be sharply reduced by using
	repeated observations of the same target field with filters that have
	overlapping passbands but differ slightly in central wavelength or
	wavelength dependence of the effective filter curve.  Such data are
routinely collected with some large field imaging cameras that use multiple
detectors and a separate filter for each of the detectors. An example is the
European Southern Observatory's VISTA InfraRed CAMera (VIRCAM). }
{We developed a method to determine more accurate redshift and line flux
	estimates from the ratio of apparent fluxes measured from observations
	in different narrowband filters and several matching broadband
	filters. A parameterized model of the line and continuum flux is used
	to predict the flux ratios as a function of redshift based on the known
filter curves. These model predictions are then used to determine the most
likely redshift and line flux.}
{We tested the obtainable quality of parameter estimation for the example of
	H$\alpha$ in the VIRCAM NB118 filters both on simulated and actual
	observations, where the latter were based on the \ultvis{} DR2 data
	set. We combined the narrowband data with deep broadband data in Y, J,
	and H. We find that by using this method, the errors in the measured
lines fluxes can be reduced up to almost an order of magnitude.}
{We conclude that existing narrowband data can be used to derive accurate line
	fluxes if the observations include images taken with sufficiently
	different filter curves. For the \ultvis{} survey, the best suited
	narrowband filter combinations allow to achieve an accuracy in
	wavelength of better than $1\,\mathrm{nm}$, and in flux of better than
	15\% at any redshift within the bandpass of the filters. By contrast,
	analyzing the data without exploiting the difference in filter curves
	leads to an uncertainty in wavelength of 10\,$\mathrm{nm}$, and up to
an order of magnitude errors in line flux estimates. }

\date{\today}

\keywords{Methods: observational -- Techniques: photometric -- Galaxies: photometry -- Galaxies: distances and redshifts -- Galaxies: star formation -- Galaxies: high-redshift}

\maketitle

%##############################
\section{Introduction}
%##############################
Strong emission lines, which are the key signature in the spectra of
star-forming galaxies, are important tools. The hydrogen Balmer lines are
useful for determining the instantaneous star formation rate (SFR), as their
strength is directly
proportional to the SFR after correcting for dust
\citep[e.g.][]{Kennicutt:1998:189}. Other strong emission lines, like
$[\ion{O}{ii}]\,\lambda3727$, are less accurate SFR indicators due to their
metallicity dependence \citep{Moustakas:2006:775,Kewley:2004:2002}, but certain
ratios between these lines and the metallicity-independent Balmer lines can be
gauged as proxies for the gas-phase metallicity
\citep[e.g.][]{Pagel:1979:95,Kewley:2002:35}.

Wide-field surveys with narrowband (NB) filters provide large line flux limited
samples down to low equivalent widths at well defined redshifts. The idea
behind the NB selection is to identify objects through an excess of the
filter-averaged NB flux density over the underlying continuum flux density,
with the latter being inferred from one or more suitable broadband (BB) filters
\citep[e.g.][]{Djorgovski:1985:L1,Moller:1993:43,Pascual:2007:30}. When the NB
observations are in a field with extensive multi-wavelength data, photometric
redshifts allow to discern between the different lines which could be the cause
for the NB excess \citep[e.g. in the COSMOS
field:][]{Ilbert:2013:55,Muzzin:2013:8}.

A vast number of NB surveys have been performed, targeting \ha{}, \fion{O}{ii},
and Ly$\alpha$ at different redshifts. While many studies have been using
observed-frame optical NB filters
\citep[e.g.][]{Fujita:2003:L115, Dale:2010:L189, Takahashi:2007:456,
  Ly:2012:63, Rhoads:2000:L85, Ouchi:2003:60, Nilsson:2007:71}, currently a
  strong focus is put on the exploitation of the airglow windows in the
  near-infrared (NIR)
\citep[e.g.][]{Best:2010,Ly:2011:109,Kochiashvili:2015}.  This is essential for
high redshifts, as the important rest-frame optical lines shift into this
wavelength regime. Among the deepest wide-field NIR surveys are the NB118
observation by \citet{Milvang-Jensen:2013:94} and the NB118 part of \ultvis{}
(Ultra Deep Survey with VISTA;
\citealt{McCracken:2012:156}), both of which are performed with the
near-infrared camera VIRCAM (VISTA InfraRed CAMera,
\citealt{Dalton:2006:62690X}) at ESO's 4.1m survey telescope, VISTA (Visible
and Infrared Survey Telescope for Astronomy,
\citealt{Emerson:2006:41}).

All these observational efforts are eventually used to estimate line fluxes
from the measured NB excesses and, with the knowledge of the redshift, line
luminosities. Obtaining these for large samples of galaxies at well defined
redshifts allows to get important insights into galaxy evolution, e.g. by means
of determining \ha{} luminosity functions, which can be converted to SFR
densities. Understanding can also be gained by relating other properties like
galaxy mass, environment, and spatial clustering to the line luminosities
\citep{Sobral:2010:1551,Sobral:2011:675}, and line based SFR estimates
might be compared to SFR estimates obtained by other means, allowing
to characterize the galaxies' star formation histories
\citep{Dominguez:2015}.

% TODO: Here maybe some more citations
Unfortunately, line flux measurements obtained from NB filters can have
substantial uncertainties. As the transmittance curves are often far from flat,
a flux measured in the filter is consistent with a range of intrinsic line
fluxes. While for sample based statistics it is in many cases possible to
overcome this problem to some extent by statistical means, other applications,
like the identification of high redshift candidates, need the best possible
flux determination for individual objects.

In this paper we discuss a method which both overcomes this flux measurement
problem and allows at the same time for a wavelength resolution an order of
magnitude below the NB width. This can be achieved by the use of slightly
different filters, which allow to break the degeneracy between central
wavelength and line flux.
\citet{Hayashi:2014ec} recently used redshift estimates from Subaru/Suprime-Cam
imaging with two different NB filters. 
% TODO: Do I still want to mention that 
In this paper, we formalize and generalize this approach to include several NB
filters in the emission line measurement, and crucially, exploit information
from broadband imaging to further constrain the redshift model. As strong
emission lines cause a flux excess even in wide filters
\citep[e.g.][]{Guiderdoni:1987wn,Zackrisson:2001gs,Schaerer:2009eq,
Shivaei:2015ij}, broadband photometry contains very useful additional
information. In the following, we refer to our method as the
throughput-variations method (TPV).

We investigate the method carefully for the VISTA NB118 filters with a special
focus on \ha{}, both with simulations and application to data. In VIRCAM, there
is one individual copy of the NB118 filters above each of its 16 non-contiguous
detectors. Although produced to be as similar as possible, the transmittance
curves of the individual filters are unavoidably slightly different from each
other, and hence useful for the proposed TPV. The extremely deep and
homogeneous BB data also available from the \ultvis{} survey are well suited to
constrain the continuum and to measure the broadband excess.

The paper is organized as follows: In sec. \ref{sec:nb118:multnb}, we motivate
the throughput-variations method (TPV). After describing estimation algorithm
and estimation model in sec. \ref{sec:nb118:algorithm}, we test usefulness and
caveats of the technique based on simulations, which emulate \ultvis{} DR2
observations of \ha{} emitters (sec. \ref{sec:nb118:simobs}).  An application
of the method to actual \ultvis{} DR2 data is presented in sec.
\ref{sec:nb118:ultravista}.  Finally, we discuss a possible modification to the
\ultvis{} NB118 observing pattern for the purpose of the TPV (sec.
\ref{sec:nb118:ultravista:turning}).

Where needed, a (flat) standard $\Lambda CDM$ cosmology with $H_0 =
70\;\mathrm{km}\,\mathrm{s}^{-1}\,\mathrm{Mpc}^{-1}$, $\Omega_\mathrm{m,0} =
0.3$ and $\Omega_{\Lambda,0} = 0.7$ was assumed.  Furthermore, we used
throughout this paper AB-magnitudes
\citep{Oke:1974:21}.  All numbers referring to specific VISTA NB118
filters are in line with the standard VISTA detector (filter)
numbering scheme \citep[cf.~][]{Ivanov:2009,Milvang-Jensen:2013:94}.

All stated wavelengths are in vacuum, except when we use common identifiers
like \fion{O}{iii}$\lambda 5007$, which are based on wavelengths in air.

All assumed VISTA/VIRCAM filter curves include quantum-efficiency and mirror
reflectivities, and the broadband filters curves include in addition an
atmosphere with $PWV = 1\;\mathrm{mm}$ at airmass 1.  For the broadband
filters, Y, J, and H, we used the same filter curve for all 16 detectors, as
available from the ESO webpage.\footnote{Filter curves, detector QE and mirror
reflectivities were downloaded from
  \url{http://www.eso.org/sci/facilities/paranal/instruments/vircam/inst/Filters_QE_Atm_curves.tar.gz}
} Details about the individual NB118 filter curves are given in Appendix
\ref{app_nb118}.  Parts of the field covered by NB118 observations include data
from a single filter/detector, while other parts include data from two filters
(cf. sec. \ref{sec:nb118:observingpattern:ultravista}).  Throughout this paper
we will refer to 'combined effective filters', which are the effective filter
responses, if data from two similar filters is combined into a single stack.

%##############################
\section{Method}
\label{sec:nb118:multnb}
%##############################

%##############################
\subsection{Estimating line fluxes from NB observations}
\label{sec:nb118:mot}
%##############################

%%%%%%%%%%%
\begin{figure*}
\centering
\resizebox{0.7\hsize}{!}{\includegraphics{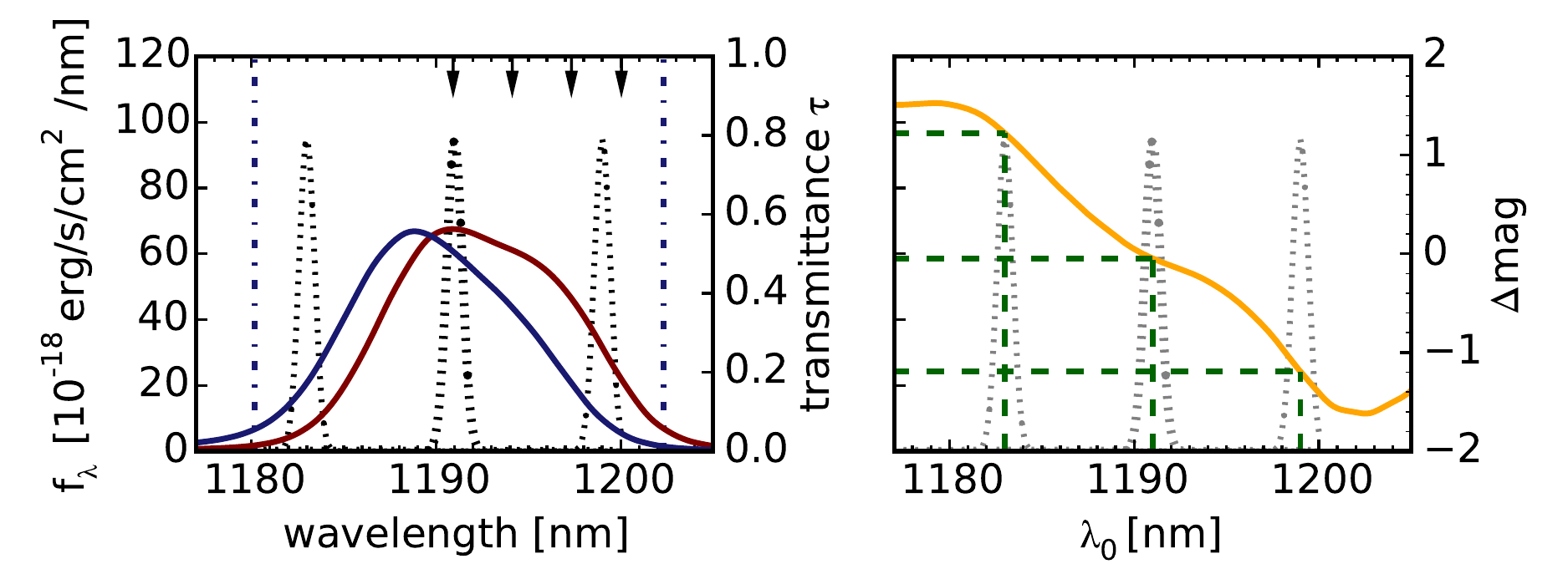}}
\caption{ Illustration of how to obtain accurate line fluxes by the use of two
	NB filters.  The left panel shows the throughput curves of two similar NB
	filters, superimposed on an emission line shifted to different wavelengths.
	The right hand panel shows the corresponding differences between the
	magnitudes measured with the two filters, $\Delta mag$, as a function of
	the wavelength of the emission line, $\lambda_0$. Here, the green dashed
	lines indicate how a measurement of $\Delta mag$ can be used to determine
	the wavelength of the line, which in turn can be used to estimate its flux
	using the throughput curves on the left. The shown filter combination is
	the filter pair 14 \& 15 of the NB118 filters, with 15 being the bluer. The
	meaning of the small arrows and the two vertical lines are described in
	sec. \ref{sec:simu_diff_pfree} and Appendix
	\ref{sec:appendix:quantitativeall}, respectively.}
\label{fig:nb118:fig_sol_deltab}
\end{figure*}
%%%%%%%%%%%

Narrowband surveys, which aim at identifying emission line galaxies
and measure their line fluxes, typically use one NB filter in
combination with one or two broadband (BB) filters at wavelengths
similar to the NB.  In the simplest case the NB passband is at the
center of a BB passband, minimizing the impact of a sloped continuum.
Then, a $BB - NB > 0$ indicates the presence of an emission line, as
the impact of a line on the filter-averaged flux density is
significantly larger in the narrower filter.

When relating the measured magnitudes to an emission line, it is useful to describe the line spectrum through the line's flux, $f_0$, its observed-frame equivalent width, $EW_{obs}$, and its central wavelength, $\lambda_0$, or equivalently its redshift. Using these three quantities, the object's spectrum can be written as:\footnote{Throughout the paper the subscripts $\lambda$ and $\nu$ to $f$ indicate that the flux densities are either per unit wavelength or per unit frequency interval.}

\begin{equation}
	f_\mathrm{\lambda}(\lambda;f_0,EW_{obs},\lambda_0)=
	\zeta_{\lambda}(\lambda;\lambda_0)\frac{f_0}{EW_{obs}} + f_0 \cdot
	\mathcal{L}_\lambda (\lambda;\lambda_0) \label{eq_model_simple}
\end{equation}

\noindent

 $\zeta_{\lambda}(\lambda)$ is the dimensionless spectral
shape of the continuum and $\mathcal{L}_\lambda (\lambda)$ is the
emission line spectrum in units of
$\mathrm{nm}^{-1}$. $\zeta_{\lambda}(\lambda)$ is normalized at the
wavelength of the relevant line and $\mathcal{L}_\lambda (\lambda)$
can include additional lines, but it is scaled so that the integral
over the relevant line is one.
%TODO; implement clearly into the context
$f_\lambda$ is related to the observed magnitudes through
(e.g.~\citealt{Buser:1986:799}):

\begin{eqnarray}
	\mathrm{f}_{\nu;\mathrm{filter}} &=&   \frac{\int \spectrum{\lambda} \filter{\lambda} \lambda \, d\lambda}{\int  \frac{c}{\lambda^2}\filter{\lambda} \lambda \, d \lambda} \label{eq:fnu} \\
	\mathrm{m}_{\mathrm{AB;filter}}        &=& -2.5 \log_{10} \mathrm{f}_{\nu;\mathrm{filter}} - 48.6
\label{eq_abmagnitudes}
\end{eqnarray}

However, there is a problem when trying to estimate line fluxes from a
single NB observation because the observed flux depends on the
wavelength of the line.  This problem can be understood from the left
panel of Fig.  \ref{fig:nb118:fig_sol_deltab} for the hypothetical
example of a line without continuum.  It is not possible to constrain
both $f_0$ and $\lambda_0$, leading to order of magnitude
uncertainties on the flux measurement. Further, the accuracy of the
central wavelength estimation is limited to 'somewhere within the
passband', which might be for 1\% filters, depending on the redshift,
not much improvement over state of the art photometric redshifts.

Only if the NB filter's passband was top-hat and if it was wide
compared to the typical line width, the precise value of $\lambda_0$
could be ignored for the flux estimation.  Effective top-hat filters
are however in the fast convergent beams of large survey telescopes
such as VISTA, CFHT, and UKIRT physically impossible (cf. also
Appendix \ref{app_nb118}).

%##############################
\subsection{Observations with several NB Filters}
\label{sec_mno}
%##############################

%%%%%%%%%%%
\begin{figure}
\resizebox{1.0\hsize}{!}{\includegraphics{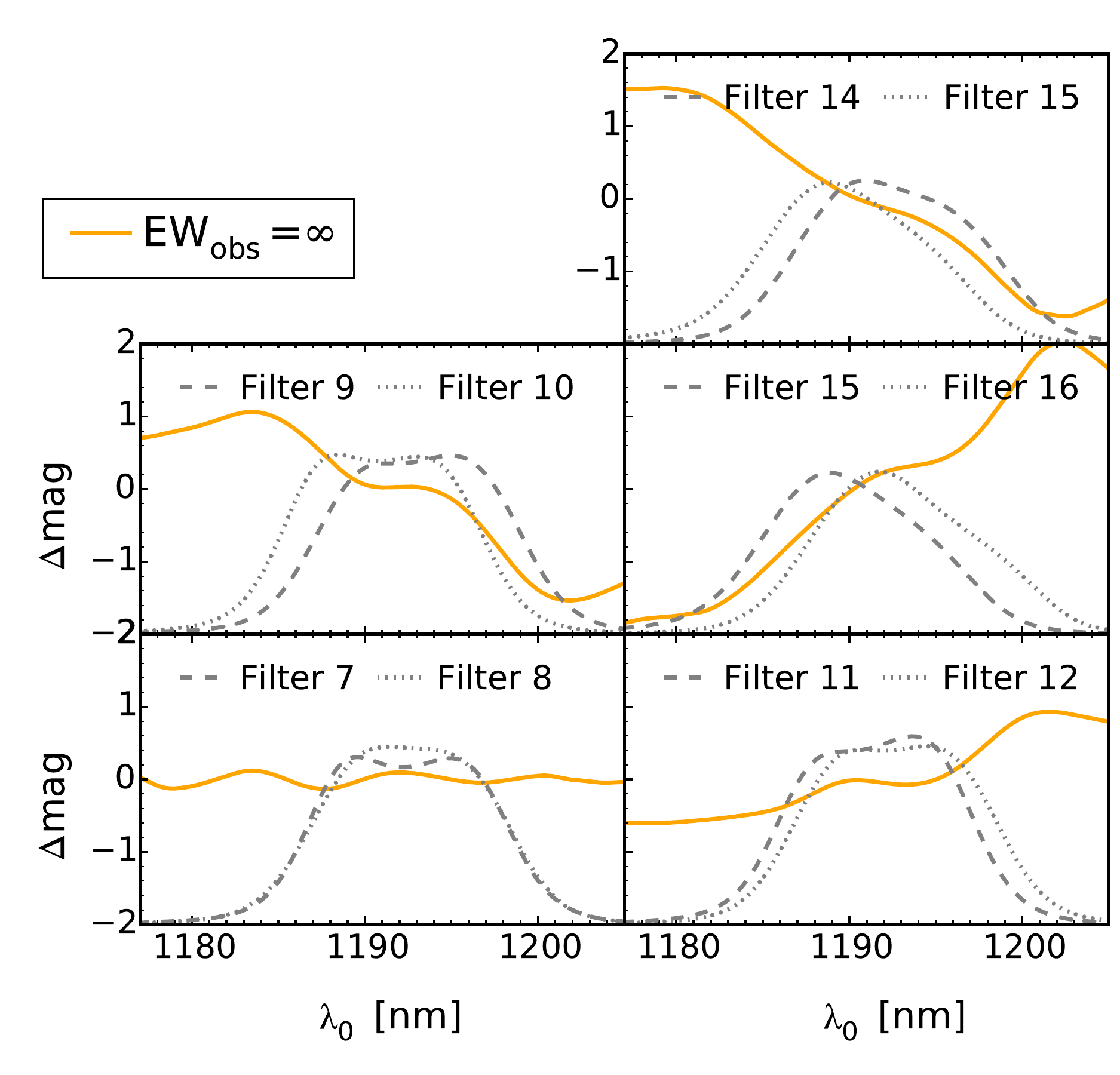}}
\caption{\label{fig:nb118:diff_nb118_combinations}
  $\Delta mag \mbox{--} \lambda_0$ curves for five different NB118
  combinations assuming an infinite $EW_\mathrm{obs}$ emission
  line. The passbands of the respective NB118 filters, for which the
  $\Delta mag \mbox{--} \lambda_0$ are shown, are also indicated in
  the different panels. The axis scaling for the filter passbands is
  linear.  }
\end{figure}

One way to solve the problem described in sec. \ref{sec:nb118:mot} is
to use observations in more than one, slightly differing NBs. Then,
two magnitude equations (eq. \ref{eq_abmagnitudes}) are available for
determining the two unknowns, $f_0$ and $\lambda_0$. $f_0$ can be
eliminated from the equation, which can subsequently be numerically
solved for $\lambda_0$ as a function of the measured magnitude
difference between the two filters, $\Delta\mbox{mag}$, as illustrated
in Fig.~\ref{fig:nb118:fig_sol_deltab}.

With a well matched pair of NB filters, it is possible to determine
the wavelength of a line within the range of wavelengths covered by
the passband of the NB filter very accurately.  Suitable filter pairs
are those that result in $\Delta mag$\mbox{--}$\lambda_0$ relations
that are monotonous and steep, allowing for a good wavelength
resolution, even when considering realistic uncertainties in
$\Delta mag$.

By means of such $\Delta mag \mbox{--} \lambda$ curves we
characterized the suitability of the various combinations between the
16 UltraVISTA NB118 filters for our method to determine the line
fluxes from two filters.  As discussed in more detail in
sec. \ref{sec:nb118:observingpattern:ultravista}, for 12 out of the
120 theoretically possible combinations of these 16 filters,
observations become directly available as part of the standard
UltraVISTA observations. One of these combinations is 14 \& 15, which
was used for illustration in Fig.  \ref{fig:nb118:fig_sol_deltab}.

While this specific pair is very well suited for the method, not all
the possible NB118 combinations are so in the same way. In Fig.
\ref{fig:nb118:diff_nb118_combinations} four further of the 12
relevant NB118 combinations are shown, where 15 \& 16 is as good, and
9 \& 10 almost as good as 14 \& 15.  The two filters 7 \& 8 are too
similar to be useful for the presented method, whereas in the
$\Delta mag$-$\lambda_0$ curve of 11 \& 12 still some information is
contained.  A quantitative characterization of the
$\Delta mag$-$\lambda_0$ curves for all 120 combinations is given in
Appendix \ref{sec:appendix:quantitativeall}.

\subsection{Continuum estimation}

In sec. \ref{sec_mno}, we have discussed the throughput variation
method neglecting the continuum. But the continuum contributes for
typical emission line galaxies significantly to the flux in NB
filters.  Hence, the magnitude differences measured for emission lines
with the same $f_0$ and $\lambda_0$ also depend on the equivalent
width, $EW_\mathrm{obs}$.

A measurement of the contribution of the continuum to the NB fluxes is
required; consequently, continuum-corrected NB magnitudes can be used
to estimate $f_0$ and $\lambda_0$ from the $\Delta mag$-$\lambda_0$
curves for infinite $EW_\mathrm{obs}$.  An accurate continuum estimate
at the wavelength of the NB filter could be obtained from measurements
in additional narrowband or mediumband filters bracketing the main NB
filter.  Typically, for reasons of time and cost efficiency, it is
resorted to BB filters. If a BB which has a passband covering the
emission line is included in the estimation, $f_0$ , $\lambda_0$, and
$EW_{obs}$ need to be estimated simultaneously and the
$\Delta mag$-$\lambda_0$ curves cannot be used directly. We describe a
statistical approach of fitting the parameters in the next section. 
%##############################
\section{Estimation algorithm}
\label{sec:nb118:algorithm}
%##############################

%##############################
\subsection{Concept}
\label{sec:nb118:algorithm_concept}
%##############################

$\Delta mag$ - $\lambda_0$ curves, as motivated in section
\ref{sec_mno}, allow to get a quick insight into the suitability of a
specific filter-combination for the throughput variations methods
(TPV).  In this section, we describe how to infer central wavelength,
$\lambda_0$, line flux, $f_0$, and equivalent width, $EW_\mathrm{obs}$, 
simultaneously using a statistical approach.

From the observation in the different filters, which can in principle include
more than two NB filters, a set $\textbf {M}$ of observed magnitudes,
$m_\mathrm{i}$, and estimates on their uncertainty, $\delta m_\mathrm{i}$ is
obtained. A model relates a set of parameters, $\bf{p}$, to a spectrum,
$f_\mathrm{\lambda}(\lambda;\bf{p})$, from which synthetically model magnitudes
in the relevant filters, $m_\mathrm{i;theo}[{\textbf{p}}]$, can be calculated.

Then, the parameter set being most probable under these data needs to
be found ($\bf{p}^\mathrm{est}$).  This is to minimize the Bayesian
posterior probability $\mathcal{P}(\textbf{p}\,|\,\textbf{M})$.
$\mathcal{P}(\textbf{p}|\,\textbf{M})$ is related to the Likelihood,
$\mathcal{L}(\,\textbf{M}\,;\,\textbf{p})$, and the prior on the
parameters, $\mathcal{PRI}(\textbf{p})$, by:

\begin{equation}
\mathcal{P}(\textbf{p}\,|\,\textbf{M})=
\alpha\;\mathcal{P}(\textbf{M}\, |\,\bf{p})\,\mathcal{PRI}(\textbf{p})
\end{equation}

$\alpha$ is a normalization constant.  As we are assuming the errors
in the measurements of the different magnitudes to be independent and
Gaussian, the total likelihood is given by the product of the normal
distributions for the individual measurements.  Consequently, the
negative log-likelihood is 1/2 the well known
$\chi^2$:

\begin{equation}
\chi^2 = \sum_i {\left ( \frac{m_\mathrm{i;obs} -
    m_\mathrm{i;theo}[{\textbf{p}}]}{\delta m_\mathrm{i}} \right ) }^2
\label{eq:nb118:chisquare}
\end{equation}

The weight of individual filters can be artificially decreased by
increasing the respective $\delta m_i$ in the calculation of
$\chi^2$. In practice we chose a minimum $\delta m_i$ for the
broadband filters, $\minestbb{}$, which we default to
$0.01\;\mathrm{mag}$. The implications from this choice are analyzed
in sec.  \ref{sec:nb118:simu_diff_est}.
 %As the uncertainties in the continuum estimation are expected to be low \cf{
 %}

%##############################
\subsection{Choice of input model}
\label{sec_method_inputmodel}
%##############################

%##############################
\subsubsection{Continuum shape}
\label{sec_method_broadband}
%##############################

Both the continuum shape and the ratios between the various emission lines will
differ from galaxy to galaxy, even so they are selected by the combination of
NB excess and photometric redshifts to be star-forming galaxies at a well
defined redshift.  If $f_\mathrm{\lambda}(\lambda;\bf{p})$ is
parametrized by eq. \ref{eq_model_simple}, educated guesses need to be made for
$\zeta_{\lambda}$ and $\mathcal{L}_{\lambda}$.  Ideally, these generic choices
for $\zeta_{\lambda}$ and $\mathcal{L}_{\lambda}$   approximate the range of
actual spectral energy distributions (SED) so well that the estimation of the
free parameters is not impacted.

While the simplest form for $\zeta_{\lambda}(\lambda)$ is a continuum
flat in $f_\nu$ or $f_\lambda$, a useful first order correction is to add an
additional parameter in form of the power law slope $\beta$
($\zeta_{\lambda}(\lambda;\beta) \propto \lambda^\beta$). This inclusion of the
slope is especially relevant when the NB is off-centre from the BB passband,
as it is the case for NB118 and J (cf. Fig. \ref{fig:nb118_y_j}).

A power law continuum will not be sufficient if the wavelength range around the
considered line includes a strong spectral break, like the $4000\,\mbox{\AA{}}$
break in the case of \fion{O}{ii}$\,\lambda3727$, but is a good assumption for
\ha{} in the NB118 filters, as discussed in sec.
\ref{sec:nb118:simobs:qual_cont}. We will address the problematic of the break
on selection and measurement of \fion{O}{ii} emitters in the NB118 data as part
of a forthcoming publication.

\subsubsection{Line shape and \fion{N}{ii} contribution}
\label{subsec:specshape}
%##############################

%%%%%%%%%%%
\begin{figure}[h]
\resizebox{1.0\hsize}{!}{\includegraphics{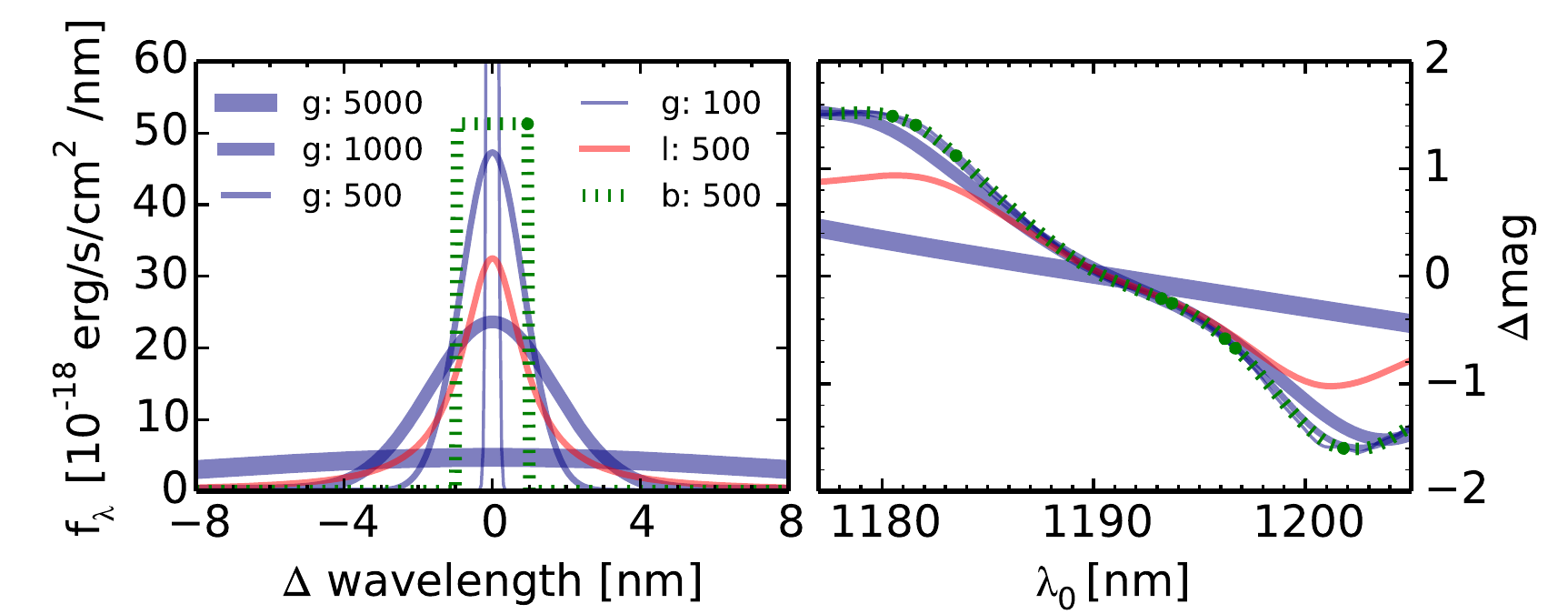}}
\resizebox{1.0\hsize}{!}{\includegraphics{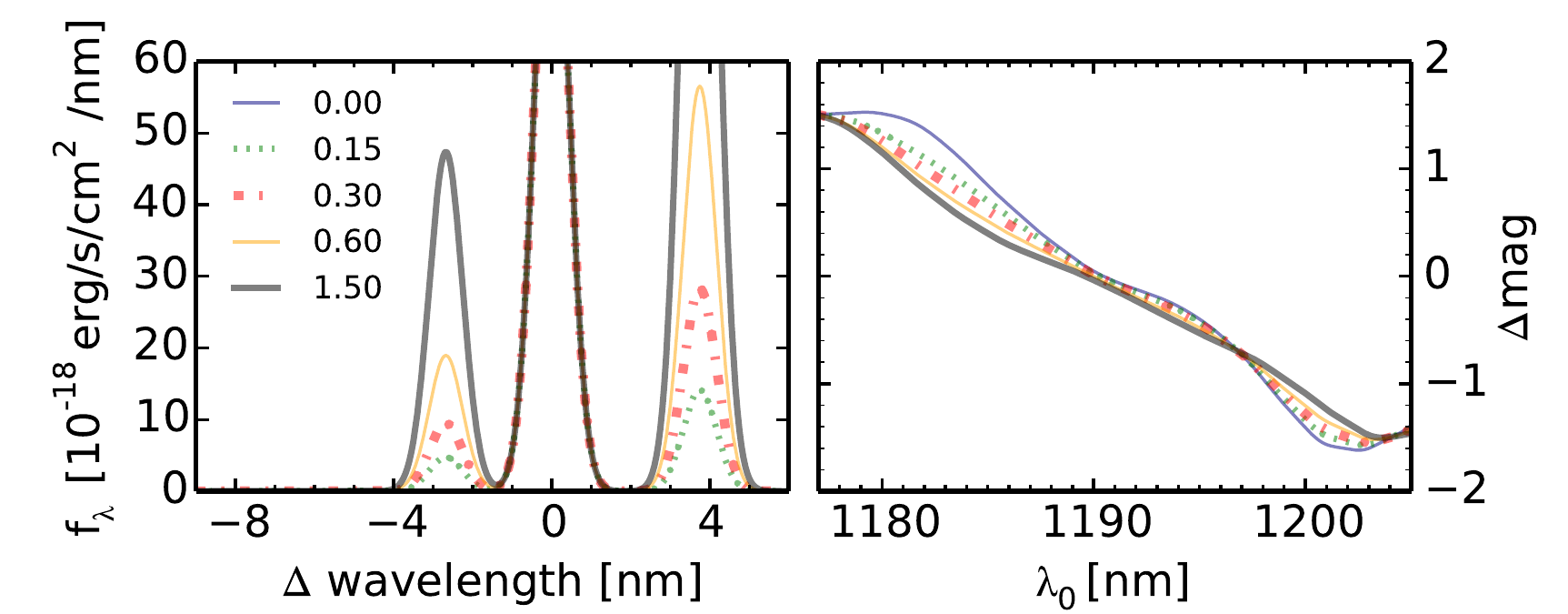}}
\caption{\label{fig_deltamag_diff} Impact of the line shape (upper
  panels) and the presence of multiple lines (lower panels) on the
  predicted $\Delta mag$ - $\lambda_0$ curves. The left panels show
  different input spectra, and the right panels the corresponding
  $\Delta mag$ - $\lambda_0$ curves.  'g', 'l', and 'b' in the legend
  refer to Gaussian, Boxcar, and Lorentzian, respectively, with the
  succeeding numbers stating the line $FWHM$ in
  $\mathrm{km}\;\mathrm{s}^{-1}$.  The presence of multiple lines is
  demonstrated for the relevant example of the \fion{N}{ii} doublet
  bracketing \ha{}, with the values in the legend corresponding to the
  assumed \fion{N}{ii}$\,\lambda6583$ to \ha{} ratios.  All shown
  results are based on the same filter combination as used for Fig.
  \ref{fig:nb118:fig_sol_deltab}.}
\end{figure}
%%%%%%%%%%%

When observing \ha{} in a NB filter, there is contamination from the
collisional excited, forbidden \fion{N}{ii} lines at $655.0$ and
$658.5\,\mbox{nm}$, which have to be included in
$\mathcal{L}_{\lambda}$.  In the NB118 filters, where \ha{} is
observed at a redshift of $z \sim 0.81$, the lines are at a difference
of $2.7\,\mbox{nm}$ and $3.6\,\mbox{nm}$ from \ha{}, respectively.
While the ratio between the fluxes in the two \fion{N}{ii} lines,
$f_{\mathrm{[NII]}\lambda6583}/f_{\mathrm{[NII]}\lambda6548}$, is
theoretically fixed to $\sim3$ \citep[][p.61]{Osterbrock:1989}, the
flux ratio $w_{6583} = F_{\mathrm{[NII]}\lambda 6583}/F_{H\alpha}$
does substantially differ from galaxy to galaxy. Values vary for pure
star formation depending on the metallicity and the ionization
parameter between 0.0 and almost 1.0 \citep[e.g.][]{Kewley:2002:35},
with even higher ratios possible for spectra with AGN contribution
\citep[e.g.][]{Kauffmann:2003:1055}. Based on low redshift data, a
typical value has been shown to be $w_{6583} = 0.3$ (e.g.
\citealt{Pascual:2007:30} and references therein).  In the absence of
knowledge about the metallicities, correlations between $w_{6583}$ and
$EW$ or mass can be used for a more sophisticated estimate.  Using
such relations in our estimation might bring some improvement, which
we consider for further investigation. Throughout this work we assume
a ratio of 0.3.

A wrong assumption on $w_{6583}$ naturally impacts both the $f_0$ and
$\lambda_0$ estimation, where the impact on the latter can be examined
by comparing $\Delta mag \mbox{--} \lambda_0$ curves for different
$w_{6583}$.  Curves for five different $w_{6583}$ between $0.0$ and
$1.5$ are shown for the example filter combination 14 \& 15 in the
lower panel of Fig.  \ref{fig_deltamag_diff}. We can conclude, while
the impact is not negligible, the resulting systematic wavelength
errors are even in the worst case not larger than
$\sim3\,\mathrm{nm}$, compared to the assumption of $w_{6583} = 0.3$.

In addition to $w_{6583}$, also the shape of the \ha{} line could
differ.\footnote{We assume that the \fion{N}{ii} lines have the same
  width as the \ha{} line.} Whereas line widths beyond
$1000\,\mathrm{km}\,\mathrm{s}^{-1}$ are not expected for solely
star-forming galaxies, line widths of several
$1000\,\mathrm{km}\,\mathrm{s}^{-1}$ are possible when originating
from a type I AGN. AGNs can be recognized by the use of the extensive
multi-wavelength data available in fields like COSMOS.

We tested the impact of the line width on the $\lambda_0$ estimation
by determining $\Delta mag \mbox{--} \lambda_0$ curves for Gaussians
with $FWHM$ of 100, 500, 1000, $5000\,\mathrm{km}\,\mathrm{s}^{-1}$,
as shown in the upper part of Fig.  \ref{fig_deltamag_diff}. While a
line with a width of several $1000\,\mathrm{km}\,\mathrm{s}^{-1}$
skews the result as expected, the difference all the way between $100$
and $1000\,\mathrm{km}\,\mathrm{s}^{-1}$ does not strongly impact the
parameter estimation.  Only, if the line had broader wings than a
Gaussian, the deviations will be visible at lower $FWHM$.  A
$500\,\mathrm{km}\,\mathrm{s}^{-1}$ boxcar line gives essentially an
identical $\Delta mag \mbox{--} \lambda_0$ curve as a Gaussian with
the same $FWHM$, but a Lorentzian will cause stronger deviations from
the estimated wavelength especially at low filter transmittances.
In the following, we assume throughout a default $FWHM$ of $250\,\mathrm
{km}\,\mathrm{s}^{-1}$. The impact of this specific choice is not expected to be large, as can be concluded from the results in this section.

\subsection{Choice of broadband filters}
\label{sec:nb118:choiceofbroad}

In order to put an observational constraint on the continuum slope
$\beta$, at least two flanking BBs need to be used.  This is in the
case of the VISTA NB118 filters in addition to J naturally Y, but also
H might be included, allowing for a stronger constraint on
$\beta$. For the analysis presented in the following we use all three
filters.

As Y and H alone are theoretically enough to constrain a power law
continuum, and any strong emission line in the NB118 filter also
contributes to the flux in J, J can in principle be used as an
estimator of the emission line flux.  Whether it is actually feasible
to obtain an accurate estimate from the J excess depends on two main
criteria:

First, the contribution of the line to the filter-averaged signal needs to be
high enough compared to the total noise in J. For the
\ultvis{} DR2 data the S/N for an infinite EW line is in J a factor 4.1 lower
than e.g. at the peak of NB118 filter 15 (For more details see Appendix
\ref{sec:nb118:app:sn}).  Therefore, for the faintest lines detectable in the
NB filter, the J-based estimate will not be very useful.

Secondly, it needs to be possible to determine the continuum contribution to
$J$ with very high precision. For example, an $EW_\mathrm{obs}=10\,\mathrm{nm}$
line has a J excess of only $0.06\,\mathrm{mag}$, while the same object causes
an excess of $0.64\,\mathrm{mag}$ at the peak of NB118 filter 15.  Hence,
deviations between estimation power law and actual continuum SED are for fluxes
measured through the broadband excess more critical.

%##############################
\subsection{Implementation of estimation code}
\label{sec:nb118:paraest::method}
%##############################

%%%%%%%%%%%
\begin{table}
  \caption{\label{tab:parapregrid} Range of parameters used for the pre-grid in the parameter estimation with the Nelder-Mead implementation of our code.}
\begin{center}
\begin{tabular}{ccccc}
\hline
&  $\lambda_0$ \tablefootmark{a} & $\log_{10}(f_0$\tablefootmark{b} ) & $log_{10}(EW_{obs}$\tablefootmark{a} ) & $\beta$ \\
\hline
\hline
start & 1170 & -17 & -1   &  -3 \\
end   & 1210 & -15 & 2.5  &   5 \\
steps & 15   &  10 & 10   &   8 \\
\hline
\end{tabular}
\end{center}
\tablefoot{
\tablefoottext{a}{$[\mathrm{nm}]$}
\tablefoottext{b}{$[\mathrm{erg}\;\mathrm{s}^{-1}\;\mathrm{cm}^{-2}]$}
}
\end{table}
%%%%%%%%%%%

Finding the right, global, maximum in a complicated probability landscape is
not always easy. We implemented two versions of a \emph{python} code to
determine the parameter-set ${\bf p}^{est}$ maximizing the posterior
probability for a given set of measured magnitudes, $\textbf{M}$.

One version of our code makes use of the implementation of the Nelder-Mead
algorithm
\citep{Nelder:1965:308} within
\emph{python/scipy.minimize}. Using the downhill simplex method, it
allows to efficiently determine the maximum.\footnote{More precisely, the code
searches for the minimum of the negative log-probability.} In order to improve
the success rate of finding the global maximum, we combined it with a
pre-calculation on a coarse grained grid. The parameter-range of this pre-grid
is given in Table \ref{tab:parapregrid}.  Then, we run the Nelder-Mead
algorithm three times, starting from the grid points with the three highest
probabilities. After comparing the found posterior probability from each run,
we assumed in case of differing maxima the found maximum with the highest
probablility as the global maximum and designated the corresponding parameters
as ${\bf p}^{est}$. If all three sub-runs were not converging, or if
$\lambda_0$ for the best-fit parameter set was within $1\,\mathrm{nm}$ of the
prior boundaries, we assumed the parameter estimation as failed.

The Nelder-Mead implementation is relatively efficient, but it does not allow
for a direct assessment of the credibility intervals.  When having actual
observations, where the experiment cannot be repeated as in simulations, the
correct way to state uncertainties is to determine the posterior probabilities,
e.g. based on a Markov chain Monte Carlo (MCMC) approach.  Therefore, we
implemented also a version based on the Metropolis-Hastings algorithm
\citep{Metropolis:1953:1087,Hastings:1970}. The code makes use of adaptive
proposal distributions based on repetitively recalculated covariance matrices,
allowing for optimal efficiency \citep{Roberts:2001:351,Rosenthal:2014:93}.  We
ran six separate chains with 100k proposal steps in each chain, where the
initial proposal distribution was chosen wide enough to explore the complete
parameter space.

Altogether, the set of four free fitting parameters, $\textbf{p}$,
does include the central wavelength of the main line, $\lambda_0$, its
integrated flux $f_0$, the observed frame equivalent width,
$EW_\mathrm{obs}$, for the main line, and the slope $\beta$ of the
continuum.  Within the prior, we constrained $\lineflux$ to a
physically reasonable range in order to exclude combinations of
unreasonably high $f_0$ and $\lambda_0$ at low transmittances of the
filters.

The range of acceptable $\bf{p}$ was set for our specific test case of
\ha{} in the NB118 filters to
$1171\,\mathrm{nm} < \lambda_0 < 1206 \,\mathrm{nm}$ and
$0.5\times10^{-17}\,\mathrm{erg}\,\mathrm{s}^{-1}\,\mathrm{cm}^{-2} <
f_0 < 100 \times
10^{-17}\,\mathrm{erg}\,\mathrm{s}^{-1}\,\mathrm{cm}^{-2}$,
respectively.  Further, we constrained
$0\,\mathrm{nm} < EW_\mathrm{obs} < 300\,\mathrm{nm}$.

Assuming a flat prior for the line flux is from a rigorous Bayesian point of
view not the right choice, as this does not reflect our complete state of prior
knowledge.  The line-luminosities are known to be approximately distributed by
a Schechter luminosity function (LF) (\citealt{Schechter:1976:297}; z=0.8 \ha{}
LFs: e.g. \citealt{Villar:2008:169,Sobral:2009:75, Ly:2011:109}). However, for
clarity of the results we still use flat priors in this work.

%##############################

%##############################
\section{Application to simulated observations}
%##############################
\label{sec:nb118:simobs}

%##############################
\subsection{Mock observations}
\label{sec:nb118:mockobs}
%##############################

For testing the proposed TPV method systematically, we were using in
this first part simulated observations for a range of spectra.  The
inputs into the simulation were chosen to closely resemble the
available \ultvis{} DR2 data.  This means that we used as inputs
VIRCAM filter curves, the characteristics of the VIRCAM IR-arrays
(gain, zeropoint (ZP)), and realistic sky brightnesses in the NB118 and the three
VIRCAM BB filters.

The sky-brightnesses in the individual NB118 filters were taken from 
\citet{Milvang-Jensen:2013:94} and significantly differ between the individual copies,
ranging from $21.2$ to $51.5\,\mathrm{e}^{-}\,\mathrm{s}^{-1}\,\mathrm{pixel}^
{-1}$. For Y, J, and H we assumed 150, 650, $4700\,\mathrm{e}^{-}\,\mathrm{s}^
{-1}\,\mathrm{pixel}^{-1}$, respectively, values typical for the \ultvis{}
observations.  We set the detector gains to $4.2\,\mathrm{e}^{-}\,\mathrm{ADU}^
{-1}$ and the ZPs on the AB system to 21.78, 24.12, 24.73, and 25.29 for NB118,
Y, J, and H, respectively.\footnote{Based on Vega ZPs from 
\url{http://casu.ast.cam.ac.uk/surveys-projects/vista/technical/photometric-properties} and Vega to AB corrections calculated by us for the assumed filter curves.}

Further, we were assuming point sources observed in 2\arcsec{}
diameter circular apertures.  The corresponding enclosed flux fraction
within the aperture is about $75\%$ for the \ultvis{} NB118 and J
PSFs. For simplicity, the same enclosed fraction was also used for Y
and H, even though the PSFs in these filters slightly differ for the
actual observations.

Based on a chosen spectrum and these inputs, we synthetically calculated with
eq. \ref{eq_abmagnitudes} the expected aperture magnitudes,
$m_\mathrm{expect}$, and from the corresponding signal-to-noise ratio (S/N) we
derived by $\Delta m_\mathrm{expect} \approx\,1.086 \frac{1}{S/N} \mathrm
{mag}$ the expected magnitude errors.

The $S/N$ was estimated from the CCD equation
\citep[e.g.~][]{Howell:2000,Chromey:2010}.  Neglecting justifiably
uncertainties from read-out noise, dark-current, linearity corrections, flat
fielding, background subtraction, and converting electrons through the gain,
$g$, to digital-units, $DN$, the $S/N$ can be calculated as:

\begin{equation}
  S/N =  \frac{\dot{N}_\mathrm{DN}^* \sqrt{g}\sqrt{t}}{\sqrt{\dot{N}_\mathrm{DN}^* + n_\mathrm{pix} \dot{b}_\mathrm{DN}^*}}
\label{eq_snrfinal}
\end{equation}
\noindent
Here $\dot{N}_\mathrm{DN}^*$ is the total number of
$DN$ per second created within the aperture due to the source, while
$\dot{b}_\mathrm{DN}^*$ is the number of $DN$ produced per pixel and
second by the sky-background. $n_\mathrm{pix}$ is the number of pixels
constituting the aperture and $t$ is the exposure time.

We assumed an observation time of $11.4\,\mathrm{hr}$ in each of two
NB118 filters through which an object is simulated to be observed,
which sums to the typical per-pixel integration time available in the
\ultvis{} DR2. The expected per-pixel integration time in the finished
survey will be $112\,\mathrm{hr}$, meaning that the DR2 NB118 data
includes only $\sim20$\% in time or 45\% in depth of the final
\ultvis{} survey goal.  Similarly, we assumed for Y, J, H the 53.2,
34.9, 29.4 hrs available in the DR2 for the same field as the NB118
data, with the time in the finalized survey expected to be
$210\,\mathrm{hr}$ in each of the BB filters, all per pixel.

%%%%%%%%%%%
\begin{table}
        \caption{Grid of spectral energy distributions used in this work for testing biases from the continuum estimation and for SED fitting. The grid used for the SED fitting is given in brackets.   }

        \begin{tabular}{ll}
                \hline
                Parameter & Values \\
		\hline
		\hline
                Stellar population: & BC03 \citep{Bruzual:2003:1000} based on \\
		& STELIB \citep{LeBorgne:2003:433}\\
		& (BC03 based on BaSeL3.1 \\
		& \citep{Westera:2002:524}) \\
                Nebular emission:  &  Recipe based on \citet{Schaerer:2009eq}; \\
                & \citet{Ono:2010:1524} \\
		IMF & Salpeter $0.1-100\,\mathrm{M}_\odot$ \\
		Metallicities Z\tablefootmark{a}& 0.008, 0.02, 0.05 (0.004, 0.008, 0.02, 0.05)\\
		$\log_{10}({Age[yr]})$\tablefootmark{b}  &   $6\mbox{--}9.9$ with 14 steps \\
		 &  ($7\mbox{--}9.8$ in steps of 0.1)\\
		$f_{cov}$\tablefootmark{c} &  0 (0.7,1)  \\
                $E_\mathrm{S}(B-V)$\tablefootmark{d}      &   0.2  ($0.0\mbox{--}1.0$ in steps of 0.02)      \\
                SFH: $\log_{10}(\tau[yr])$   &  $8,\infty$ ($7.9\mbox{--}10.5$ in steps of 0.2)              \\
        $\alpha$\tablefootmark{e} & None (1.75) \\
                \hline
        \end{tabular}

	\label{tab:sspinput}
	\tablefoot{
	    \tablefoottext{a}{$Z_\odot = 0.02$}
		\tablefoottext{b}{Age of the universe at $z=0.81$ was $6.59\;\mathrm{Gyr}$  ($\log_{10}({Age[yr]}) = 9.82$)}
		\tablefoottext{c}{We assume that the covering fraction of the gas, $f_\mathrm{cov}$, is related to the escape of the ionizing radiation, $f_\mathrm{esc;ion}$, through $f_\mathrm{esc;ion} = 1-f_\mathrm{cov}$}
		\tablefoottext{d}{Extinction assuming \citet{Calzetti:2000:682}
		extinction law. The stellar extinction is assumed to be a factor 0.7
		lower than the nebular extinction.}
		\tablefoottext{e}{Parameter $\alpha$ of \citet{Dale:2002:159} models used for dust emission.}
	   }
\end{table}
%%%%%%%%%%%

While we tested correctness and stability of our estimation code by using
spectra as input models, which could be exactly matched by the estimation
model, the mock observations used in the following were based on realistic
galaxy SEDs. For the tests we used the high resolution BC03 models
\citep{Bruzual:2003:1000} based on the $R\sim2000$ STELIB
\citep{LeBorgne:2003:433} library assuming a Salpeter IMF ($M =
0.1-100\,M_\odot$, \citealt{Salpeter:1955:161}).

%##############################
\subsection{Quality of continuum estimation from simulations}
\label{sec:nb118:simobs:qual_cont}
%##############################

%%%%%%%%%%%
\begin{figure*}
\centering
\resizebox{0.8\hsize}{!}{\includegraphics{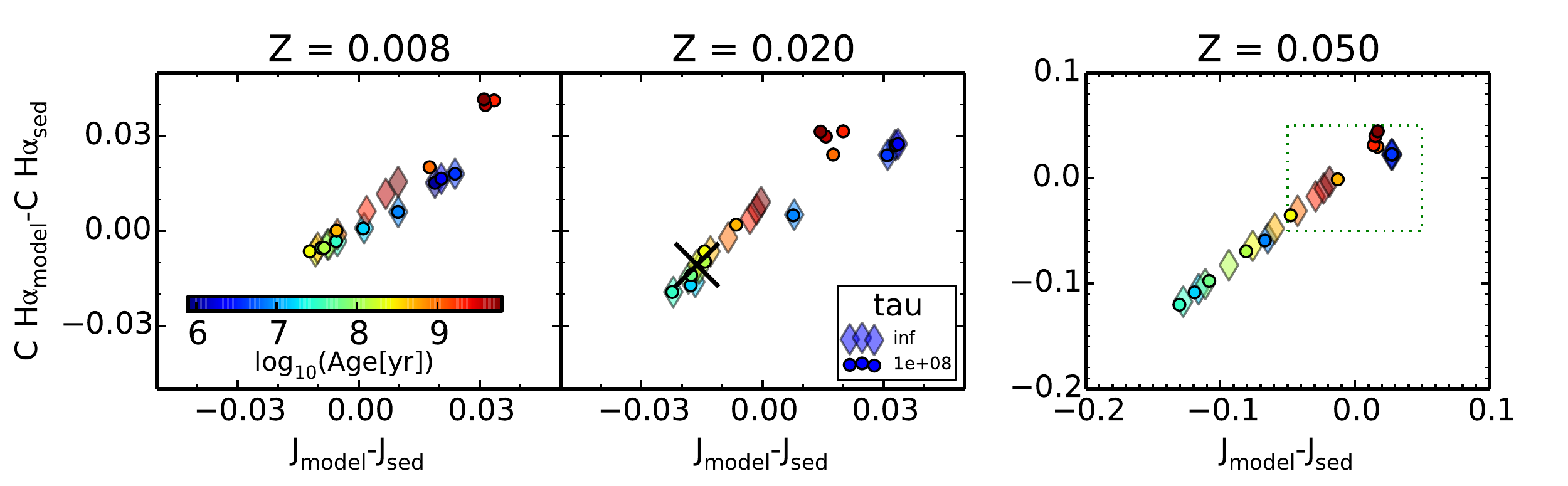}}
\caption{\label{fig:nb118:contfit} Difference between the
  continuum of synthetic spectral energy distributions (SED) and the estimate
  obtained from a power law fit to these SEDs (model) using Y, J, and H
  filters.  The magnitude difference between $J_{model}$ and $J_{sed}$ is
  plotted against the magnitude difference between SED and model directly at
  the wavelength of \ha{} (cf. sec.  \ref{sec:nb118:simobs:qual_cont}).  The
  range of SED parameters, for which results are shown, is summarized in Table
  \ref{tab:parapregrid}.  For the $Z=0.05$ panel, a wider
  axis scale was chosen than in the other two cases. The narrower
  range used for $Z=0.008$ and $Z=0.020=Z_\odot$ is indicated as dotted box in
  the $Z=0.05$ case. The cross in the $Z=0.020$ plot represents the SED which
  was used for the parameter estimation test described in sec. \ref{sec_test}.}
  \end{figure*}
%%%%%%%%%%%

In the TPV estimation we use three BB filters, Y, J, H. This means that a
relatively large wavelength range is included. The range from the blue end of Y
to the red end of H corresponds at $z = 0.81$ to a rest-frame wavelength range
from $540$ to $990\,\mathrm{nm}$.

When estimating $f_0$ and $\lambda_0$ from the throughput variations, the
constraints on the continuum at the wavelength of the NB filter need to be
precise. Further, as \ha{} is also contributing to J, an excess in J impacts
the estimation results with the algorithm described in sec.
\ref{sec:nb118:algorithm_concept}. Hence, the continuum needs also to be
precisely estimated averaged over the complete J passband.

Therefore, it is important to carefully assess the expected deviations between
the power law fit and the actual continua both in NB118 and
J. Features in the spectral energy distributions (SEDs) could in principle
results in such deviations. These are not necessarily of the
same extent exactly at the wavelength of \ha{} in the NB118 filter and
averaged over J, especially as the NB118 passband is at the blue end of the
J passband.

While no strong spectral features are expected over the covered wavelength
range, only a formal test can give a clear answer. Therefore, we fit power law
continua to synthetic magnitudes calculated for a grid of model SEDs without
nebular emission included (Table \ref{tab:sspinput}). For consistency, we had
scaled all input model SEDs to $f_\lambda = 1\times10^{-17}
\,\mathrm{erg}\;\mathrm{s}^{-1}\;\mathrm{cm}^{-2}\;\mathrm{nm}^{-1}$
at the wavelength of \ha{}, before calculating input magnitudes and
uncertainties as in sec. \ref{sec:nb118:mockobs}.

In Fig. \ref{fig:nb118:contfit} the difference between the continuum magnitude
at \ha{} from the best-fit power law continuum, $C\,H\alpha_{model}$, and that
measured directly from the input SED, $C\,H\alpha_{sed}$, is shown for a range
of SEDs. $C\,H\alpha_{sed}$ was determined by averaging the flux-densities over
two $8\,\mathrm{nm}$ wide intervals at $1172\,\mathrm{nm}$ and
$1205\,\mathrm{nm}$, corresponding to rest-frame wavelengths of
$647.5\,\mathrm{nm}$ and $665.7\,\mathrm{nm}$, respectively. These were chosen
to exclude \ha{} absorption. On the other axis, the differences between input
and estimation are shown for $J$.  We conclude that the deviations averaged
over the J passband are of similar extent as those for the continuum magnitudes
directly at \ha{}.

Both for stellar populations with solar and sub-solar (Z=0.4$\,Z_\odot$)
metallicity these deviations are very small. Except for very young ages and for
populations without ongoing star formation, the differences between fit and
input magnitude are below $0.02\,\mathrm{mag}$.  On the other hand, strong
deviations exist for super-solar metallicities, exceeding $0.1\,\mathrm{mag}$
for expected population ages.  While this is cause for some concern,
star-forming galaxies with stellar metallicity as high as $Z=2.5\,Z_\odot$ are
expected to be very rare \citep[e.g.][at $z = 0.7$]{Gallazzi:2014:72}, even at
highest masses.  Most of the NB selected galaxies will have stellar masses
$\lesssim 10^{10}\,M_\odot$ \citep[e.g.][]{Kochiashvili:2015} and hence not be
among the most metal rich systems according to the mass-metallicity relations
(e.g. \citealt{Tremonti:2004:898}; at $z\sim0.7$:
\citealt{Savaglio:2005:260, Lamareille:2009:53}).  \footnote{Typically stellar
  metallicities are, at least for local galaxies, $0.5\,\mathrm{dex}$ below the
  gas-phase metallicity
  \citep{Gallazzi:2005:41}.}

We can conclude that the continuum estimate is expected to be excellent, making
both the TPV and, at least to some extent, also the use of the J excess
feasible.

%##############################
\subsection{Quality of full parameter estimation from simulations}
\label{sec_test}
%##############################

We assessed the expected accuracy of \ha{} estimations at different wavelengths
for different $f_0$ and $EW$, different filter combinations, and different
assumptions in the estimation algorithm. For this purpose we created in each
considered case 500 realizations of observed magnitudes: we calculated for a
given input spectrum the synthetic magnitudes and uncertainties, and randomly
perturbed each of the magnitudes.\footnote{More precisely, we added the random 
noise to the flux densities.}

Then, we ran our Nelder-Mead TPV code (cf. sec.
\ref{sec:nb118:paraest::method}) on each of the realizations.  In this way, we
found the expected distribution of best fit parameters for hypothetical
repeated observations of the same object.  Finally, the determined 4d
distributions in the $\lambda_0\mbox{--}f_0\mbox{--}EW_{obs}\mbox{--}\beta$
space
were reduced to 1d distributions for each of these four parameters by
marginalization over the three other parameters.

In all cases we used as input continuum a constantly star-forming solar
metallicity SED with $E(B-V)=0.2$ and an age of $3\times10^{8}\,\mathrm{yr}$.
This SED might be considered as a typical example for the \ha{} emitters
selected in the UltraVISTA data. The expected offset between power law
continuum and this model continuum is marked in Fig. \ref{fig:nb118:contfit} as
a cross.  It is important to note, that this chosen continuum SED has an \ha{}
absorption $EW_{obs}$ of $0.7\,\mathrm{nm}$ at $z = 0.81$.\footnote{Measured
from the SED with
  \emph{IRAF}/\emph{splot}} Therefore, even perfectly estimated \ha{}
emission fluxes, $f_{0;est}$, are too low by 16\%, 8\%, 5\%, 3\%, and
1\% for \ha{} emission with $EW_{obs}$ of 4, 7, 10, 20, and
$100\;\mathrm{nm}$, respectively, as the TPV code does not correct for
the \ha{} absorption.\footnote{The factors were calculated assuming an
  additional \fion{N}{ii} contribution with $w_{6583}=0.3$.  As the
  measured contribution of the \fion{N}{ii} lines is for the actual
  NB118 filters redshift dependent, the expected underestimation
  factors also depend somewhat on the redshift.}  We added an \ha{}
line with chosen $f_0$ and $w_{6583}=0.3$ to this continuum. The
continuum was scaled so that the added line has a specific \ha{}
$EW_{obs}$. We remark that adding \ha{} to a somewhat arbitrary
continuum is not completely self-consistent, but using the same
continuum allows for a clear comparison of the results.

Line fluxes refer in the simulation parts of this paper to total fluxes, even
though all measurements and estimations are simulated to be performed within
the 2\arcsec{} apertures.

%##############################
\subsubsection{Different line parameters}
\label{sec:simu_diff_pfree}
%##############################

%%%%%%%%%%%
\begin{table*}
  \caption{\label{tab:est_simu_results}
    Mean and standard deviations of best estimates for \ha{} fluxes $f_{0;est}$
    and wavelengths $\lambda_{0;est}$ using the TPV on repeated simulations of
    observations of the same \ha{} emitters at $z=0.81$. More precisely, the
    values are for the distributions of $(f_{0;est}-f_{0;in})/f_{0;in}$ and
    $(\lambda_{0;est}-\lambda_{0;in})$, where 'in' is referring to the true
    input values. Results are stated for different $f_{0;in}$, different
    $EW_\mathrm{obs;in}$ (cf. sec. \ref{sec:simu_diff_pfree}), different
    assumptions in the estimation (cf. sec. \ref{sec:nb118:simu_diff_est}), and
    different filter combinations (cf. sec. \ref{sec:nb118:simu_diff_filter}).
    For each of the assumptions, results are given at the combined effective
    filter's mean wavelength (A) and three different wavelengths redwards of it
    (B, C, D). At 'C'  and 'D' the combined effective filter's wavelength has
    dropped to 50\% and 20\% of its peak value, respectively. 'B' is halfway
    between the 'A' and 'C' wavelengths. Results for input parameters, where
    objects are not selected by the NB118 selection criteria in sec.
    \ref{sec:selcriteria} are in brackets. The full distributions, to which the
    stated mean and standard deviation correspond, are shown for 'A' and 'C' in
    Fig. \ref{fig:nb118:results_standard_test_diff_flux} $\mbox{--}$ Fig.
    \ref{fig:results_standard_test_diff_filters}. }
\setlength{\tabcolsep}{3pt}
\begin{tabular}{r|cccc|c|cccc}
\hline

  & &\multicolumn{2}{c}{$(f_{0;est}-f_{0;in})/f_{0;in}$} & & & \multicolumn{4}{c}{$(\lambda_{0;est}-\lambda_{0;in})\;[nm]$}\\
    %  & \multicolumn{8}{c}{$\lambda_{0;\mathrm{in}} - \lambda_{0;\mathrm{peak}}$ [nm]}\\
      \hline
      \hline
  &	A  & B  & C & D &  & A & B & C & D \\
   \hline
   \hline
   \\[-1Ex]
   & \multicolumn{9}{c}{Different input fluxes (Fig. \ref{fig:nb118:results_standard_test_diff_flux})} \\[1Ex]
   \hline
3\tablefootmark{a}    &	($0.07\pm0.28)$  & ($-0.01\pm0.32)$  & ($-0.11\pm0.41)$  & ($-0.18\pm0.45)$   &  &$(0.11\pm2.88)$  & $(-1.50\pm3.47)$  & $(-2.68\pm4.63)$  & $(-5.09\pm7.37)$   \\
5\tablefootmark{a}    &	$-0.04\pm0.13$  & $-0.11\pm0.19$  & ($-0.18\pm0.26)$  & ($-0.27\pm0.32)$   &  &$-0.34\pm1.79$  & $-1.27\pm2.35$  & $(-1.78\pm3.07)$  & $(-2.84\pm5.40)$   \\
8\tablefootmark{a}    &	$-0.07\pm0.07$  & $-0.14\pm0.12$  & $-0.24\pm0.16$  & ($-0.32\pm0.23)$   &  &$-0.27\pm1.13$  & $-0.96\pm1.49$  & $-1.14\pm1.38$  & $(-9.12\pm9.05)$   \\
15\tablefootmark{a}   &	$-0.08\pm0.04$  & $-0.14\pm0.07$  & $-0.23\pm0.10$  & $-0.33\pm0.12$   &  &$-0.15\pm0.67$  & $-0.70\pm0.89$  & $-0.81\pm0.70$  & $-0.70\pm0.70$   \\
30\tablefootmark{a}   &	$-0.07\pm0.02$  & $-0.08\pm0.04$  & $-0.14\pm0.06$  & $-0.29\pm0.07$   &  &$0.00\pm0.39$  & $-0.10\pm0.43$  & $-0.25\pm0.36$  & $-0.53\pm0.31$   \\
\hline
\\[-1Ex]
   & \multicolumn{9}{c}{Different input EW (Fig. \ref{fig:nb118:results_standard_test_diff_ew})} \\[1Ex]
\hline
4\tablefootmark{b} &	$-0.16\pm0.06$  & $-0.22\pm0.11$  & ($-0.36\pm0.14)$  & ($-0.64\pm0.14)$   &  &$-0.08\pm1.10$  & $-0.70\pm1.37$  & $(-1.14\pm1.40)$  & $(-2.35\pm2.63)$  \\
7\tablefootmark{b} &	$-0.10\pm0.05$  & $-0.18\pm0.09$  & $-0.30\pm0.13$  & ($-0.44\pm0.16)$   &  &$-0.32\pm0.90$  & $-1.00\pm1.17$  & $-1.22\pm1.14$  & $(-1.38\pm2.34)$  \\
10\tablefootmark{b} &	$-0.08\pm0.05$  & $-0.15\pm0.10$  & $-0.23\pm0.13$  & ($-0.32\pm0.17)$   &  &$-0.22\pm0.92$  & $-0.99\pm1.18$  & $-0.96\pm1.06$  & $(-1.02\pm2.58)$  \\
20\tablefootmark{b} &	$-0.03\pm0.06$  & $-0.07\pm0.11$  & $-0.12\pm0.14$  & ($-0.16\pm0.17)$   &  &$-0.07\pm0.96$  & $-0.47\pm1.16$  & $-0.50\pm0.89$  & $(-0.69\pm2.41)$  \\
100\tablefootmark{b} &	$0.01\pm0.07$  & $-0.01\pm0.11$  & $-0.03\pm0.15$  & ($-0.05\pm0.18)$   &  &$-0.01\pm1.10$  & $-0.24\pm1.21$  & $-0.27\pm0.90$  & $(-1.04\pm3.64)$  \\
\hline
\\[-1Ex]
   & \multicolumn{9}{c}{Different assumptions in the estimation (Fig. \ref{fig:results_standard_test_diff_biases}) } \\[1Ex]
\hline
$ 0.01$\tablefootmark{c} &	$-0.08\pm0.05$  & $-0.15\pm0.10$  & $-0.23\pm0.13$  & ($-0.32\pm0.17)$    &  &$-0.22\pm0.92$  & $-0.99\pm1.18$  & $-0.96\pm1.06$  & $(-1.02\pm2.58)$   \\
$ 0.05$\tablefootmark{d} &	$-0.04\pm0.08$  & $-0.05\pm0.14$  & $-0.05\pm0.25$  & ($-0.24\pm0.28)$    &  &$0.19\pm1.26$  & $-0.13\pm1.38$  & $-0.14\pm1.35$  & $(-0.97\pm2.46)$   \\
eff.\tablefootmark{e} &	$-0.10\pm0.05$  & $-0.24\pm0.11$  & $-0.31\pm0.17$  & ($-0.32\pm0.19)$    &  &$-1.37\pm0.58$  & $-4.65\pm1.96$  & $-6.09\pm5.30$  & $(-6.40\pm8.42)$   \\
\hline
\\[-1Ex]
   & \multicolumn{9}{c}{Different filter pairs (Fig. \ref{fig:results_standard_test_diff_filters})} \\[1Ex]
\hline
14/15 \tablefootmark{f}  &	$-0.08\pm0.05$  & $-0.15\pm0.10$  & $-0.23\pm0.13$  & ($-0.32\pm0.17)$   &  &$-0.22\pm0.92$  & $-0.99\pm1.18$  & $-0.96\pm1.06$  & $(-1.02\pm2.58)$ \\
7/8\tablefootmark{f} &	$-0.08\pm0.06$  & $-0.17\pm0.08$  & $-0.32\pm0.19$  & ($-0.31\pm0.19)$   &  &$-1.43\pm0.99$  & $-4.99\pm1.83$  & $-5.63\pm5.33$  & $(-6.61\pm7.92)$ \\
11/12\tablefootmark{f} &	$-0.06\pm0.06$  & $-0.15\pm0.08$  & $-0.27\pm0.14$  & ($-0.32\pm0.17)$   &  &$-1.32\pm1.47$  & $-2.83\pm2.33$  & $-1.07\pm1.79$  & $(-1.88\pm4.45)$ \\
\hline
\end{tabular}
\tablefoot{
\tablefoottext{a}{$f_{0;\mathrm{in}}\;[10^{-17}\,\mathrm{erg}\;\mathrm{s}^{-1}\;\mathrm{cm}^{-2}]$}
\tablefoottext{b}{$EW_\mathrm{obs;in}\;[\mathrm{nm}]$}
\tablefoottext{c}{$\minestbb{}>0.01\;\mathrm{mag}$}
\tablefoottext{d}{$\minestbb{}>0.05\;\mathrm{mag}$}
\tablefoottext{e}{one NB118 filter; assuming effective combined NB118 filter 14 + 15}
\tablefoottext{f}{Filter pair}
}
\end{table*}
%%%%%%%%%%%
%%%%%%%%%%%
\begin{figure}[ht]
\centering
\setlength{\tabcolsep}{0cm}
	\begin{tabular}{rl}

	%	Wavelength estimation\hspace{3Ex} & \hspace{3Ex} Flux estimation \\
		\resizebox{0.5\hsize}{!}{\includegraphics[]{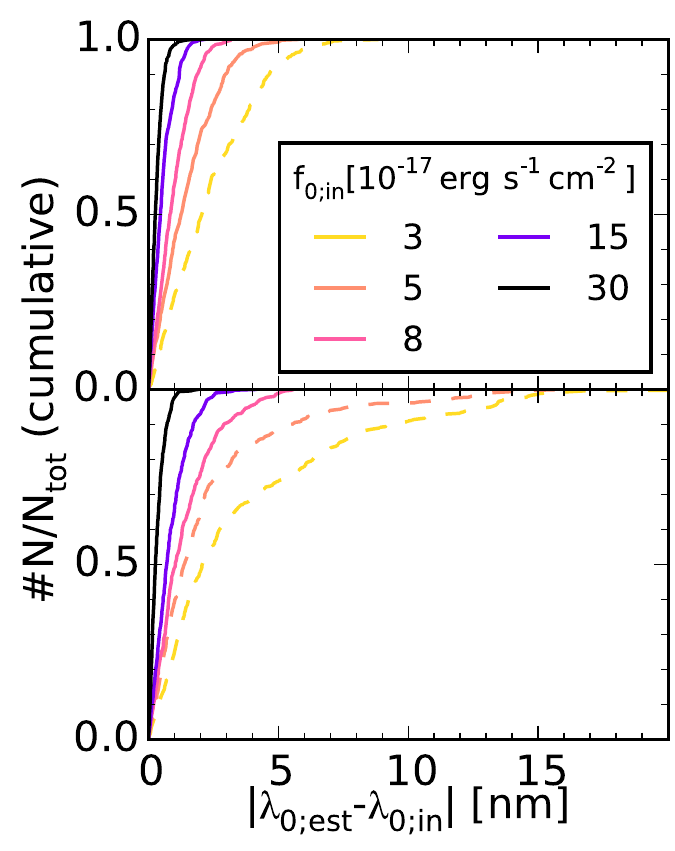}} &
	\resizebox{0.5\hsize}{!}{\includegraphics[]{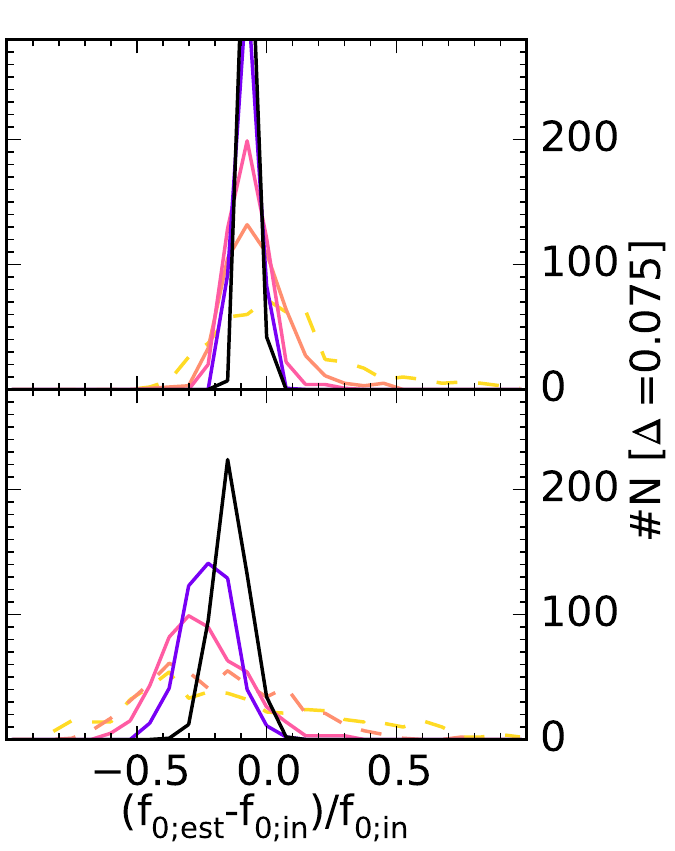}}\\
	\end{tabular}
        \caption{\label{fig:nb118:results_standard_test_diff_flux}
          Distributions of the best fit wavelength
          $\lambda_{0;\mathrm{est}}$ (left panels; cumulative) and
          flux $f_{0;\mathrm{est}}$ (right panels) for input spectra
          with an emission line at the effective wavelength (top
          panels) and at 50\% transmittance (bottom panels) of the two
          NB118 filters' combined passband.  In each panel,
          distributions are shown for five different Ha fluxes. The
          respective values are listed in the legend in units of
          $\mathrm{erg}\;\mathrm{s}^{-1}\;\mathrm{cm}^{-2}$. The input
          $EW_\mathrm{obs;in}$ was in all five cases $10\,\mathrm{nm}$.
          Fits were obtained at both transmittances for 500
          realizations of simulated observations in NB118 14\&15, Y,
          J, and H. Dashed lines indicate that the objects would not
          be selected as NB excess objects.  }
\end{figure}
%%%%%%%%%%%

%%%%%%%%%%%
\begin{figure}[ht]
\setlength{\tabcolsep}{0cm}
	\begin{tabular}{rl}
		\resizebox{0.5\hsize}{!}{\includegraphics[]{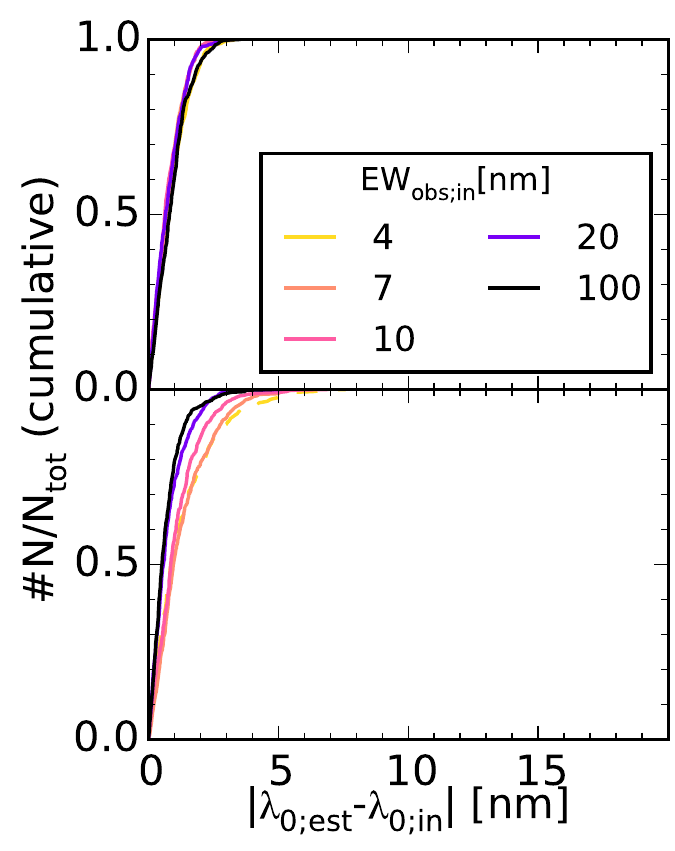}} &
	\resizebox{0.5\hsize}{!}{\includegraphics[]{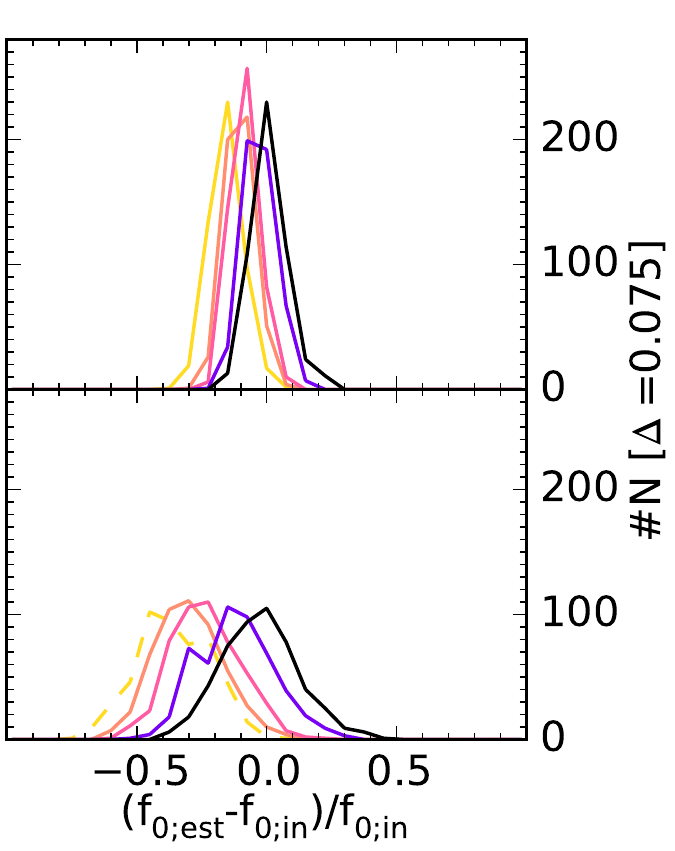}}\\
	\end{tabular}
	\caption{\label{fig:nb118:results_standard_test_diff_ew} Distributions of best
	fit wavelength and flux similar to Fig. \ref{fig:nb118:results_standard_test_diff_flux}. Here, distributions for five different $EW_\mathrm{obs;in}$ are shown. $EW_\mathrm{obs;in}$ values stated in the legend are in units of $\mathrm{nm}$. The input flux was in all cases  $10\times10^{-17}\mathrm{erg}\;\mathrm{s}^{-1}\;\mathrm{cm}^{-2}$.}
%%%%%%%%%%%
\end{figure}

Ideally, a robust parameter estimation is possible for the complete range of
$f_0\mbox{--}EW_{obs}\mbox{--}\lambda_0$ existing in objects selected to be
\ha{} NB118 emitters.  Therefore, we performed tests based on mock observations
for a set of five input line fluxes, $f_{0;\mathrm{in}}$, over the complete
relevant flux range from
$3.0\mbox{--}30.0\times10^{-17}\;\mathrm{erg}\;\mathrm{s}^{-1}\;\mathrm{cm}^{-2}$
and for five input $EW_\mathrm{obs;in}$ over the range from
$4.0\mbox{--}100.0\,\mathrm{nm}$.  In the two cases, we fixed the respective
other quantity to $EW_\mathrm{obs;in}=10\;\mathrm{nm}$ and $f_\mathrm{0;in} =
10.0\times10^{-17}\;\mathrm{erg}\;\mathrm{s}^{-1}\;\mathrm{cm}^{-2}$.
Throughout this section we were assuming observations in filter combination 14
\& 15, which is well suited for the TPV.

The resulting marginalized distributions for the best fit $f_\mathrm{0;est}$
and $\lambda_{0;est}$ are presented for each of the input
$EW_\mathrm{obs;in}\mbox{--}f_\mathrm{0;in}$ combinations at two different
input wavelengths in Fig. \ref{fig:nb118:results_standard_test_diff_flux} and
\ref{fig:nb118:results_standard_test_diff_ew}.  Mean and standard
deviation of these distributions are stated in Table
\ref{tab:est_simu_results}.

The two wavelengths correspond to the mean wavelength of the two filter's
combined effective passband and the wavelength where the transmittance of this
passband is 50\% of its peak value. The mean wavelength was calculated as in
eq. 6 of \citet{Milvang-Jensen:2013:94}.  Table \ref{tab:est_simu_results}
includes results for two additional wavelengths. One is in the middle of the
interval between mean wavelength and the 50\% transmittance, and the other is
at 20\% transmittance.  The four wavelengths are indicated as small arrows in
Fig. \ref{fig:nb118:fig_sol_deltab}.

We note that not for all $EW_{obs}\mbox{--}f_0\mbox{--}\lambda_0$ combinations
objects would also be selected as NB excess objects when applying selection
criteria (cf. sec. \ref{sec:selcriteria}), either because of not having enough
S/N or not a high enough NB excess. These cases are indicated both in the plots
and the table.

Several important things can be inferred from this analysis.  The bias in the
$\lambda_0$ estimation is $\le1\,\mathrm{nm}$ for all those among the tested
$EW_{obs}\mbox{--}f_{0}$ combinations, which would be selected as NB excess
objects. Further, the spread in the estimation, $\sigma_{\lambda_0;est}$, is in
all cases with selection below $3\,\mathrm{nm}$. For
$f_{0;in}=10.0\times10^{-17}\,\mathrm{erg}\;\mathrm{s}^{-1}\;\mathrm{cm}^{-2}$
the spread is even $\le1\,\mathrm{nm}$ over the complete relevant
$EW_\mathrm{obs}$ range, with little dependence on the $EW_\mathrm{obs}$.

While the $f_{0;\mathrm{est}}$ seems to significantly change with
$EW_{obs}$, it is important to keep here the \ha{} absorption in our
assumed model SED in mind. At the mean wavelengths of the combined
effective filter almost all apparent bias is only for this reason,
whereas at lower transmittances there is some additional bias. This
additional bias is resulting from the mismatch in the continuum.

%##############################
\subsubsection{Different estimation assumptions}
\label{sec:nb118:simu_diff_est}
%##############################

%%%%%%%%%%%
\begin{figure}
\centering
\setlength{\tabcolsep}{0cm}
	\begin{tabular}{rl}
		\resizebox{0.5\hsize}{!}{\includegraphics[]{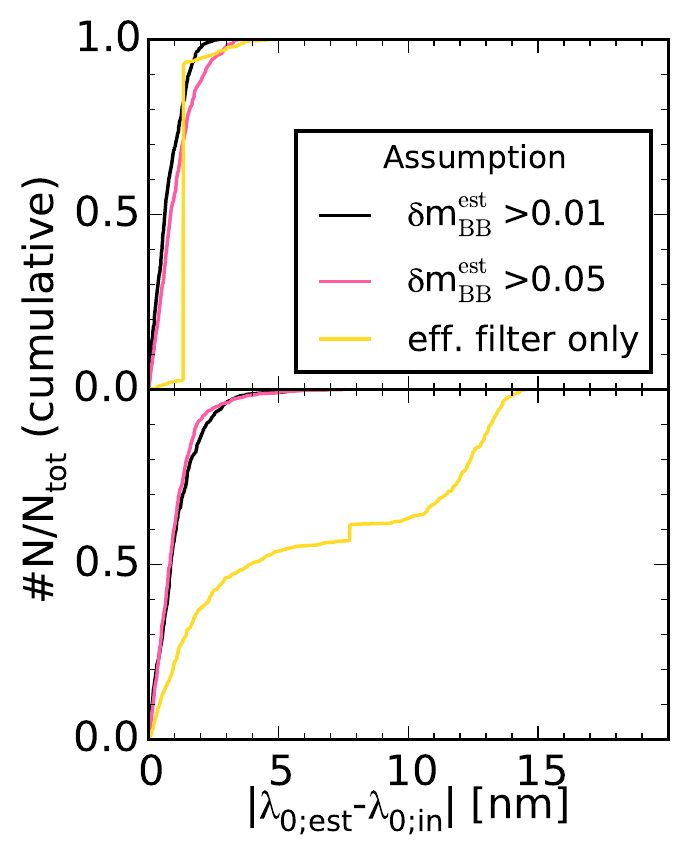}} &
	\resizebox{0.5\hsize}{!}{\includegraphics[]{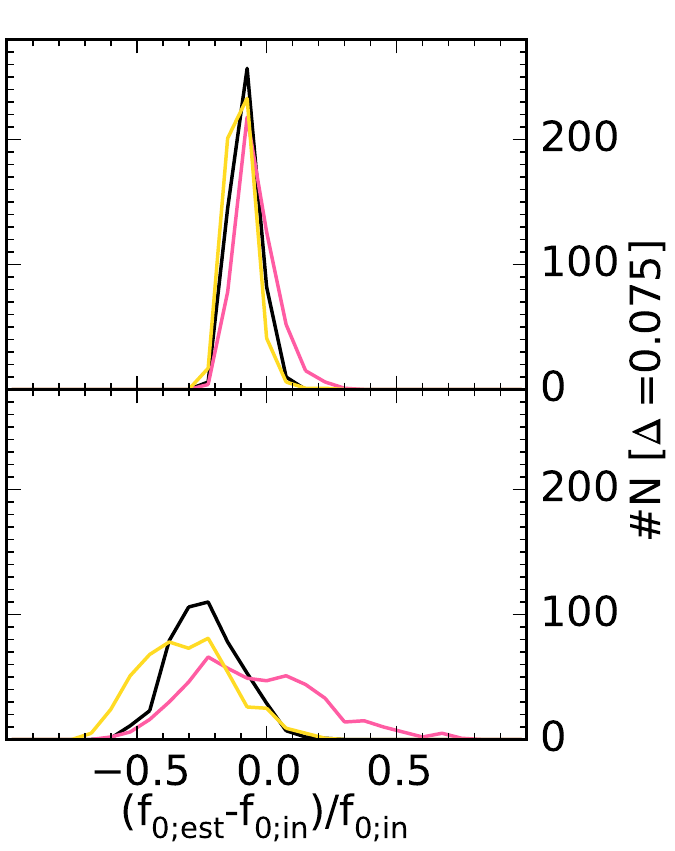}}\\
	\end{tabular}
	\caption{\label{fig:results_standard_test_diff_biases}
	Similar to Fig. \ref{fig:nb118:results_standard_test_diff_flux}, but here estimation results for three different assumptions in the TPV algorithm are presented.
	First, results are shown both for assuming the default minimal BB estimation uncertainty (cf. eq. \ref{eq:nb118:chisquare}), $\minestbb{} = 0.01$, and an alternative $\minestbb = 0.05$.
	  Secondly, results are presented for the case of using the data from the combined NB118 stack in the estimation (eff. filter. only), instead of the data from the two NB118 filters separately.
}
\end{figure}
%%%%%%%%%%%

As discussed in sec. \ref{sec:nb118:choiceofbroad}, the J excess is
providing a flux estimate, as Y and H alone allow for a very good
continuum estimate. In the case of a single NB118 filter, this would
be supplemented by an additional lower limit on the flux.  Therefore, one might
wonder how much additional estimation power is really coming from the
use of two NB filters.

In order to asses this, we performed the parameter estimation
simulation for a galaxy with $EW_{obs}$ of $10\;\mathrm{nm}$ and an
\ha{} line with $f_0$ of
$10\times10^{-17}\,\mathrm{erg}\;\mathrm{s}^{-1}\;\mathrm{cm}^{-2}$
using the three BB filters either combined with the pair 14 \& 15 or a
single NB118 filter.  For the latter we assumed the combined effective
filter for 14 \& 15.  While this does not correspond to an actual
filter, it is the applicable wavelength response when analyzing a
joint stack of data coming from both 14 \& 15.

Estimation histograms for both cases are shown in Fig.
\ref{fig:results_standard_test_diff_biases}.  Clearly, the use of the
throughput variations between the two NB118 filters allows for an
excellent $\lambda_0$ estimation over the complete relevant wavelength
range.  By contrast, no robust estimation is possible when using only
the single filter.

This is due to two main reasons.  First, while in principle a
wavelength resolution is possible when combining one NB filter with
the flux measured based on a BB filter, a bimodality between a blue
and a red solution is unavoidable.  Secondly, at 50\% transmittance
another effect is obvious.  As for the specific continuum SED, the
continuum magnitude is estimated slightly too bright, the \ha{} flux
which is estimated from the J excess alone is underestimated.
Therefore, even for a line several $\mathrm{nm}$ away from the peak,
the lower limit from the NB filter indicates a stronger flux than
estimated from the excess in J.  Hence, the best possible
reconciliation with the NB excess is for the estimation algorithm a
solution being at the peak of NB filter.  This is why the cumulative
histograms in Fig.  \ref{fig:results_standard_test_diff_biases} jumps
at around $7\;\mathrm{nm}$.

The difference between using the two NB filters separately and using the single
combined effective filter is not as dramatic for the flux estimate, but still
results in a huge improvement.  At 50\% transmittance, where the line still
contributes significant signal to the NB data, the reduction of the bias due to
the two NB filters corresponds for our standard $\minestbb{}$ to a factor of
1.3. At higher NB transmittances the effect is even larger. Halfway between the
combined filter's mean wavelength and 50\% transmittance (cf. 'B' in Table
\ref{tab:est_simu_results}) it is a factor of 1.6.  The difference is at $20\%$
transmittance negligible (cf. 'D' in Table
\ref{tab:est_simu_results}), as at the corresponding low
transmittances in the NB filters the flux estimation is mainly relying
on the J excess.  Objects at this wavelength would not be part of our
NB excess sample at the given $EW_\mathrm{obs}$.

We also investigated the consequences of increasing $\minestbb$ to
$0.05$, i.e.  giving less weight to the BB filters. Clearly, the bias
is significantly reduced. Even at $50\%$ transmittance, the bias is
close to zero, keeping in mind the \ha{} absorption. On the other
hand, the scatter is significantly increased.

%##############################
\subsubsection{Different filter combinations}
%##############################
\label{sec:nb118:simu_diff_filter}

%%%%%%%%%%%
\begin{figure}
\centering
\setlength{\tabcolsep}{0cm}
	\begin{tabular}{rl}
		\resizebox{0.5\hsize}{!}{\includegraphics[]{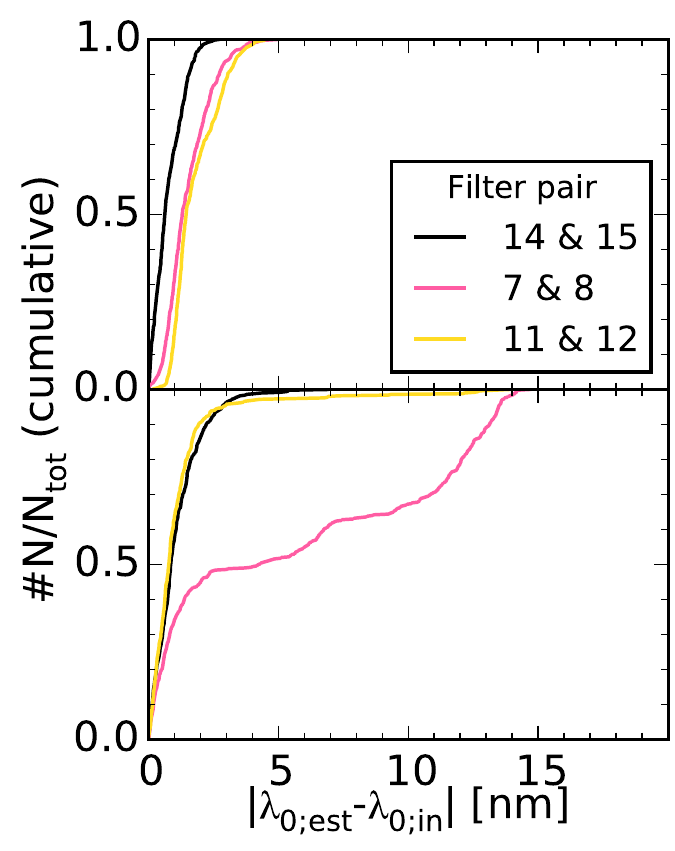}} &
		\resizebox{0.5\hsize}{!}{\includegraphics[]{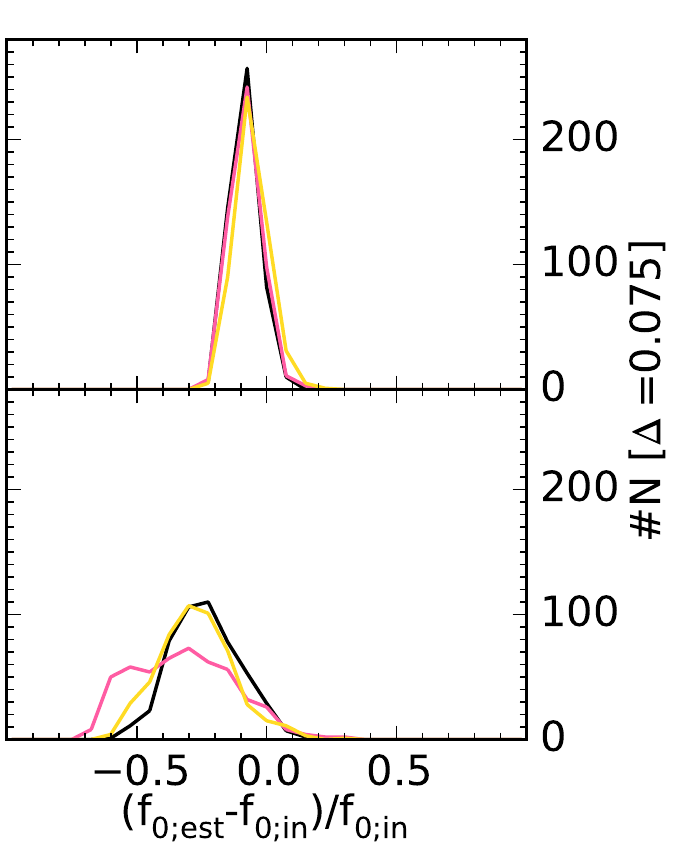}}\\
	\end{tabular}
	\caption{\label{fig:results_standard_test_diff_filters}
	Similar to Fig. \ref{fig:nb118:results_standard_test_diff_flux}, but here results are shown for different filter combinations.
	An $EW_\mathrm{obs;in} = 10\;\mathrm{nm}$ and a $f_\mathrm{0;in} =
	10\times10^{-17}\;\mathrm{erg}\;\mathrm{s}^{-1}\;\mathrm{cm}^{-2}$ were assumed.
	}
\end{figure}
%%%%%%%%%%%

As a third test, we compared the estimation quality for three
different filter combinations, including 14 \& 15, which was used
throughout sections \ref{sec:simu_diff_pfree} and
\ref{sec:nb118:simu_diff_est}, 7 \& 8 and 11 \& 12. All three
combinations were discussed through $\Delta mag \mbox{--} \lambda_0 $
curves in sec. \ref{sec_mno}. The pairs are well suited for a
fair comparison, as the sky-brightnesses are with 36 \& 32, 38 \& 30,
33 \& $28\,\mathrm{e}^{-}\,\mathrm{s}^{-1}\,\mathrm{pixel}^{-1}$,
respectively, similar in the three pairs.

The resulting distributions for the three different filter
combinations, simulated for an $EW_\mathrm{obs;in} = 10\,\mathrm{nm}$
line with
$f_\mathrm{0;in} =
10\times10^{-17}\,\mathrm{erg}\;\mathrm{s}^{-1}\;\mathrm{cm}^{-2}$,
are shown in Fig. \ref{fig:results_standard_test_diff_filters}.  It
needs to be noted that both 7 \& 8 and 11 \& 12 are more top-hat than
14 \& 15, which means that a one filter estimation is correct over a
larger wavelength range.

We are showing in Fig. \ref{fig:results_standard_test_diff_filters}
the histograms for the mean wavelength and the 50\% transmittance for
the respective combined effective filters.

Combination 7 \& 8 behaves as expected overall similar to a single
effective filter, meaning that it is not very useful for an improved
wavelength resolution. The wavelength resolution from 11 \& 12 is also
not as precise around the peak as for 14 \& 15, as the
$\Delta mag \mbox{--} \lambda_0$ curve is relatively flat there. At
intermediate transmittances, where the slope is similar to that of 14
\& 15, the wavelength resolution is on the other hand similar good.

%##############################
\section{Application to UltraVISTA data}
\label{sec:nb118:ultravista}
%##############################

%##############################
\subsection{UltraVISTA NB118 observing pattern}
\label{sec:nb118:observingpattern:ultravista}
%##############################

%%%%%%%%%%%
\begin{figure}
	\includegraphics[width = \columnwidth]{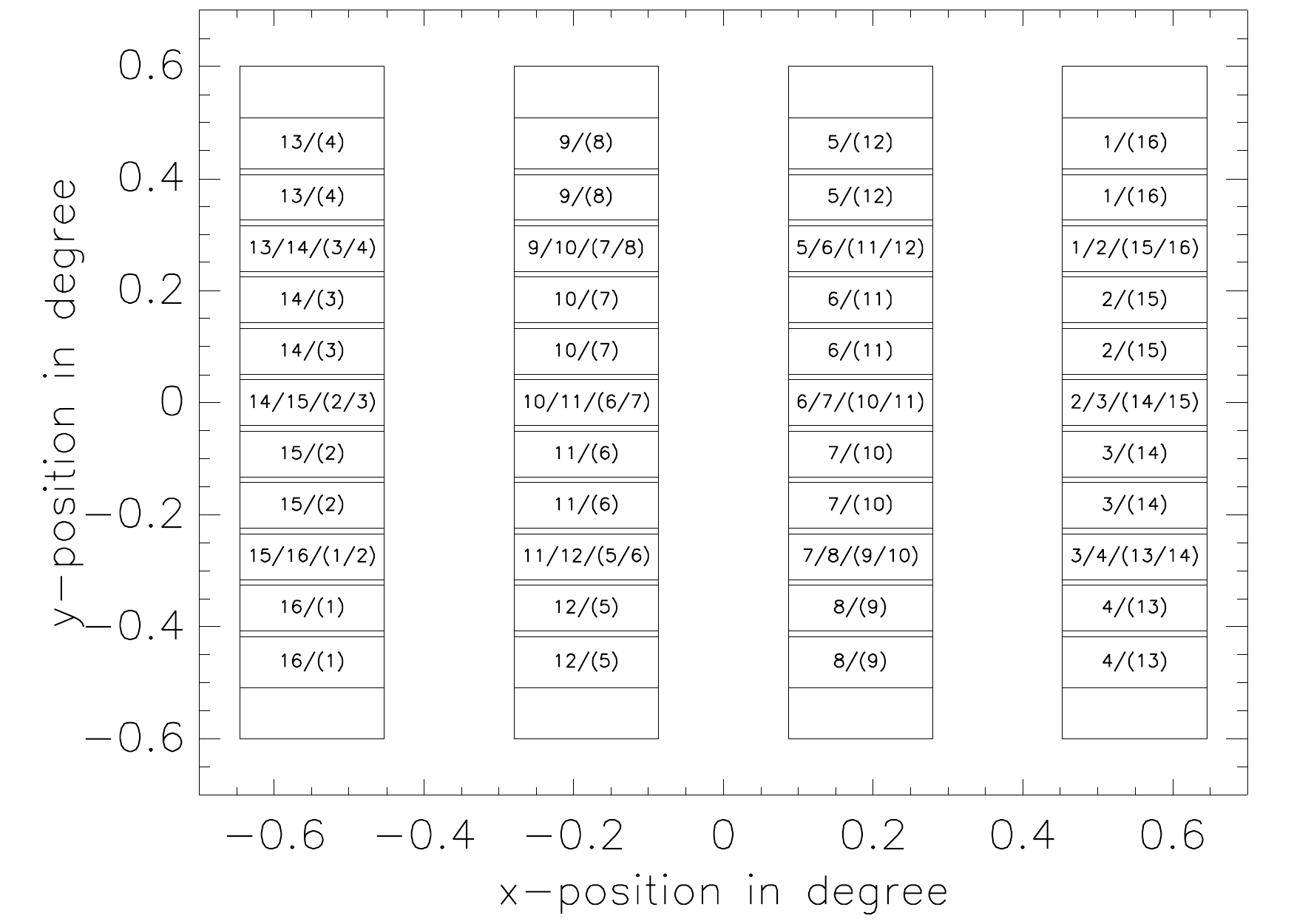}
	\caption{\label{fig:nb118:uvista:filters} Diagram showing
          which copies of the NB118 filters contribute to the
          different parts of the field covered by the UltraVISTA NB118
          tile.  The numbers in brackets are only relevant, when using
          the suggested modification of the observing pattern, as
          discussed in sec.  \ref{sec:nb118:ultravista:turning}.  }
\end{figure}
%%%%%%%%%%%

VISTA NB118 narrowband observations are already available from the
NB118 GTO observations \citep{Milvang-Jensen:2013:94} and the
intermediate \ultvis{} data releases
\citep{McCracken:2012:156}\footnote{\citet{McCracken:2012:156} are
  describing the UltraVISTA DR1. The current release is DR2:
  \url{www.eso.org/sci/observing/phase3/data_releases/uvista_dr2.pdf}},
and more data is continuing to become available within the ongoing
\ultvis{} observations.

Interestingly, some parts of the covered field are becoming directly applicable
to our method.  VIRCAM covers with a single pointing a non contiguous area on
the sky that consists of 16 separate patches, corresponding to the individual
detectors. They total $0.59\,\mathrm{deg}^2$ \citep[p.11]{Ivanov:2009}.  This
single-pointing field coverage is referred to as a pawprint.  In order to cover
an area on the sky contiguously, a so called tile consisting of six pawprints
is required.

The six contributing pawprints are three steps in one direction of the sky (y),
which are performed for two steps in the perpendicular direction (x).  For the
NB118 part of \ultvis{}, only one of the two x-positions is observed, resulting
in four stripes (cf. Fig. \ref{fig:nb118:uvista:filters}).

Each pointing in the y-direction is separated by 47.5\% of a detector (or
5.5\arcmin).\footnote{Each of the 16 detectors is a 2048x2048 Raytheon VIRGO
HgCdTe array.  100\% of a detector corresponds with the average pixel-scale of
$0.34 \arcsec\,\mathrm{pixel}^{-1}$ to 11.6\arcmin.}  Consequently,
observations of at least two pawprints contribute to the covered field with the
exception of the outermost parts. The filter numbers in the different patches
of the pawprint can bee seen in Fig. \ref{fig:nb118:uvista:filters}.  Most
important for our method, in 20.5\% percent of the stripes two pawprints
contribute with two different filters.  In addition for 6 tiny patches, which
total 4.8\% percent of the stripes, two pawprints contribute with one filter,
while one pawprint contributes with a second filter.  Due to the random jitter
within a $2'\times2'$ box, the regions are somewhat smeared out.

%##############################
\subsection{Data}
%##############################

The controlled environment of our simulations demonstrated that the
TPV is expected to work.  Here, we apply the method to the actual
\ultvis{} DR2 data.  A stack of NB118 data is available as part of
this data release.  In regions of overlap, this stack includes data
from both contributing NB118 copies.  For the purpose of the TPV, we
need these data separately.  Therefore, we produced 16 custom NB118
stacks, each of which including only the data from one
filter/detector.  Reduction, stacking, and flux calibration were
basically done in the same way as for the publicly available joined
NB118 DR2 stack (cf.
\citealt{Milvang-Jensen:2013:94,McCracken:2012:156}\footnotemark[9])
and the same observations were included.

Employing SExtractor's \citep{Bertin:1996:393} double-image mode, we
obtained in each of the individual NB118 filters photometry in the
same 2\arcsec{} circular apertures as in the detection image, where the
latter was the joined NB118 DR2 stack including data from all 16
filters.  Matching dual image photometry was also obtained for the Y,
J, and H DR2 stacks. We corrected all SExtractor aperture flux and
magnitude errors for correlation by means of empty aperture
measurements. This was necessary, as the stacks were produced on the
non-native 0\arcsec{}.15 pixel scale and interpolation was required in the
reduction as a consequence of the dithering
strategy.\footnote{Correlation corrections were previously determined
  based on \ultvis{} DR1 data.} Stated observed magnitudes are
aperture magnitudes in 2\arcsec{} and are written in italic, where
$NB118[x]\;(x \in [1,16])$ is referring to the magnitude in an
individual NB118 filter and $NB118$ is referring to the magnitude
in the joint stack.

We corrected the aperture magnitudes in Y, H, and NB118 to the J aperture based
on the enclosed fractions for point-sources. The NB118 per-detector aperture
magnitudes were corrected taking into account that different pawprints
contribute to different parts of the NB118 per-detector stacks. I.e., we make
sure that the remaining small seeing and ZP variations in data from different
pawprints are corrected.

%##############################
\subsection{Sample selection}
%##############################

%###############
\subsubsection{Selection criteria}
\label{sec:selcriteria}
%###############

We used following criteria to select NB excess objects in the regions
of overlapping filters.

\begin{itemize}
	\item Position in field:
		\begin{equation}
		\textnormal{Observed in at least two different filters}
		\label{eq:pos}
		\end{equation}

	\item Color-cut, which must be satisfied in at least one of the two contributing NB118 filters.
   		\begin{equation}
   			J_\mathrm{corr} - NB118[\mathrm{i}] > 0.2
   			\label{eq:col_crit}
   		\end{equation}
   		 The index i refers generically to the number of this filter and the second filter in the pair is referred to as j.  
 	\item Significance of NB excess at the four $\sigma$ level ($\kappa = 4$) at least in one filter, which satisfies also eq. \ref{eq:col_crit}
 		\begin{equation}
			f_{NB118[\mathrm{i}]} - f_{J_{\mathrm{corr}}} > \kappa \times \delta(f_{NB118[\mathrm{i}]} - f_{J_{\mathrm{corr}}})
			\label{eq:colsig}
		\end{equation}
		This criterion corresponds to the often used $\Sigma$ criterion \citep{Bunker:1995:513}.
		$\delta(f_{NB118[\mathrm{i}]} - f_{J_{\mathrm{corr}}})$ is the one sigma uncertainty on the flux difference.
		A justification of the choices in eq. \ref{eq:col_crit} and \ref{eq:colsig} is given in Appendix \ref{app:selcriteria}.
		
	\item Significance of NB118 detection in the second filter at the $2.5\,\sigma$ ($\kappa = 2.5$) level.
    		\begin{equation}
   			 f_{NB118[\mathrm{j}]}  > \kappa \times \delta f_{NB118[\mathrm{j}]}
			 \label{eq:detsig}
	        \end{equation}
 	\item Mask
     		\begin{equation}
     		(\alpha,\delta) \not\subset \mathcal{M},
     			%\textnormal{We were excluding regions ...}
     			\label{eq:mask}
		\end{equation}
                where $(\alpha,\delta)$ are the coordinates of the
                object, and $\mathcal{M}$ are regions which are
                excluded due to bright stars, reflections, being close
                to detector boundaries, and a defect region in
                detector 16.  We require additionally that the
                SExtractor flags in both contributing NB118 filters
                and Y and J are smaller than 4.
\end{itemize}

$f_{J_\mathrm{corr}}$ is the broadband flux density corrected to the
position of the NB118 filter and $J_\mathrm{corr}$ was calculated from
$f_{J_\mathrm{corr}}$ in the usual way by eq. \ref{eq_abmagnitudes}.
$f_{J_\mathrm{corr}}$ was approximated, depending on the S/N in Y, and the $Y-J$ color, as:

\begin{eqnarray}
\textnormal{if }f_{Y}/{\delta f_{Y}} < 2.0\textnormal{ then } & f_{J_\mathrm{corr}} & = f_{J}\\
\textnormal{else if $Y - J > 0.5$ then } & f_{J_\mathrm{corr}} & = f_{J}\cdot10^{-0.4\cdot0.125}, \label{eq:colcorr:flux3}\\
\textnormal{else } & f_{J_\mathrm{corr}} & = f_{J}^{0.75} f_{Y}^{0.25}.
\label{eq:colcorr:flux4}
\end{eqnarray}

\noindent

% TODO Language, otherwise ok
Color correction and selection criteria are based on those of
\citet{Milvang-Jensen:2013:94}, but adjusted for the use of observations in two
NB filters. Additional alterations include the use of flux uncertainties
instead of magnitude uncertainties and a change of slope in the color
correction.
% The latter modification was necessary to take account for the
% improved broadband zeropoints used for the \ultvis{} DR2. \citet
% {Milvang-Jensen:2013:94} used the \ultvis{} DR1. 
% Further, the change compensates for a different Vega-to-AB conversion used in
% our work. 
We justify the selection in some more detail in Appendix
\ref{app:selcriteria}.

%%%%%%%%%%%
\begin{figure}
\includegraphics[width = \columnwidth]{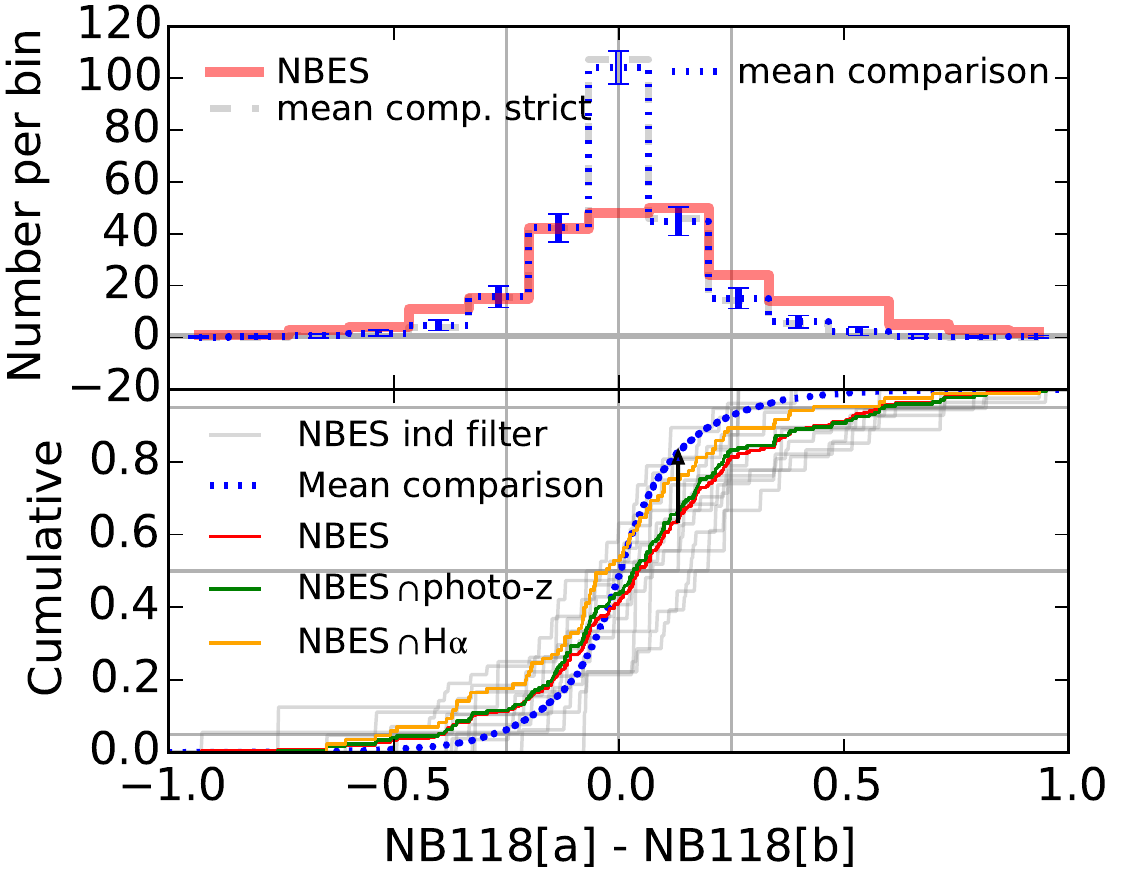}
\caption{ Statistical distribution of magnitude differences for
  objects with observations available in two different NB118 filters.
  Histograms are shown for the sample of 239 objects with NB excess
  (NBES), and for 100 comparison samples. The comparison samples have
  in each of the 12 contributing filter combinations the same number
  of objects and distribution of stack magnitudes as the NBES, but no
  NB excess was required.  \textbf{Upper:} Both the histogram for the
  NBES and the histograms obtained from the mean of the 100 comparison
  samples are shown. In the latter case, the standard deviation
  between the different samples is indicated by errorbars.
  \textbf{Lower:} Cumulative distribution for the mean of the
  comparison samples and for the NBES sample.  The maximal differences
  between the two samples is indicated as arrow.  In addition, for the
  NBES, the cumulative histograms are included for the individual
  filter combinations.  Further, two subsets of the NBES based are
  plotted.  }
\label{fig:stat_test}
\end{figure}
%%%%%%%%%%%

\subsubsection{The NBES sample}

Our main NB-excess sample (NBES) was selected based on the criteria in
eq.~\ref{eq:pos} -- \ref{eq:mask}, resulting in 239 objects.  Matching
to the photo-z catalog of \citet{Ilbert:2013:55}, we can identify as
the cause of the NB excess in 86 cases \ha{}+\fion{N}{ii} or
\fion{S}{ii}, 56 H$\beta$ or \fion{O}{iii}, 28 \fion{O}{ii}, and six
\fion{S}{iii}, where the redshift cuts of $0.6<z<0.95$, $1.25<z<1.55$,
$2.1<z<2.4$, and $0.2<z<0.4$ were applied, respectively.  The
remaining 63 objects could either not be matched to the
\citet{Ilbert:2013:55} catalog or do not have a photo-z within the
four intervals. Based on a full simulation of the NB118 observing
pattern, as presented in Appendix \ref{sec:ultravista:expnumber}, we
expect for the selection criteria about $100\mbox{--}250$ \ha{}
emitters, where the number depends on the chosen literature \ha{}
luminosity function, the equivalent width distribution, and the
$w_{6583}$ distribution. In the final UltraVISTA data we expect to
select about twice as many \ha{} emitters within the same sub-field. 

In addition to the NBES, we picked 100 comparison samples, each of
which having in each of the filters the same number of objects as the
NBES with the same stack $NB118$ distribution. More precisely, we
split the range between $19<NB118<24$ into 40 bins. The objects in
the comparison samples (CS) were randomly drawn from a selection,
where we did by contrast to the NBES not impose a color-significance
or color-cut. In order to completely avoid NB excess objects in the
comparison sample, we also created stricter versions, where we imposed
$J_{corr}-NB118 < 0$ (SCS).

%##############################
\subsection{Statistical analysis of throughput variations}
%##############################

%%%%%%%%%%%
\begin{figure*}[ht]
\centering
\includegraphics[width = 1.0\textwidth]{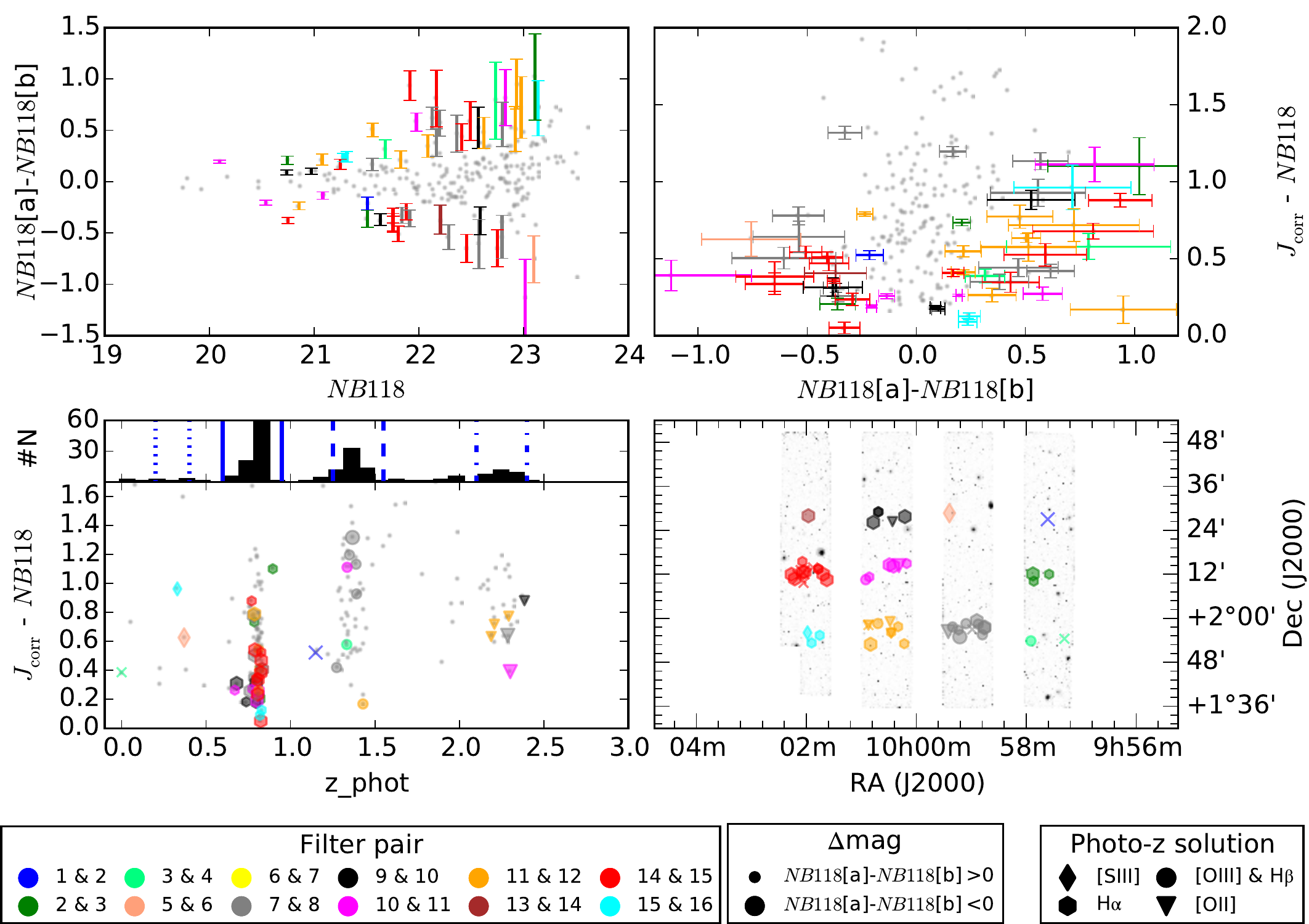}
\caption{\label{fig:nb_excess_selection} Magnitude, redshift, and
  field distribution for our sample of NB excess objects with observations
  available in two different NB118 filters (NBES).  Those NBES objects with
  their magnitudes in the two contributing NB118 filters differing by more than
  $2.5\sigma$ are shown in color, with the colors referring to the different
  filter pairs. All other NBES objects objects are shown in the background in
  gray.  \textbf{Upper left:} The difference in magnitude between the two
  individual contributing filters is plotted against the magnitude in the
  stack. Magnitudes were measured in 2\arcsec{} diameter apertures and
  errorbars are $1\sigma$ uncertainties.  \textbf{Upper right}: The
  $J-NB118_{corr}$ color excess is plotted against the magnitude difference
  between the individual NB118 filters. $J-NB118_{corr}$ is corrected for the
  continuum slope by means of the $Y-J$ color (see sec. \ref{sec:selcriteria})
  \textbf{Lower left}: $J-NB118_{corr}$ plotted against the photometric
  redshift \citep{Ilbert:2013:55}. The three main groups are \ha{}
  ($z\sim0.8$), \fion{O}{iii}+ H$\beta$ ($z\sim1.4$), and \fion{O}{ii}
  ($z\sim2.2$). Different symbols refer to membership in these groups. The X
  symbol (not in legend) marks objects with significant throughput variations,
  where no association with a specific line was possible. A histogram of the
  photometric redshifts is included in the upper part of this panel. The
  vertical lines indicate the redshift intervals which are used for the
  classification into \fion{S}{iii} (dotted), \ha{} (solid), \fion{O}{iii}+
  H$\beta$ (dashed), and \fion{O}{ii} (dash-dotted), respectively.
  \textbf{Lower right:} Position of the objects in the field of view. The DR2
  NB118 stack is shown in the background. Larger symbols have a $NB118[a] -
  NB118[b] > 0$ and smaller symbols have $NB118[a] - NB118[b] < 0$.  }
\end{figure*}
%%%%%%%%%%%

We tested whether we see at all NB excess objects showing throughput
differences beyond the statistical fluctuations.  In the upper panel
of Fig.  \ref{fig:stat_test}, the histogram of magnitude differences,
$\Delta mag = NB118[a] - NB118[b]$, is shown as solid red curve for
the NBES.  All 12 filter combinations are included in the same
histogram, where the identifiers $a$ and $b$ generically refer to the
filter numbers in the pairs, with $a < b$.  Also included in the
figure is the mean histogram of the 100 CS and the corresponding
standard deviation. The SCS sample is indicated as a dashed light gray
line, which is hardly visible as it is basically identical with the CS
histogram.

Including predominantly objects without emission in the NB118 filters,
the spread in the CS should be caused by noise only. By contrast, the
spread in the histogram for the NBES is expected to be caused both by
noise and actual throughput variations, and indeed, it clearly differs
from the comparison sample.  The difference can also be evaluated in
the lower panel of Fig.  \ref{fig:stat_test}, where the cumulative
histograms are shown.  Applying the two-sample KS test to the two
histograms, we can formally rule out the null-hypothesis that both
samples are originating from the same distribution.\footnote{At the
  99.999997\% level; more precisely, the comparison sample used in the
  KS test was the combination of the 100 realizations, having
  effectively 100*239 objects.} The maximum difference between the two
cumulative curves is marked as an arrow in the figure.

One relevant concern is that objects with unusually large random or
systematic errors in one of the two filters could show the required NB
excess exactly for that reason, which might lead to a
misidentification of such objects as NB excess objects and hence an
inclusion in the NBES, biasing this sample to objects with large
throughput variations.  However, as discussed above and further shown
in Fig. \ref{fig:nb_excess_selection}, photometric redshifts of 74\%
of the NBES objects can be well identified with actual lines,
indicating a relatively clean sample.  Reassuringly, the average
cumulative magnitude difference curve does not change much using only
this sub-sample (green curve in Fig. \ref{fig:stat_test}).  For the
subsample of \ha{} emitters the difference to the general population
is slightly smaller, being explainable as the most extreme
$EW_{\mathrm{obs}}$ NB excess objects are predominately \fion{O}{iii}
emitters.

As a next step, we selected those objects from the main NB-excess
sample (NBES), which show a flux difference between the two filters
differing from zero at least at the $2.5\sigma$ level, have a
$\Delta mag >0.05$, and are fainter than $NB118 = 20$. The latter
two criteria are meant to avoid objects, where the difference might be
caused by small remaining zeropoint errors or by small PSF
differences, being especially relevant for merging objects.

The resulting 55 objects, 8 of which are so close to a neighboring
object that SExtractor marked them as de-blended\footnote{SExtractor
  FLAGS = 2 or 3}, are shown in four different plots in Fig.
\ref{fig:nb_excess_selection}.  While all objects of the NBES are
included in the plots as gray dots, objects in the throughput
difference sample are color-coded by the relevant filter pairs.
Adding to the confidence that those objects with strong TPV are indeed
NB excess objects caused by emission lines, a comparison of this
subsample to the photometric redshifts of \citet{Ilbert:2013:55}
allowed in an even larger fraction than in the full NBES for an
identification with one of the four main redshift solutions (85\% vs
74\%).\footnote{For one of two [SIII]$\lambda9533$ candidates, visual
  inspection of the SED allowed for an unambiguous identification with
  an extreme EW object at $z=0.8$, showing a $z'$-filter excess
  corresponding to a rest-frame [OIII] + H$\beta$ equivalent exceeding
  $100\;\mathrm{nm}$} The 8 remaining objects are classified in the
\citet{Ilbert:2013:55} catalog as either masked (2), star (1), XMM
detected (1), a photo-z not in the intervals (1), or we could not find
a match within a radius of 0.5\arcsec{} (3).

While the number of objects is too small to make strong statistical
conclusions for the individual filters, we find that strong throughput
variations are indeed mainly found in those pairs for which they were
expected (14 \& 15, and also 15 \& 16 and 9 \& 10). However, there is
one major exception. Filter pair 7 \& 8 shows surprisingly
(cf. Fig. \ref{fig:nb118:diff_nb118_combinations}) a relatively large
number of objects with strong differences.  With some of them being
brighter in filter 7 and others in filter 8, an erroneous ZP can be
ruled out as reason for the behavior. A visual inspection of the
objects also does not indicate obvious problems.  Therefore, we need
to conclude that one of the two filters seems to substantially differ
from our expectations (cf. also Appendix \ref{app_nb118}).

%##############################
\subsection{NB118 H$\alpha$ measurements for individual objects}
\label{sec:data_paraest}
%##############################

\begin{table*}
\caption{\label{tab:nb118:table_results1} $\lambda_0$ and $f_0$ estimates for \ha{} both from our TPV method and other methods. The latter are based on conversions from zCOSMOS \fion{O}{ii} and H$\beta$ fluxes (cf. sec. \ref{sec:estimation_zcosmos}), on \ha{} estimates from SED fitting (cf. sec. \ref{sec:estimation_sed}), and on SFR estimates from the combination of UV + IR (cf. sec. \ref{sec:estimation_total}).  As an example, cutouts for objects 121 and 135 are shown in Fig. \ref{fig:nbexcess_examples}. Properties of the best-fit SEDs and the $EW_{obs}$ from the TPV are given in the Appendix as part of the supplementing Table \ref{tab:nb118:table_results2}.
}
\begin{tabular}{lcrrrrrrrrrr}
\hline
\hline
ID \tablefootmark{a} & {\tiny zCOSMOS} \tablefootmark{b} & pair\tablefootmark{c} &  $\lambda_0$\tablefootmark{d} (TPV)  &
$\lambda_0$\tablefootmark{d} (spec) & $\lambda_0$\tablefootmark{d,f} (phot-z) & $f_{H\alpha}$\tablefootmark{e} (TPV) & $f_{H\alpha}$\tablefootmark{e} ($\fion{O}{ii}$) & $f_{H\alpha}$\tablefootmark{e} (H$\beta$) &    $f_{H\alpha}$\tablefootmark{e} (total) &  $f_{H\alpha}$\tablefootmark{e} (SED) \\
\hline
7 & 810332 & 15/16 & $1202.2_{-0.2}^{+0.2}$ & 1201.3 & 1201 & $38.5_{-2.1}^{+2.0}$ & $16.5{\scriptstyle \pm2.6}$ & $34.2{\scriptstyle \pm5.2}$ & $55.6_{-18.4}^{+70.8}$ & $19.4_{-2.2}^{+4.4}$ \\
14 &  & 15/16 & $1185.5_{-1.4}^{+2.1}$ &  & 1180 & $4.0_{-1.1}^{+1.3}$ &  &  & $3.6_{-2.0}^{+8.1}$ & $5.2_{-1.1}^{+4.7}$ \\
33 & 810529 & 15/16 & $1201.1_{-0.3}^{+0.3}$ & 1199.1 & 1194 & $23.6_{-2.1}^{+2.1}$ & $13.2{\scriptstyle \pm2.3}$ &  & $25.4_{-10.4}^{+55.4}$ & $13.0_{-0.0}^{+10.9}$ \\
94 &  & 14/15 & $1197.5_{-0.4}^{+0.4}$ &  & 1199 & $13.6_{-1.3}^{+1.2}$ &  &  & $30.2_{-16.0}^{+43.4}$ & $1.0_{-0.4}^{+0.4}$ \\
96 & 822610 & 14/15 & $1198.9_{-0.4}^{+0.4}$ & 1197.6 & 1186 & $13.0_{-1.2}^{+1.2}$ & $11.3{\scriptstyle \pm1.8}$ &  & $7.6_{-2.6}^{+12.7}$ & $10.4_{-1.9}^{+5.1}$ \\
97 &  & 14/15 & $1194.6_{-0.6}^{+0.5}$ &  & 1204 & $14.0_{-1.0}^{+1.0}$ &  &  & $9.7_{-3.8}^{+10.2}$ & $12.4_{-1.6}^{+6.0}$ \\
99 & 822560 & 14/15 & $1186.2_{-0.9}^{+1.1}$ & 1189.9 & 1191 & $9.8_{-1.2}^{+1.6}$ & $2.4{\scriptstyle \pm0.9}$ &  & $15.0_{-5.8}^{+29.1}$ & $9.9_{-4.9}^{+0.0}$ \\
104 &  & 14/15 & $1186.6_{-1.1}^{+1.3}$ &  & 1201 & $5.0_{-0.6}^{+0.9}$ &  &  & $13.3_{-6.3}^{+15.3}$ & $3.5_{-0.6}^{+1.5}$ \\
105\tablefootmark{g,h} & 822732 & 14/15 & $1199.6_{-0.1}^{+0.1}$ & 1199.4 & AGN & $66.7_{-2.6}^{+2.7}$ & $2.4{\scriptstyle \pm0.4}$ & $24.7{\scriptstyle \pm8.6}$ & $65.0_{-19.6}^{+56.4}$ & $66.6_{-7.5}^{+0.9}$ \\
111 &  & 14/15 & $1198.1_{-0.9}^{+0.7}$ &  & 1197 & $5.8_{-1.2}^{+1.0}$ &  &  & $5.9_{-2.5}^{+10.7}$ & $5.0_{-1.9}^{+0.6}$ \\
113 & 823319 & 14/15 & $1198.4_{-0.6}^{+0.5}$ & 1197.7 & 1186 & $7.4_{-1.1}^{+1.1}$ & $5.0{\scriptstyle \pm1.1}$ & $7.1{\scriptstyle \pm2.6}$ & $13.0_{-6.4}^{+25.4}$ & $5.7_{-0.1}^{+4.5}$ \\
114 & 822686 & 14/15 & $1193.8_{-0.6}^{+0.6}$ & 1193.2 & 1190 & $13.4_{-0.9}^{+1.0}$ & $9.6{\scriptstyle \pm1.7}$ &  & $14.8_{-6.5}^{+19.0}$ & $13.0_{-2.6}^{+4.5}$ \\
117\tablefootmark{g} & 822508 & 14/15 & $1193.1_{-0.7}^{+0.7}$ & 1194.7 & AGN & $29.2_{-1.8}^{+1.9}$ & $11.2{\scriptstyle \pm2.5}$ &  & $53.0_{-24.0}^{+95.9}$ & $26.6_{-7.1}^{+17.7}$ \\
121 & 822822 & 14/15 & $1197.5_{-0.3}^{+0.3}$ & 1197.3 & 1174 & $14.7_{-1.1}^{+1.1}$ & $23.3{\scriptstyle \pm3.9}$ & $33.5{\scriptstyle \pm6.5}$ &  & $18.7_{-6.4}^{+2.4}$ \\
122 &  & 14/15 & $1200.6_{-0.4}^{+0.5}$ &  & 1197 & $12.3_{-1.5}^{+1.4}$ &  &  & $9.3_{-3.6}^{+16.7}$ & $6.3_{-0.0}^{+8.4}$ \\
124 &  & 14/15 & $1189.2_{-2.5}^{+2.9}$ &  & 1188 & $2.3_{-0.3}^{+0.6}$ &  &  &  &  \\
125 & 822496 & 14/15 & $1196.8_{-0.7}^{+0.7}$ & 1198.5 & 1190 & $11.4_{-1.6}^{+1.8}$ & $4.6{\scriptstyle \pm1.1}$ &  & $23.9_{-10.2}^{+34.7}$ & $4.1_{-2.2}^{+0.3}$ \\
126 &  & 14/15 & $1183.3_{-0.6}^{+0.6}$ &  & 1199 & $8.5_{-1.3}^{+1.3}$ &  &  &  & $5.4_{-1.5}^{+1.7}$ \\
128 &  & 14/15 & $1184.2_{-0.3}^{+0.3}$ &  & 1162 & $15.5_{-1.1}^{+1.2}$ &  &  &  & $17.9_{-4.8}^{+3.8}$ \\
131 &  & 14/15 & $1186.1_{-1.5}^{+2.2}$ &  & 1201 & $3.5_{-0.7}^{+1.1}$ &  &  &  & $3.2_{-0.7}^{+1.9}$ \\
135\tablefootmark{i} & 823097 & 14/15 & $1185.9_{-0.6}^{+0.7}$ & 1184.5 &  & $10.5_{-1.1}^{+1.3}$ & $11.8{\scriptstyle \pm1.9}$ & $25.3{\scriptstyle \pm7.5}$ & $23.5_{-9.0}^{+28.4}$ & $19.3_{-7.4}^{+3.2}$ \\
138 & 822504 & 14/15 & $1190.8_{-0.5}^{+0.6}$ & 1195.0 & 1186 & $19.2_{-0.5}^{+0.7}$ & $19.4{\scriptstyle \pm2.9}$ & $20.4{\scriptstyle \pm4.6}$ & $14.7_{-5.4}^{+17.5}$ & $12.7_{-1.7}^{+3.9}$ \\
147 &  & 14/15 & $1189.8_{-2.9}^{+2.4}$ &  & 1153 & $4.7_{-0.3}^{+0.9}$ &  &  &  & $3.3_{-0.9}^{+1.8}$ \\
150 &  & 14/15 & $1193.7_{-1.5}^{+1.0}$ &  & 1188 & $7.1_{-1.0}^{+0.9}$ &  &  &  & $4.8_{-0.3}^{+4.9}$ \\
153 &  & 14/15 & $1195.9_{-1.0}^{+0.8}$ &  & 1182 & $6.7_{-1.0}^{+1.0}$ &  &  &  & $9.5_{-2.6}^{+3.7}$ \\
161 &  & 14/15 & $1196.9_{-1.2}^{+0.8}$ &  & 1211 & $7.5_{-1.3}^{+1.1}$ &  &  &  & $4.7_{-1.1}^{+2.4}$ \\
164 &  & 14/15 & $1182.8_{-0.5}^{+0.6}$ &  & 1184 & $9.1_{-1.3}^{+1.3}$ &  &  &  & $5.9_{-0.0}^{+4.1}$ \\
167 & 823087 & 14/15 & $1190.3_{-0.3}^{+0.4}$ & 1191.7 &  & $27.4_{-0.3}^{+0.4}$ & $30.4{\scriptstyle \pm3.4}$ & $46.3{\scriptstyle \pm8.5}$ & $29.5_{-9.8}^{+29.6}$ & $22.3_{-3.9}^{+3.0}$ \\
170 &  & 14/15 & $1192.8_{-2.2}^{+1.6}$ &  & 1186 & $5.3_{-0.6}^{+0.7}$ &  &  &  & $3.5_{-0.1}^{+3.0}$ \\
172 &  & 09/10 & $1198.9_{-0.5}^{+0.5}$ &  & 1180 & $7.3_{-1.1}^{+1.1}$ &  &  &  & $3.8_{-0.9}^{+1.4}$ \\
186 & 839235 & 09/10 & $1199.3_{-0.2}^{+0.2}$ & 1197.6 & 1104 & $17.0_{-1.3}^{+1.2}$ & $9.1{\scriptstyle \pm2.7}$ & $17.8{\scriptstyle \pm6.2}$ & $20.7_{-11.5}^{+25.3}$ & $8.8_{-2.2}^{+4.6}$ \\
204 & 838539 & 09/10 & $1183.2_{-0.4}^{+0.6}$ & 1182.3 & 1141 & $23.8_{-4.2}^{+3.9}$ & $4.4{\scriptstyle \pm2.2}$ &  & $30.6_{-14.2}^{+45.0}$ & $2.8_{-1.8}^{+0.1}$ \\
205 & 838552 & 09/10 & $1182.9_{-0.4}^{+0.5}$ & 1181.6 & 1180 & $19.9_{-3.3}^{+2.9}$ & $6.9{\scriptstyle \pm1.4}$ &  & $37.9_{-17.2}^{+55.6}$ & $7.0_{-0.2}^{+4.7}$ \\
226 &  & 09/10 & $1197.0_{-0.6}^{+0.6}$ &  & 1201 & $9.6_{-1.1}^{+1.3}$ &  &  & $22.0_{-10.3}^{+29.4}$ & $9.0_{-0.0}^{+8.1}$ \\
235 &  & 09/10 & $1184.3_{-0.6}^{+1.1}$ &  & 1185 & $7.9_{-1.7}^{+1.3}$ &  &  &  & $3.5_{-0.8}^{+2.5}$ \\
\hline
\end{tabular}
\tablefoot{
	\tablefoottext{a}{NBES}
	\tablefoottext{b}{zCOSMOS bright 20k \citep{Lilly:2009:218} ID}
	\tablefoottext{c}{contributing NB118 filters}
	\tablefoottext{d}{H$\alpha$ central wavelength [$\mathrm{nm}$]}
	\tablefoottext{e}{[$10^{-17}\,\mathrm{erg}\;\mathrm{s}^{-1}\;\mathrm{cm}^{-2}$]}
	\tablefoottext{f}{redshifts from \citet{Ilbert:2013:55}}
	\tablefoottext{g}{\emph{Chandra} point source \citep{Civano:2012:30}}
	\tablefoottext{h}{\emph{zCOSMOS} confidence class 13.5 (BL AGN)}
	\tablefoottext{i}{\fion{Ne}{v}$\,\lambda3426$ detected in \emph{zCOSMOS} spectrum}
 }
\end{table*}

\begin{figure}
\setlength{\tabcolsep}{0pt}
\begin{tabular}{c|c|c|c|c}
\hline
\hline
ACS (F814W) & Y (DR2) & NB118 14 & NB118 15 & J (DR2) \\
\hline
\multicolumn{5}{c}{NBES 121 / zCOSMOS 822822} \\
\hline
\includegraphics[width = 0.2\columnwidth]{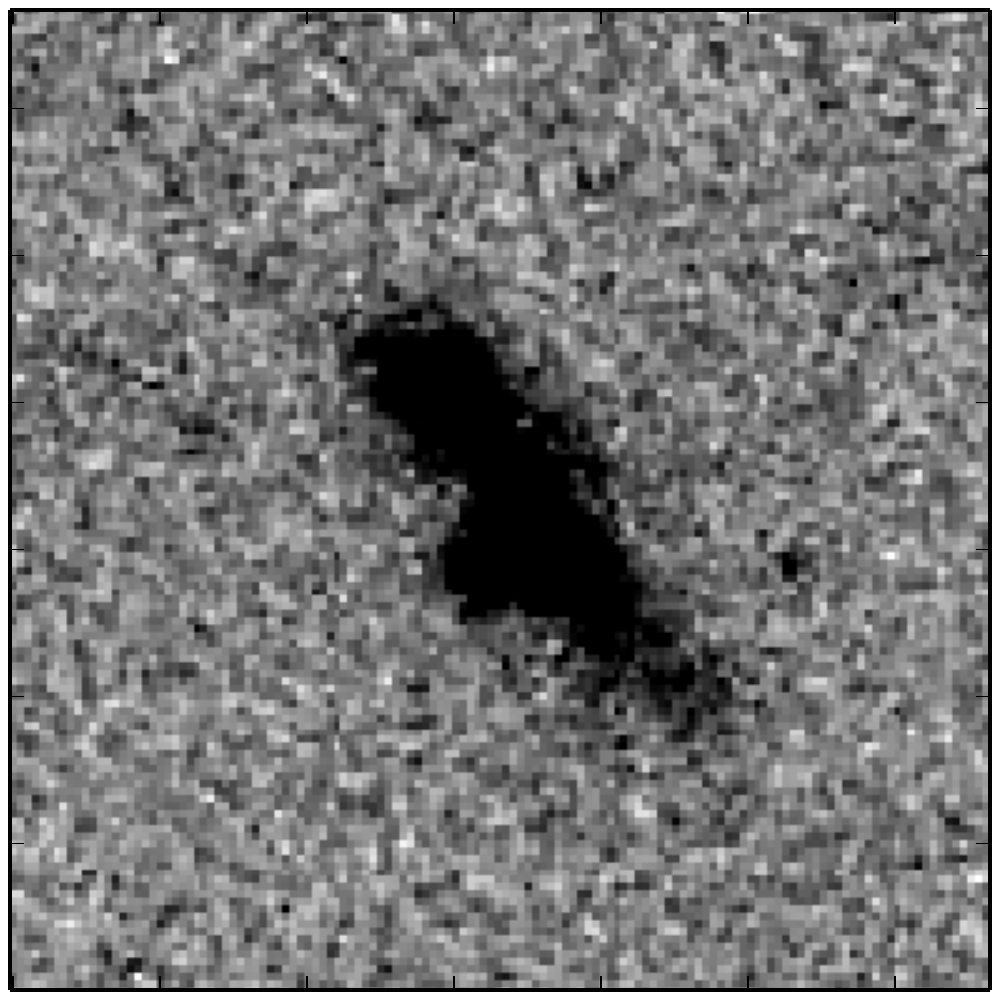} &
\includegraphics[width = 0.2\columnwidth]{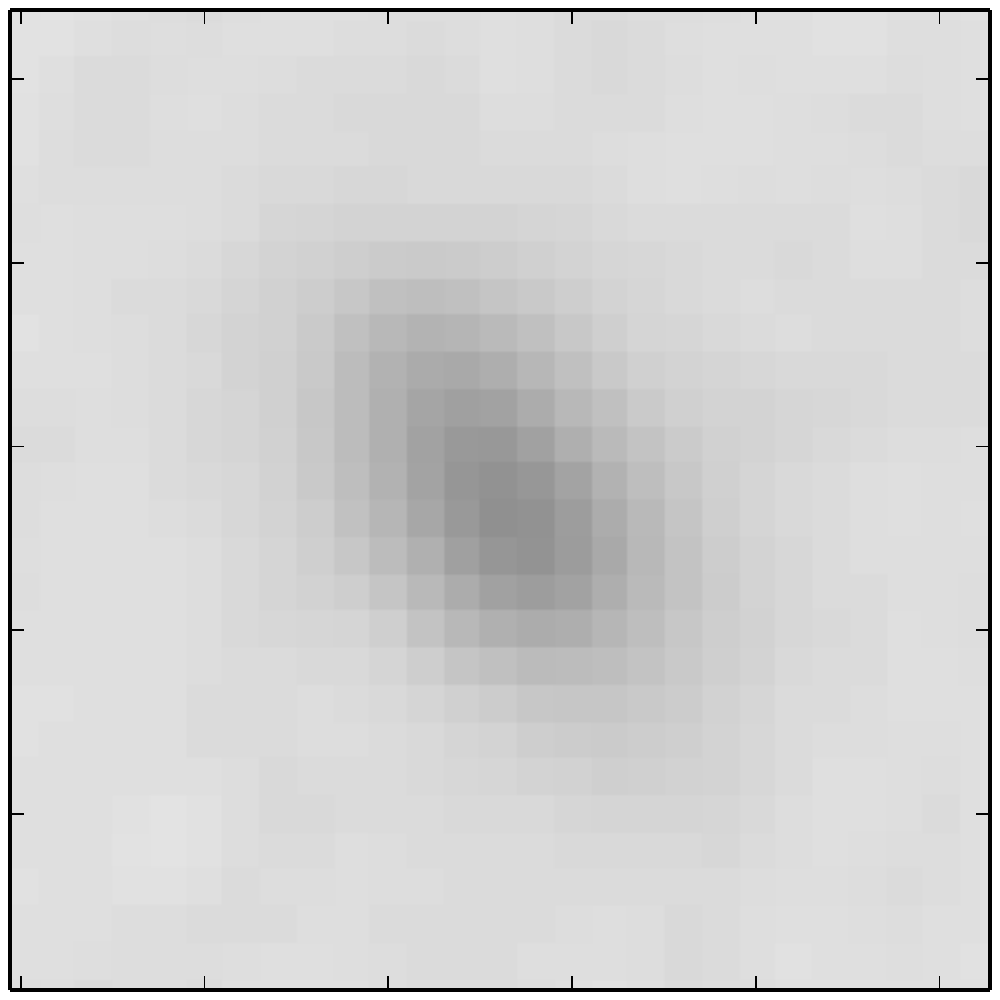} &
\includegraphics[width = 0.2\columnwidth]{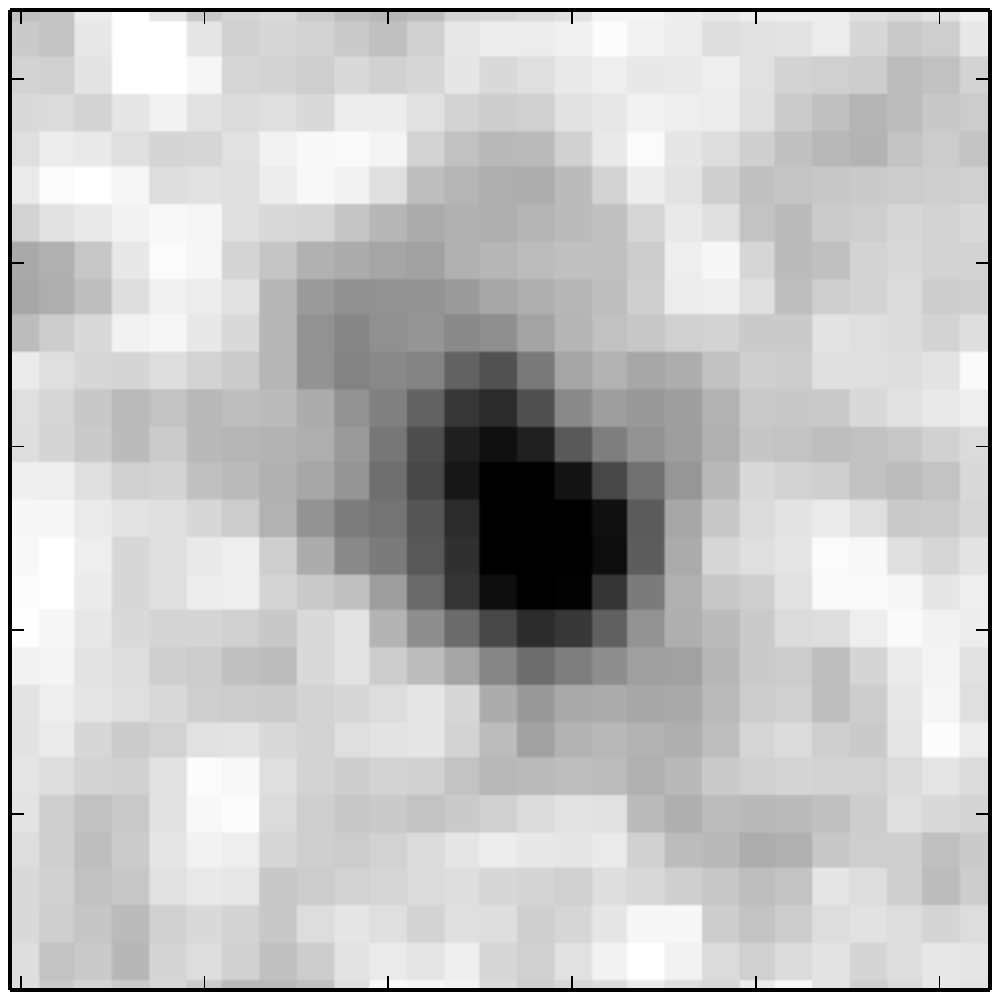} &
\includegraphics[width = 0.2\columnwidth]{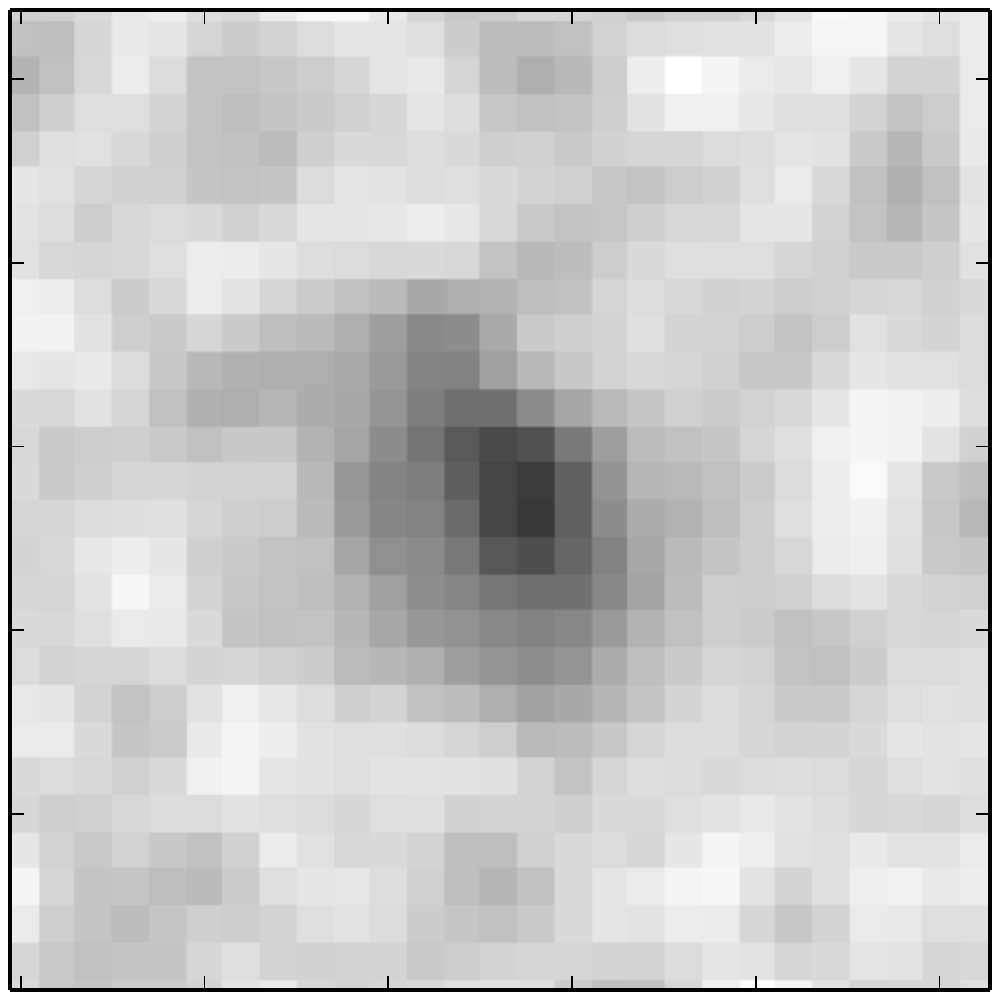} &
\includegraphics[width = 0.2\columnwidth]{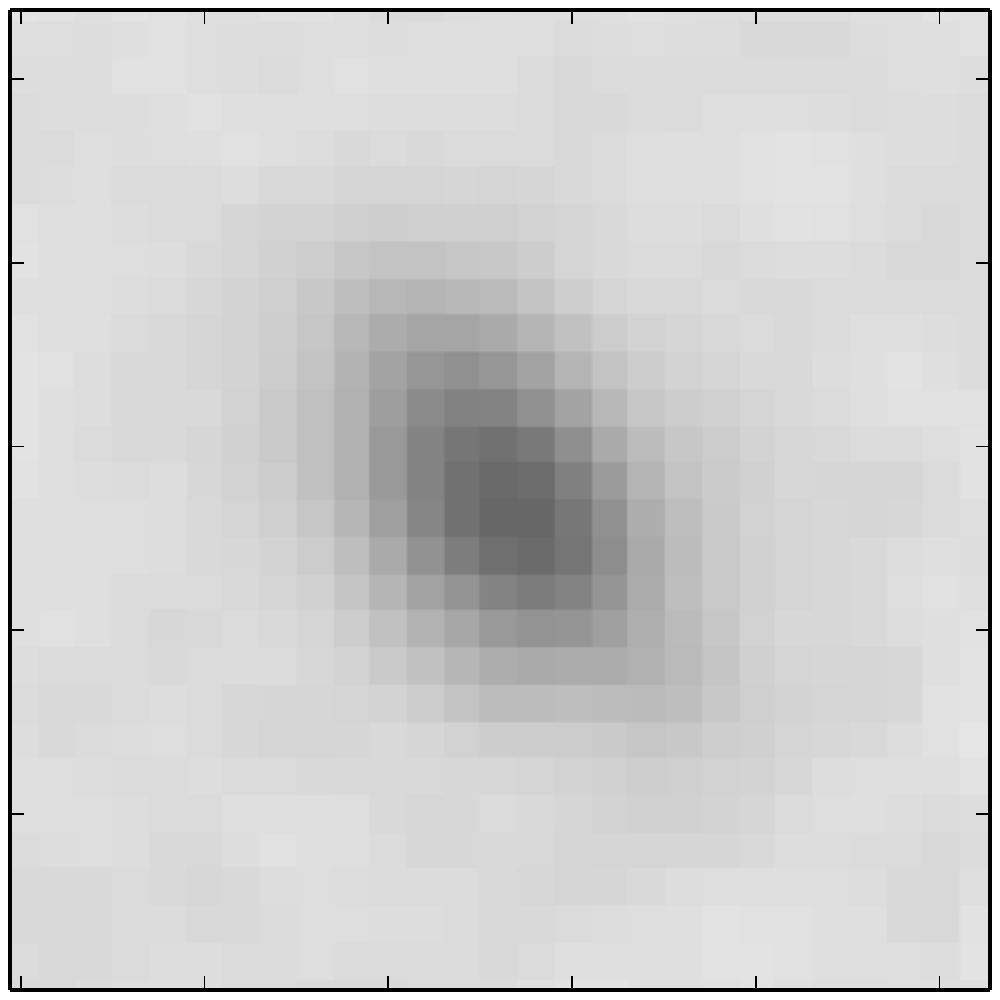}
\\
\hline
\multicolumn{5}{c}{NBES 135 / zCOSMOS 823097} \\
\hline
\includegraphics[width = 0.2\columnwidth]{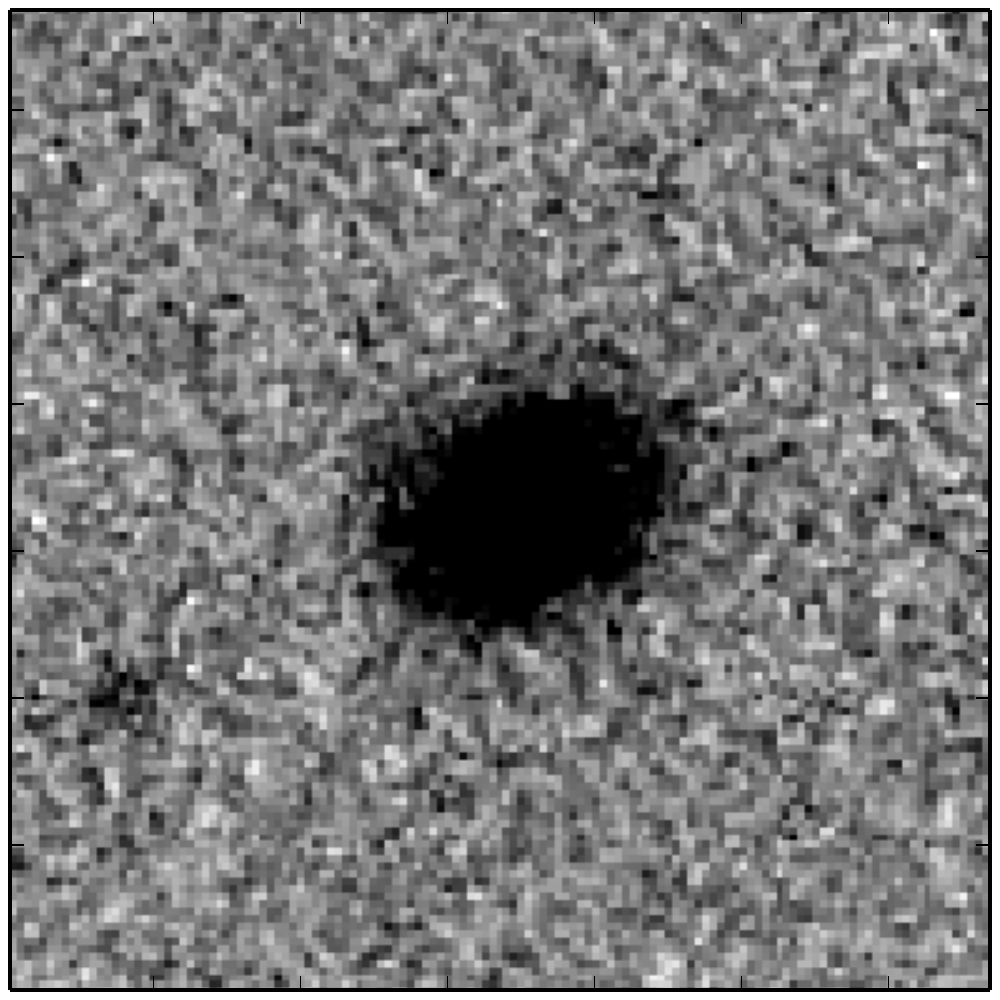} &
\includegraphics[width = 0.2\columnwidth]{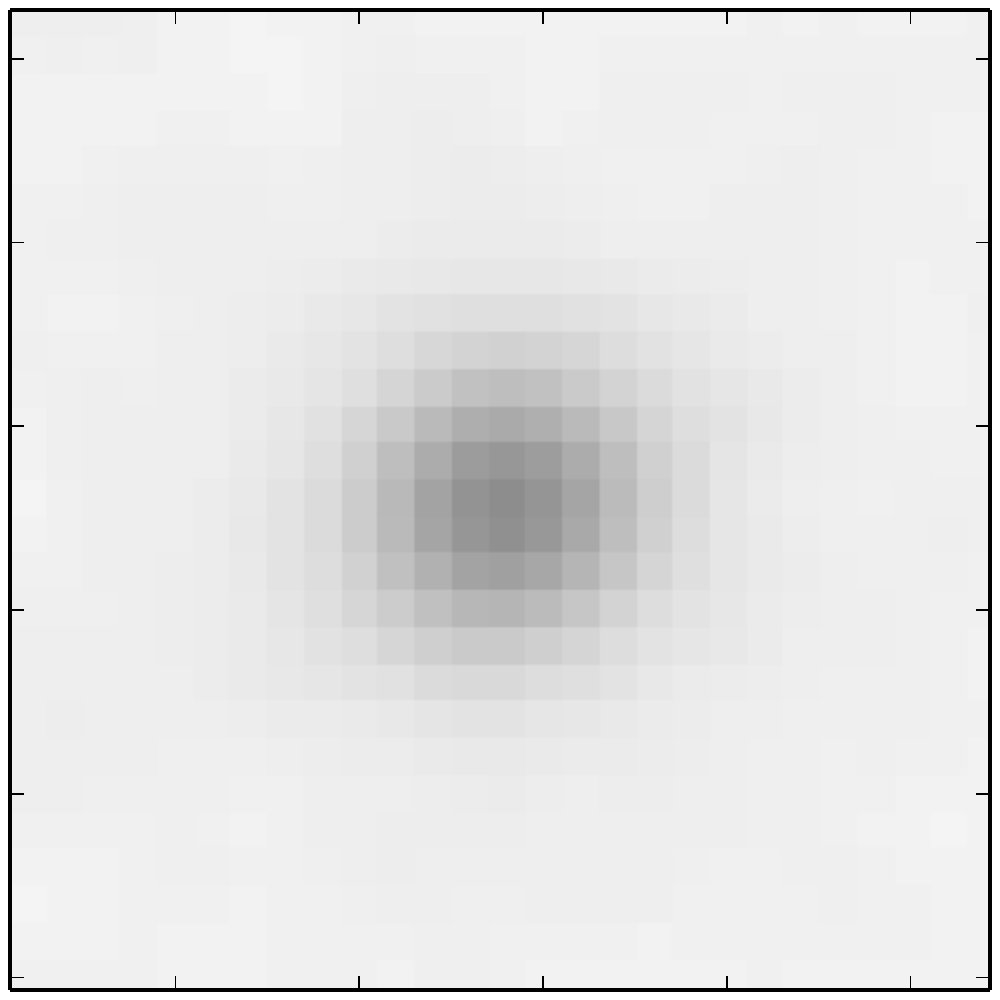} &
\includegraphics[width = 0.2\columnwidth]{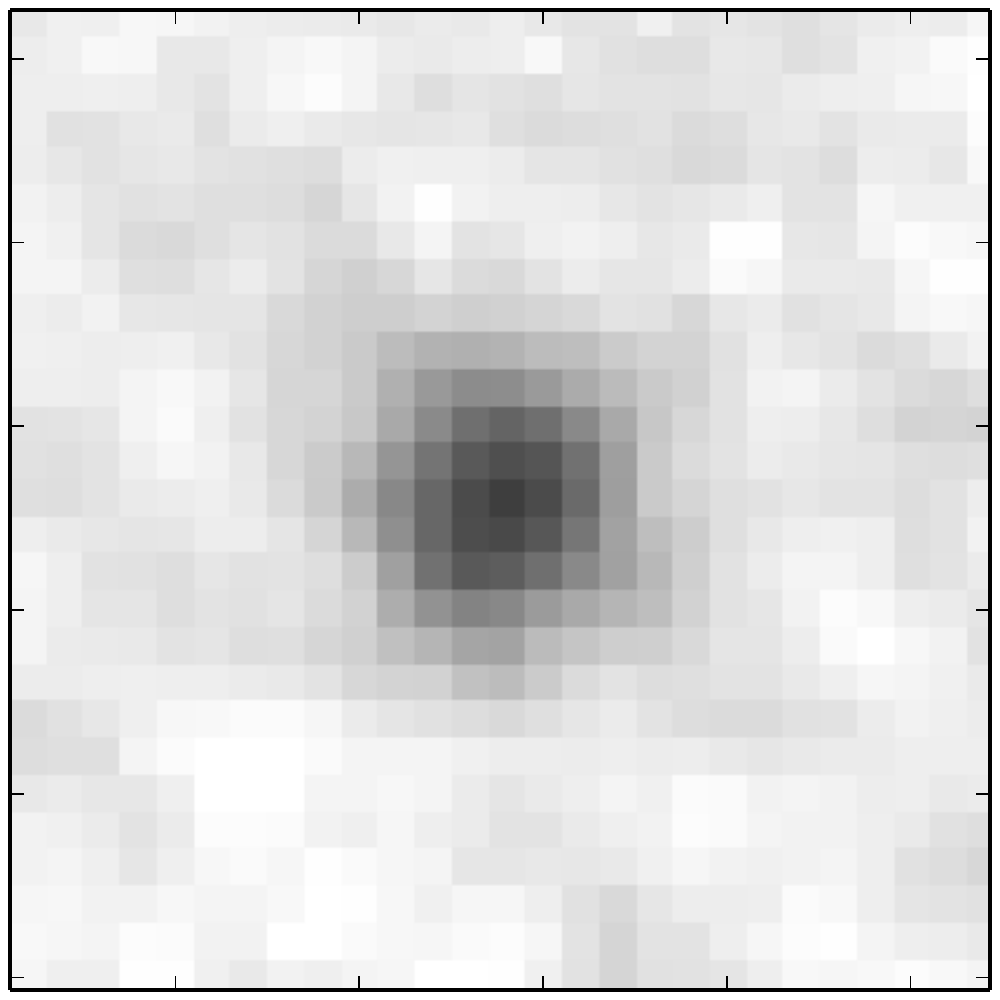} &
\includegraphics[width = 0.2\columnwidth]{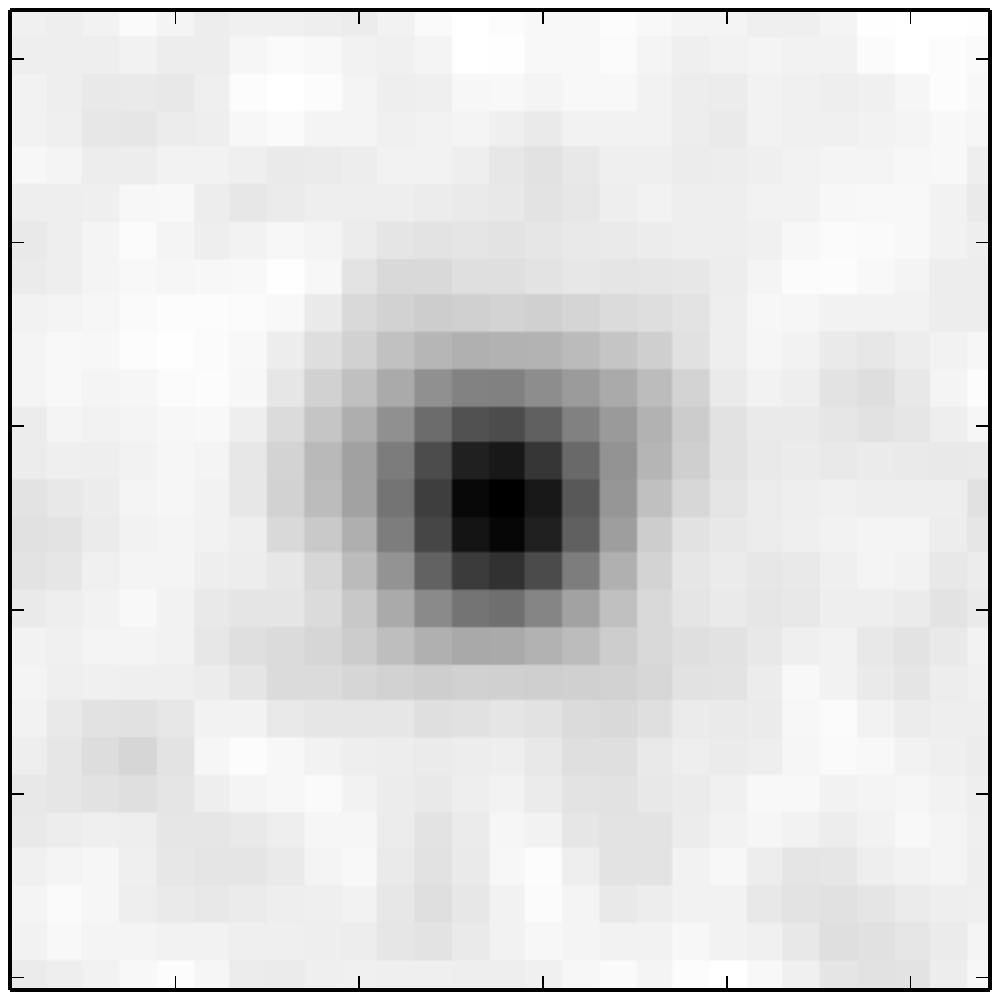} &
\includegraphics[width = 0.2\columnwidth]{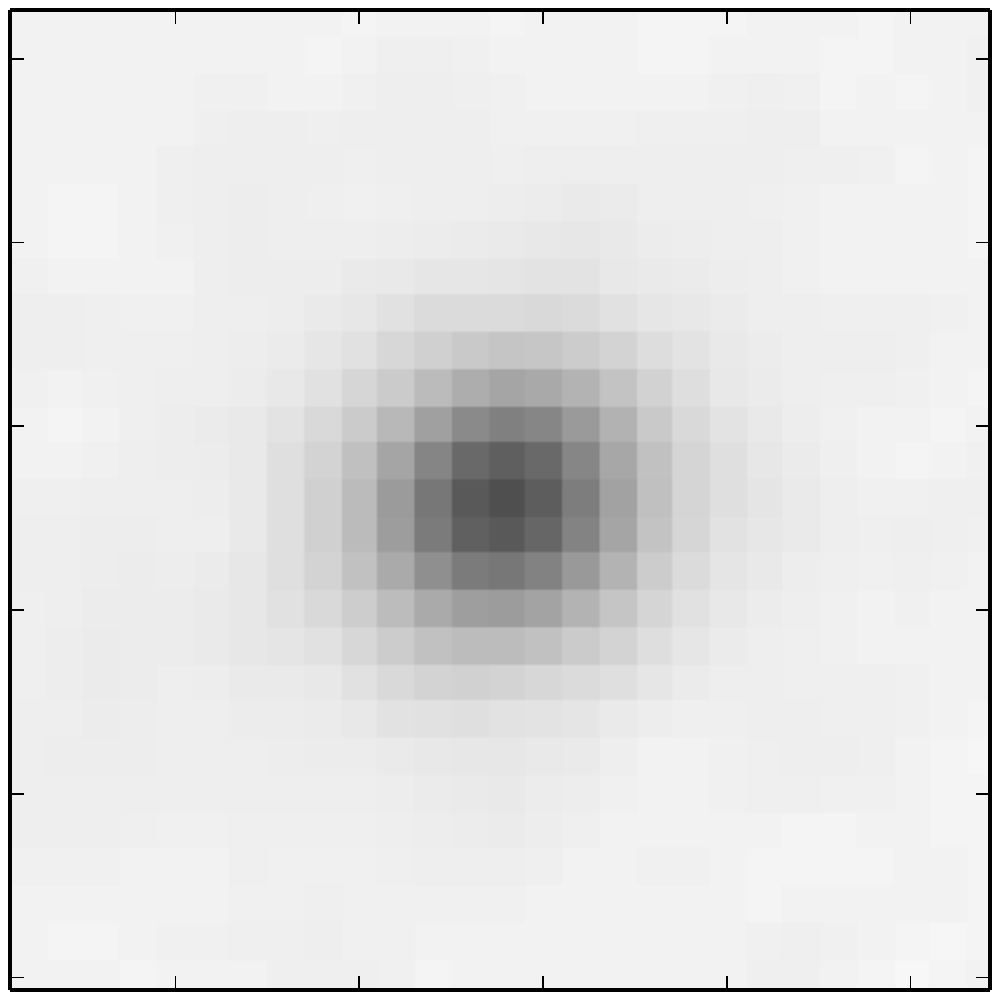}\\
\end{tabular}
\caption{4\arcsec{}x4\arcsec{} cutouts in \ultvis{} DR2 Y and J, NB118
  14 and 15, and HST/ACS/F814W for two \ha{} emitters with significant
  throughput differences between filters 14 \& 15. Both objects have
  spectroscopic redshifts from zCOSMOS \citep{Lilly:2009:218} and are
  included in Table \ref{tab:nb118:table_results1}. For each object,
  the four \ultvis{} panels are scaled to the same surface-brightness
  in $f_\nu$.}
\label{fig:nbexcess_examples}
\end{figure}

%%%%%%%%%%%
\begin{figure}[h]
	\begin{center}
		\includegraphics[width=\columnwidth]{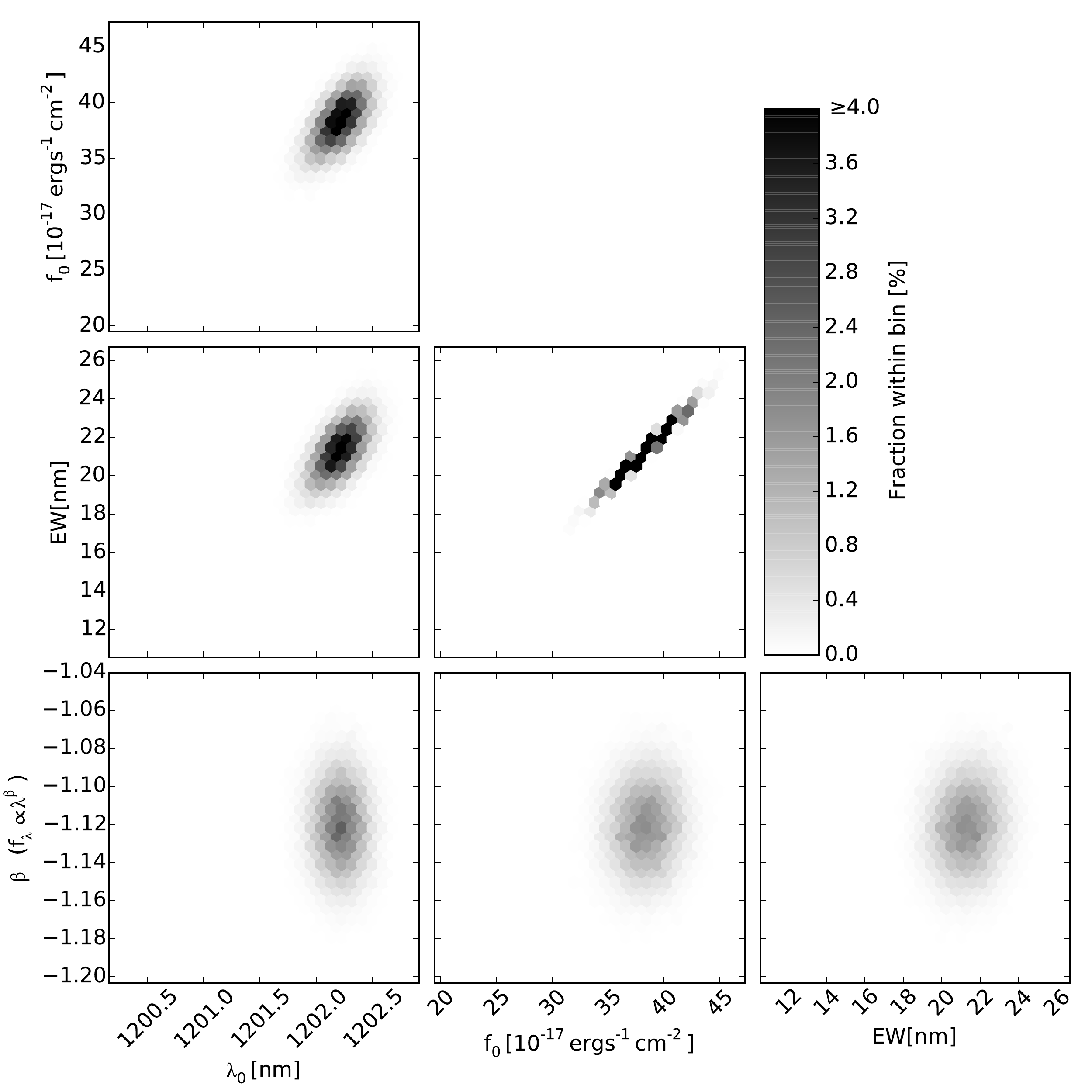}
	\end{center}
	\caption{Results from the TPV parameter estimation with our MCMC code for the example object with the NBES ID 7 (zCOSMOS ID 810332).
	Shown are the 2D histograms, indicating the correlations between the four fitted parameters.
	}
	\label{fig:mcmc_example}
\end{figure}
%%%%%%%%%%%

%###############
\subsubsection{Estimation with TPV}
\label{sec:esimation_details}
%###############

We applied our parameter estimation method to all those NB excess objects in
our sample, which are located in one of the three filter pairs suited for our
method (9 \& 10, 14 \& 15, 15 \& 16) and which we classify as \ha{} emitters
based on their photo-z's or their zCOSMOS redshifts (cf. sec.
\ref{sec:estimation_zcosmos}) being consistent with \ha{} in the NB118 filter. We
also verified the \ha{} selection through the use of a $i'-Ks$ versus $B-r'$
plot, similar to \citet{Sobral:2009:75,Sobral:2013bv}. 
For all selected objects we estimated all \ha{} flux ($f_0$), central
wavelength ($\lambda_0$), $EW_{obs}$, and continuum slope ($\beta$), and the
corresponding uncertainties based on the MCMC implementation of our algorithm
(cf. sec. \ref{sec:nb118:paraest::method}).  Estimated $\lambda_0$ and $f_0$
for all objects are listed in Table \ref{tab:nb118:table_results1}.  An example
for the resulting credibility intervals is shown in Fig. \ref{fig:mcmc_example}
for the object with NBES ID 7, an object with high \ha{} flux and being located
at the boundaries of the passband. The statistical uncertainties are very
small, but as discussed above, systematic errors are at least of the same
order.

%###############
\subsubsection{Estimation with generic NB118 method}
\label{sec:estimation_simple}
%###############

For comparison, we also estimated fluxes with a conventional NB
estimation method using the stack NB118 magnitudes. Here we used as
estimator:

\begin{equation}
	f_0 = (f_{\lambda;NB118} - f_{\lambda;J_\mathrm{corr}}) \times W_\mathrm{NB118}
\end{equation}

The effective filter width, $W_\mathrm{NB118}$, was determined for the
combined effective filters using the approach described by
\citet{Pascual:2007:30}\footnote{The relevant equations from
  \citet{Pascual:2007:30} are especially 7 and 12.}.  In the
$W_\mathrm{NB118}$ we took account for the \fion{N}{ii} contribution
following the approach of \citet{Pascual:2007:30}.  The $f_\lambda$
were determined from the AB magnitudes in the common way.  The used
estimator is the simplest possible form, which assumes that the impact
of the emission line to the broadband magnitude can be neglected.

%###############
\subsection{Independent \ha{} flux estimates}
\label{sec:comp_halpha_estimates}
%###############

In order to asses the quality of our NB118 parameter estimation, we
needed to compare to estimates of $\lambda_0$ and $f_0$ from
independent methods. While a direct comparison to J spectroscopy would
be ideal, we had to rely in the absence of such data on information
obtained from available optical spectroscopy and multi-wavelength
photometry.  This is, we obtained \ha{} flux estimates in three ways:

\begin{enumerate}
\item \ha{} fluxes obtained from SED fitting
\item Conversion of the total SFR obtained from UV+IR into \ha{} fluxes
\item Conversions between H$\beta$ and/or \fion{O}{ii} fluxes from zCOSMOS spectra into \ha{} fluxes
\end{enumerate}

%###############
\subsubsection{Estimation from SED fitting}
\label{sec:estimation_sed}
%###############

We performed SED fitting using our own python code \emph{coniecto},
which normalizes through a common $\chi^2$ minimization a set of
models with respect to mass, and allows consequently to find the model
allowing for the smallest $\chi^2$. Our parameter grid was chosen fine
enough to avoid biases due to degeneracies between different
parameters. The full range of parameters is stated in Table
\ref{tab:sspinput}.

As input we used the \citet{Muzzin:2013:8} photometric catalog and
included in total 29 filters in the fitting, from GALEX FUV to IRAC 4.
The photometry in \citet{Muzzin:2013:8} is based on 2\arcsec{}.1
diameter apertures applied to PSF homogenized images. We compared for
each of the objects our 2\arcsec{} aperture photometry for Y, J, H to
that in the \citet{Muzzin:2013:8} catalog.\footnote{The Y, J, H data in
  the \citet{Muzzin:2013:8} catalog is based on UltraVISTA DR1 data.}
On average, the difference in the magnitudes is with
$0.018\,\mathrm{mag}$ very small. We corrected all quantities obtained
based on the \cite{Muzzin:2013:8} photometry to match our apertures,
which also crudely takes care of small differences in the centroid
from our detection and that in \citet{Muzzin:2013:8}.

In addition to stellar continua based on BC03 models
\citep{Bruzual:2003:1000} using a Salpeter IMF, we also add dust and
nebular emission to the models, including lines and continuum.  For
the dust emission we use the \citet{Dale:2002:159} models under the
assumption of energy conservation, meaning that all radiation absorbed
by dust must be reemitted. We remark that emission from dust,
including PAH features, is in the IRAC bands at $z=0.81$ only of minor
importance.

Throughout this and the following sections we consistently used a
\citet{Calzetti:2000:682} extinction law.  The extinction of the stellar
continuum, $E_{S}(B-V)$, was chosen to be 0.7 times the nebular extinction,
$E_{N}(B-V)$. \citet{Calzetti:2000:682} find based on a sample of local
starburst galaxies \citep{Calzetti:1997:162} a factor 0.44 between the two
extinctions. On the other hand, under this assumption \citet{Erb:2006:128} find
a systematic discrepancy between \ha{} and UV based SFRs at $z\sim2$, with
equal extinction for both components giving more consistent results, in
agreement with some more recent studies \citep[e.g.][]{Shivaei:2015ij}. Other
studies argue for differential nebular and stellar extinction also at high
redshifts
\citep[e.g.][]{ForsterSchreiber:2009cf,Wuyts:2011da}. These discrepant
conclusions can be partially explained by a SFR dependence of the ratio
between nebular and stellar extinction, in a sense that the ratio is higher for
higher SFRs \citep{Price:2014hb,Reddy:2015ho}.
Our chosen value of 0.7 should be understood as a compromise.
For \lya{}, we use either the same $E_{N}(B-V)$ as for all other lines or
somewhat arbitrarily a ten times higher extinction.  Finally, we apply IGM
absorption to the SED models using the parameterization of
\citet{Inoue:2014:1805}.

We interpret the calculated $\chi^2$ grid in a Bayesian way
\cite[e.g.][]{Kauffmann:2003iz,Benitez:2000jr,daCunha:2008cy,Noll:2009:1793}.
Assuming Gaussian errors, the likelihood is given by $e^{-\frac{1}{2}\chi^2}$.
As a prior is naturally imposed through the sampling of the grid, we can
directly interpret the likelihoods as posterior probabilities. We determined
the posterior probability distributions (PDFs) both for the input parameters
and a range of derived parameters through marginalization over the other
input parameters. Marginalization is realized by summing the posterior
probabilities. While we determined the probabilities for the input parameters
at the sampling points of these parameters, we binned for the derived
parameters. Finally we determine the 68\% confidence intervals, by excluding
the first and last 16\% in the PDFs. If the point of minimum $\chi^2$ is
outside the 68\% interval, we extend the uncertainty interval to include the
minimum $\chi^2$ value. 
Uncertainties derived in this way are listed for the SED based $f_{H\alpha}$
estimates in Table \ref{tab:nb118:table_results1} and for other SED parameters
in Table \ref{tab:nb118:table_results2}.

As we are adding 5\% of the flux to the formal flux-uncertainties, in order to
reduce artificial impacts of possible ZP uncertainties and template mismatches,
the stated uncertainties should not be over-interpreted.

%###############
\subsubsection{Estimation from total SFR}
\label{sec:estimation_total}
%###############

For those objects with significant Spitzer/MIPS $24\,\mathrm{\mu m}$
detection in the \citet{Muzzin:2013:8} catalog, we obtained total SFRs
from the sum of UV and IR based SFRs.  While \citet{Muzzin:2013:8}
provide these values as part of their catalog, we can assume a more
precise redshift and it is hence worth to recalculate the values.

We determined total IR luminosities, $L_{FIR}$, by scaling the
\cite{Dale:2002:159} templates so that synthetic MIPS magnitudes match
the measured ones, and consequently integrating the scaled templates
over the range from $8\mbox{--}1000\,\mathrm{\mu m}$. Following
\citet{Wuyts:2008:985} we used as result the mean of the
\citet{Dale:2002:159} models for $\alpha$ between $1\mbox{--}2.5$,
where $\alpha$ is the power law index, characterizing the fractional
dust mass, $dM_{d}$, heated by a certain interstellar radiation
intensity, U, meaning $dM_{d}(U)\propto U^{-\alpha}d U$.  Upper and
lower limit of the stated uncertainties are given by the values
obtained for $\alpha=1$ and $\alpha=2.5$. It is noteworthy, that in
the case of contribution from an AGN, the determined values will not
be correct.

The total infrared luminosity can be converted into a SFR by
\citep{Kennicutt:1998:189}:
\begin{eqnarray}
	\mathrm{SFR}_\mathrm{IR}(M_\odot\;\mathrm{yr}^{-1}) = 4.5\times10^{-44} L_{FIR} [\;\mathrm{erg}\;\mathrm{s}^{-1}]
	\label{eq:convirsfr}
\end{eqnarray}

For the unobscured UV SFR we determined first the
the continuum luminosity density at a rest-frame wavelength of
$\lambda=2800\mbox{\AA}$ from our best fit SED model and converted this luminosity into a SFR by
using \citep{Kennicutt:1998:189}:

\begin{eqnarray}
	\mathrm{SFR}_\mathrm{UV}(M_\odot\;\mathrm{yr}^{-1}) = 1.4\times10^{-28} L_\nu [\;\mathrm{erg}\;\mathrm{s}^{-1}\;\mathrm{Hz}^{-1}]
	\label{eq:convluvsfr}
\end{eqnarray}

Finally, we converted the determined total
$\mathrm{SFR}_\mathrm{tot}$, being the sum of
$\mathrm{SFR}_\mathrm{UV}$ and $\mathrm{SFR}_\mathrm{IR}$, to \ha{}
fluxes, using the relation from \citet{Kennicutt:1998:189},
\begin{eqnarray}
	\mathrm{SFR}(M_\odot\;\mathrm{yr}^{-1}) = 7.9 \times 10^{-42}\;L(\mathrm{H}\alpha)[\mathrm{erg}\;\mathrm{s}^{-1}] \label{eq:kennicutt},
	\label{eq:convoiiha}
\end{eqnarray}

the generic relation between flux and luminosity for our assumed
cosmology, and the nebular extinction obtained from the SED fit.

%###############
\subsubsection{Estimation from zCOSMOS data}
\label{sec:estimation_zcosmos}
%###############

In addition, we were using redshifts and line fluxes from the zCOSMOS survey
\citep{Lilly:2007:70, Lilly:2009:218}.  Matching the coordinates of our NBES
objects to the zCOSMOS-bright 20k data-set\footnote{The publicly available
zCOSMOS-bright DR2 10k is a subset of this catalog.} revealed an overlap of 35
objects with redshift information. It is reassuring that for all matched
objects the redshifts confirmed the presence of an emission line within the
filter, which is for 31 objects \ha{} + \fion{N}{ii}. The number of those \ha{}
emitters being in the three most useful filter pairs 9 \& 10, 14 \& 15, and 15
\& 16 are 3, 11, and 2, respectively.

The zCOSMOS VLT/VIMOS spectra are covering the wavelength range from
550 to 970$\mathrm{nm}$. This means that they include for an object
with \ha{} in the NB118 filter both
\fion{O}{ii}$\lambda\lambda3727,3729$, being unresolved in the VIMOS
data, and H$\beta$ at observed-frame wavelengths of $675\,\mathrm{nm}$
and $880\,\mathrm{nm}$, respectively, assuming $z=0.81$.

We matched the \emph{zCOSMOS} spectral fluxes to our imaging apertures
by multiplying the continuum flux density at the lines' wavelength
obtained from the SED fitting with the respective zCOSMOS
$EW_\mathrm{obs}$.  In this way, we avoid slit loss and flux
calibration issues, with some remaining discrepancy expected from the
spatial distribution of \ion{H}{ii} regions, if objects are more
extended than the PSF.

The \fion{O}{ii} fluxes can be converted into SFRs, using the
calibration by \citet{Kewley:2004:2002}:
\begin{eqnarray}
	\mathrm{SFR}(M_\odot\;\mathrm{yr}^{-1}) = (6.58\pm1.65) \times 10^{-42}\;L(\mathrm{[OII]})[\mathrm{erg}\;\mathrm{s}^{-1}]
	\label{eq:convoiiha2}
\end{eqnarray}
This equation is for intrinsic, meaning reddening corrected
luminosities.  Therefore, we de-reddened the measured \fion{O}{ii}
fluxes as an intermediate step, again assuming the $E_{N}(B-V)$ from
the best fit SED. The obtained SFR was then converted into an \ha{}
flux using again eq. \ref{eq:convoiiha}.

The ratio between \fion{O}{ii} and \ha{} is depending to some degree
on metallicity and the ionization parameter
\citep[e.g.][]{Moustakas:2006:775}. By contrast, H$\beta$ allows for a
more direct conversion, with the additional advantage of a lower
difference in reddening between the wavelengths of H$\beta$ and \ha{}
than between \fion{O}{ii} and \ha{}. Unfortunately, the H$\beta$ S/N
is relatively low in the zCOSMOS spectra.  For those objects with
H$\beta$ detection at least at the 2$\sigma$ level, we obtained an
\ha{} estimate by converting between the reddening corrected
values. The intrinsic ratio between \ha{} and H$\beta$ is 2.86 for
typical conditions in \ion{H}{ii} regions, assumed to be
$n_{e} = 100\;\mathrm{cm}^{-2}$ and $T_{e} = 10000\;\mathrm{K}$
\citep[][p. 84]{Osterbrock:1989}.

%###############
\subsection{Comparison of TPV with other estimates}
\label{sec:nb118:comptpvother}
%###############

%%%%%%%%%%%
\begin{figure}[h]
	\begin{center}
		\includegraphics[width= 0.9\columnwidth]{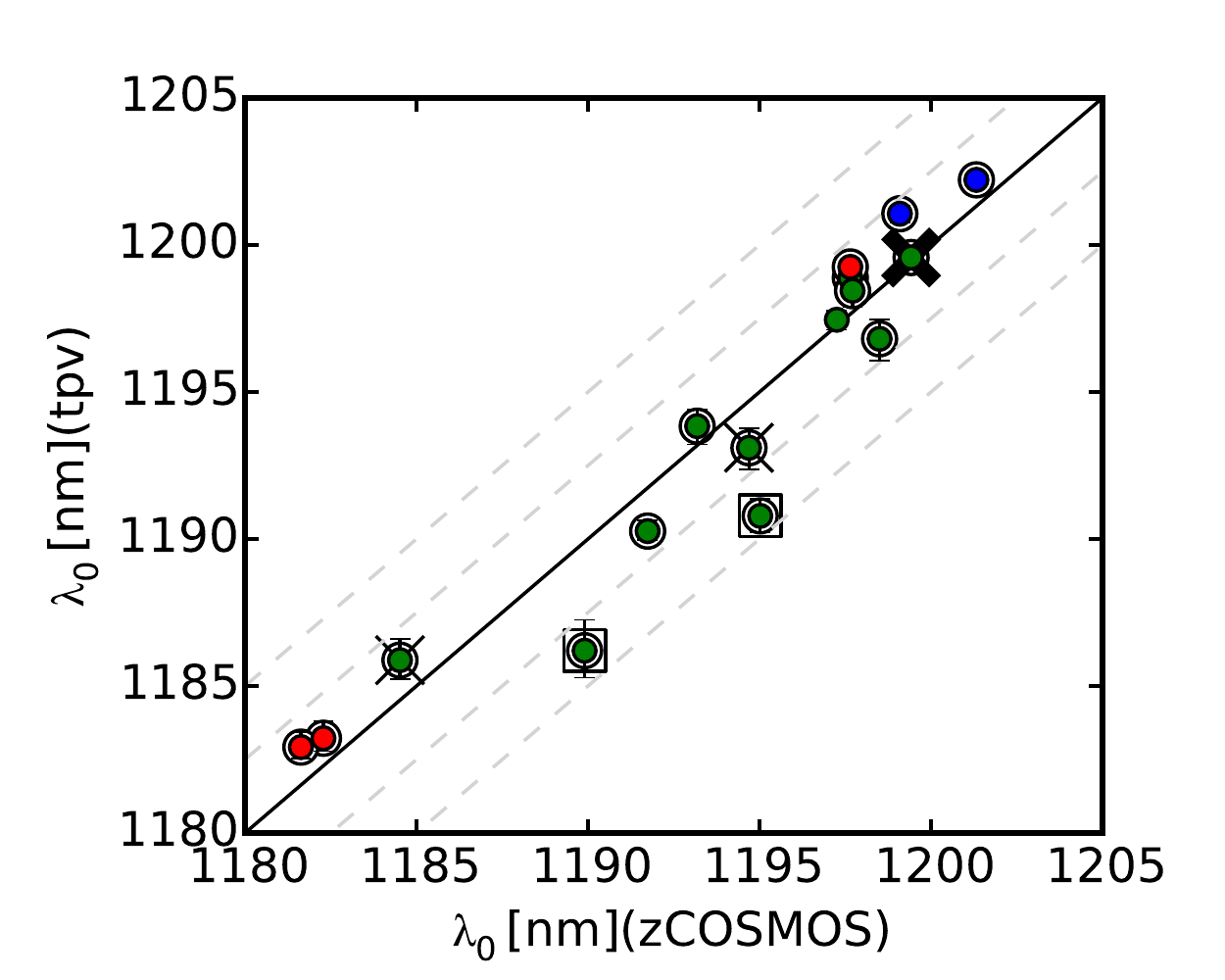}
	\end{center}
	\caption{Comparison between \ha{} central wavelengths obtained from the \emph{zCOSMOS} redshifts and those obtained from our TPV method. The dashed diagonal lines indicate differences of $2.5\;\mathrm{nm}$ and $5\;\mathrm{nm}$ from the 1:1 relation, respectively. Used symbols are explained in the legend of Fig. \ref{fig:wavelen_fluxratio_sed}.
          \label{fig:wave_est}}
\end{figure}
%%%%%%%%%%%

Using the zCOSMOS redshifts and the line flux estimates of sec.
\ref{sec:comp_halpha_estimates}, we can now directly compare the TPV
estimates with those from completely independent measurements and
thereby measure the success of the TPV method. First, we consider the
redshifts. Because of the high accuracy of the spectroscopic
redshifts, the accuracy of the TPV estimates can be directly assessed
with a straight forward comparison.  In Fig.  \ref{fig:wave_est}, the
TPV estimates of the wavelength $\lambda_0$ is plotted against the
$\lambda_0$ obtained from the spectroscopic \emph{zCOSMOS} survey.  It
can be seen that the TPV wavelength estimates closely follow
spectroscopic estimates, with a mean difference of
$-0.10\,\mathrm{nm}$ and a scatter of $1.9\,\mathrm{nm}$. All except
two objects are within $2.5\,\mathrm{nm}$. Excluding these two
outliers, the scatter decreases to $1.2\,\mathrm{nm}$. We remark that
the observed scatter is larger than the statistical error estimate
from our estimation code. This indicates that the errors are dominated
by systematic effects. Expected systematic uncertainties result from
mismatches between true and estimation continua, discrepancies between
true and assumed \fion{N}{ii} ratios, and uncertainties in the
available filter curves (cf. Appendix \ref{app_nb118}).

\begin{figure}[h]
	\begin{center}
		\includegraphics[width= \columnwidth]{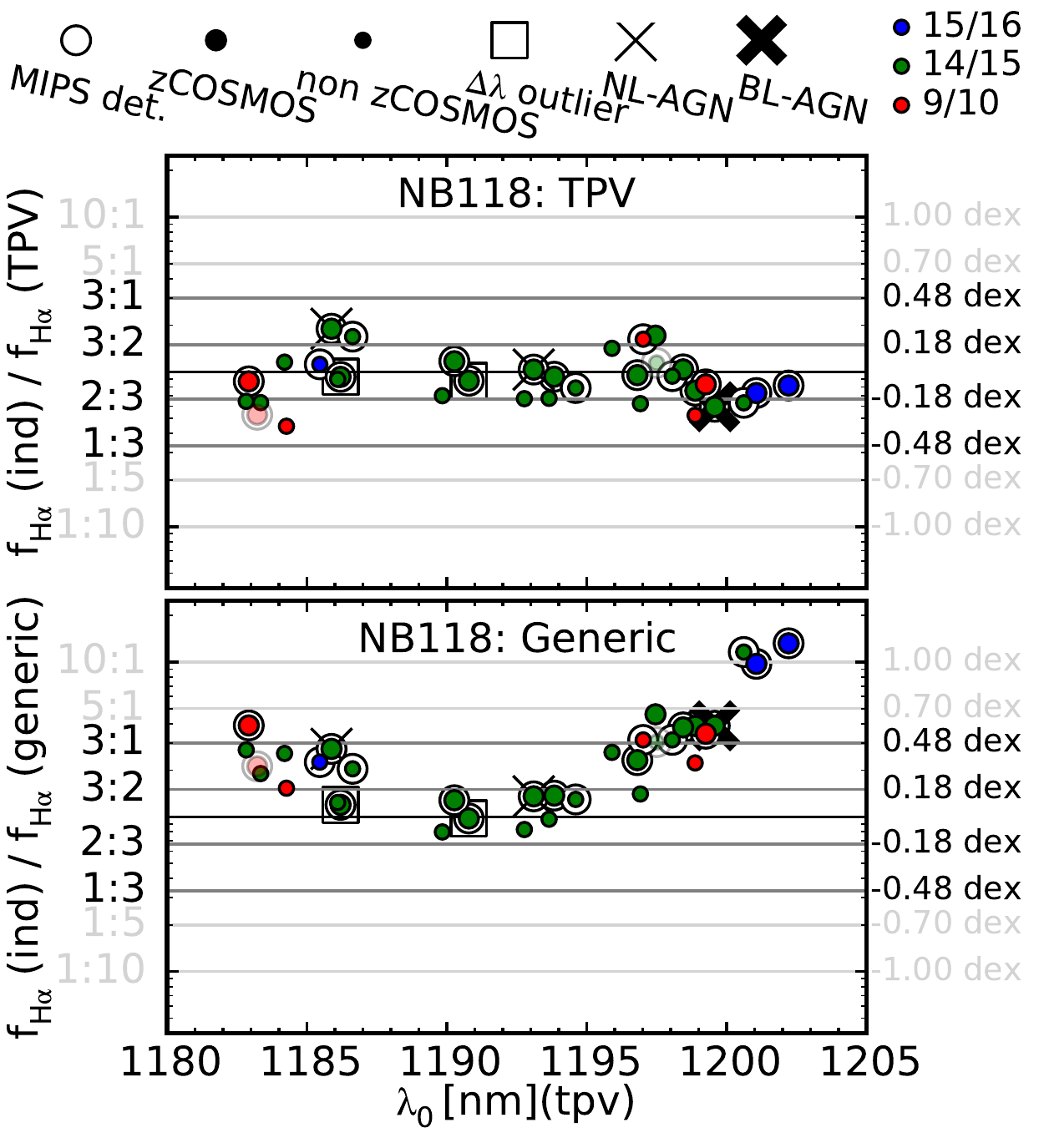}
	\end{center}
	\caption{ \label{fig:wavelen_fluxratio_sed} \textbf{Upper
            panel:} Ratios between the \ha{} fluxes estimated from a
          combination of different methods, which are independent from
          the NB118 data (ind;
          cf. sec. \ref{sec:comp_halpha_estimates} and
          \ref{sec:nb118:comptpvother}), and those estimated based on
          the NB118 data using the TPV technique. Colors refer to the
          three considered filter pairs, while symbols have meanings
          as indicated in the legend, with combinations of the
          different symbols possible. Objects with strongly discrepant
          individual independent flux estimates are shown faded.
          \textbf{Lower panel:} Similar ratios as in the upper panel,
          but now using instead of the TPV method a generic NB
          estimation method.  }
\end{figure}

\begin{figure}[h]
	\begin{center}
		\includegraphics[width= \columnwidth]{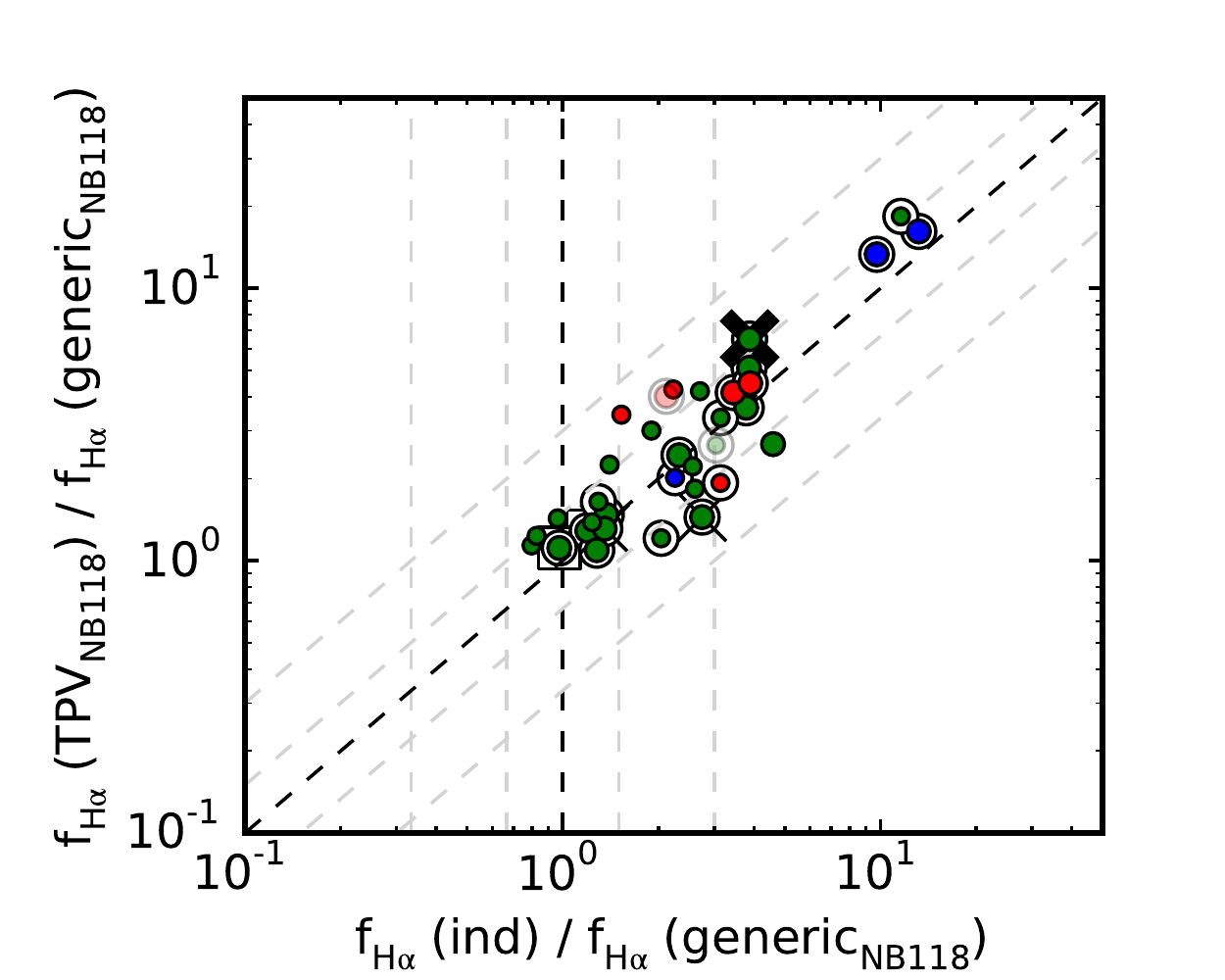}
	\end{center}
	\caption{Ratio between the TPV and a generic NB118 flux estimate is plotted against the ratio between an independent flux estimate and the generic NB118 flux estimate. For objects on the diagonal the independent estimate and the TPV NB118 estimate are identical. The dashed lines indicate factors of 1.5 and 3 between the estimates. Used symbols are explained in the legend of Fig. \ref{fig:wavelen_fluxratio_sed}.
          \label{fig:alt_ratios}
        }
\end{figure}

Next, we consider the \ha{} line flux measurements. In this case,
individual independent estimates are not necessarily more accurate
than the TPV estimates.  For that reason, we combined UV+IR, SED,
\fion{O}{ii} and H$\beta$ based estimates by taking the mean of the
individual values.  In Fig.  \ref{fig:wavelen_fluxratio_sed}, the ratio of the
NB118 line flux estimates to the combined estimates are plotted versus
wavelength.  The top panel shows the TPV estimates, while the bottom panel shows
the generic discussed in Sec. \ref{sec:estimation_simple}.  The same data is
shown in a different way in Fig. \ref{fig:alt_ratios}. In this figure the
ratios between the three different estimates can be directly assessed for
individual objects.

%%%%%%%%%%%
%%%%%%%%%%%

%%%%%%%%%%%
%%%%%%%%%%%

For several objects the UV+IR and the SED based estimates differ by more than a
factor ten. The combined line flux for these objects is therefore uncertain,
and these objects are shown faded in the figure.  Specifically marked in the
figure is one of the objects, which is according to the zCOSMOS flag 13.5 (cf.
\citealt{Lilly:2009:218}) a broad line AGN, and one object which shows
\fion{Ne}{v}$\,\lambda3426$ in its spectrum and is hence identified as hosting
a type II AGN
\citep[e.g.][]{Mignoli:2013:29}. 
About 80\% of the field with data from the filter pairs 9 \& 10, 14 \& 15, and 15 \& 16 are covered by Chandra data from \citet{Elvis:2009io}, out of which $\sim40\%$ have deep coverage.
Matching to the Chandra point source catalog \citep{Civano:2012:30} revealed one further X-ray detected object.  We classify it for the plot as NL AGN. Our independent flux estimates for these objects with certain AGN contribution are only of limited usefulness.

The figure demonstrates that the agreement between the TPV and
combined independent line flux estimates is good. The mean of the
differences is $-0.06\,\mathrm{dex}$ with a standard deviation of
$0.15\,\mathrm{dex}$. In particular, there is no trend with wavelength
over the complete wavelength range of the bandpass.  By contrast, the
generic single NB filter estimate leads to substantially biased line
flux estimates, that can underestimate the flux by as much as a factor
of 10 towards the edge of the filter band.  Such a bias would for
example have significant impact on the investigation of the structure
at wavelengths around $1198\,\mathrm{nm}$ (see sec.
\ref{sec:spatialstruc}).

Overall, we conclude that our TPV flux estimates are robust and
unbiased, and the error on the flux is by as much as a factor of 20
smaller than using generic single NB estimates to derive line fluxes.

%##############################
\subsection{Spatial and redshift distribution in example filter pair}\label{sec:spatialstruc}
%##############################

%%%%%%%%%%%
\begin{figure}[h]
\begin{center}
\includegraphics[width=
    1.0\columnwidth]{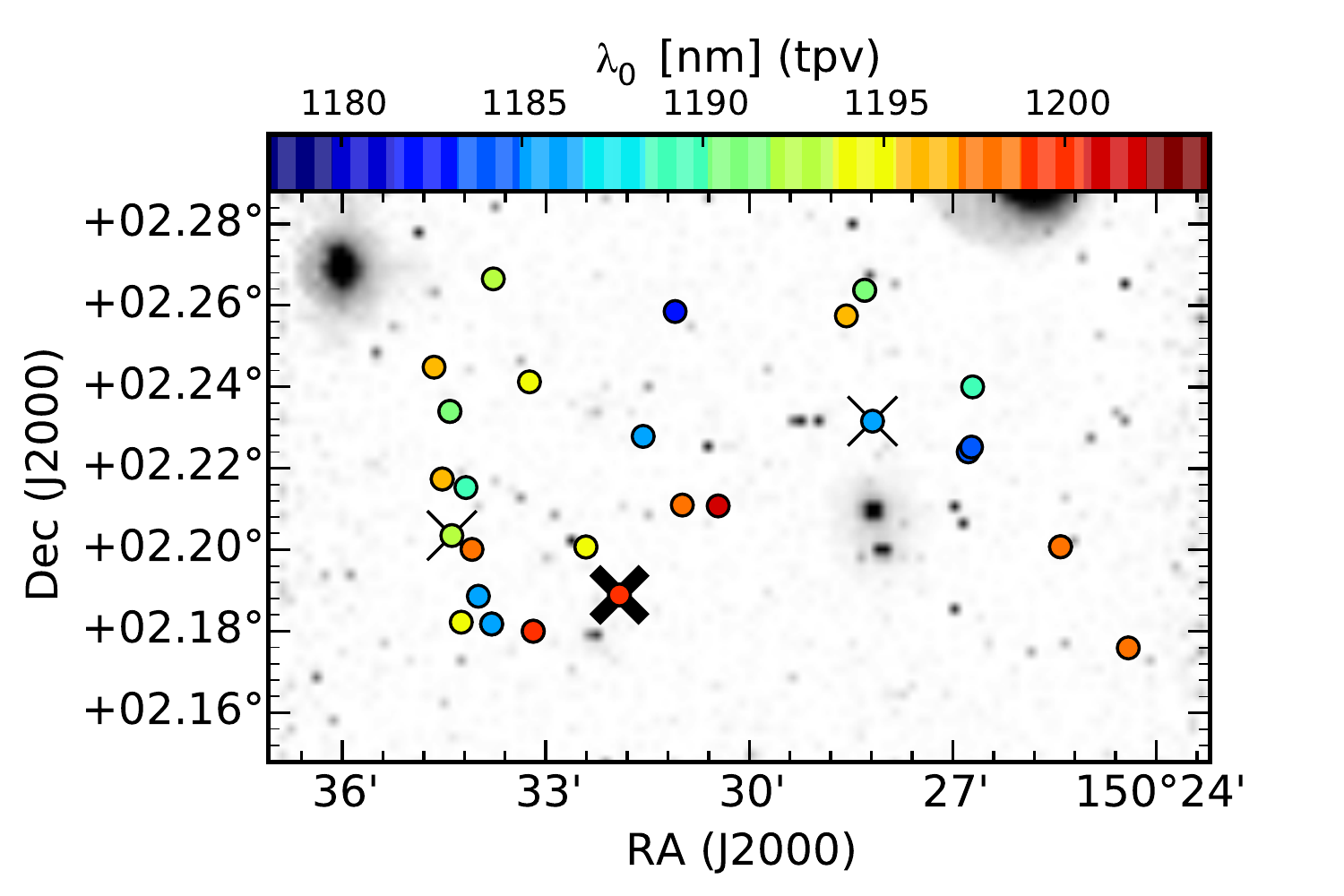}
 \end{center}
 \caption{Position and \ha{} wavelength obtained with the TPV method
   for objects in the part of \ultvis{} that is covered by the NB118
   filter pair 14 \& 15. Indicated as thick cross is a BL AGN, while
   two NL AGNs are indicated as thin crosses. The \ultvis{} DR2 NB118
   image is shown in the background.\label{fig:field_plot}}
\end{figure}
%%%%%%%%%%%

The use of the TPV method allows us to directly identify three
dimensional structures, like filaments or sheets, the detection of
which would otherwise require spectroscopic observations with
sufficient resolution.

This is demonstrated in Fig. \ref{fig:field_plot}, which shows the spatial
distribution of \ha{} emitters observed with the filter pair 14 \& 15, combined
with the wavelength information obtained from the TPV.  The size of one NB118
detector corresponds to a comoving distance of 9.7 Mpc at the redshift of \ha{}
in the NB118 filter, while the depth of the volume covered by the wavelength
range from $1180\,\mathrm{nm}$ to $1205\,\mathrm{nm}$ is much larger, namely 103
Mpc comoving.\footnote{Alternatively, expressing central wavelength differences
as redshift velocities, $1\;\mathrm{nm}$ corresponds to
$252\,\mathrm{km}\;\mathrm{s}^{-1}$.} Observations with a single NB118 filter
can not resolve the depth of the field.

By contrast, the redshift resolution obtained with the TPV is
sufficient to identify several objects that are at similar redshifts
as the BL-AGN, revealing substantial clustering associated with the
AGN.  In addition, there are also several \ha{} emitters within the
field but at the other redshift end of the volume, i.e. they are
spatially well separated from the AGN cluster.  Finally, there is a
string of objects towards the east. This feature includes objects at
various redshifts, and is likely a sheet-like structure.  While a
deeper discussion of these structures is beyond the scope of this
paper, this example demonstrates the amount of additional information
which can be gained from the TPV.

%##############################
\subsection{Optimizing the observing pattern}
\label{sec:nb118:ultravista:turning}
%##############################

%%%%%%%%%%%
\begin{figure}[t]
\centering
\includegraphics[width=0.9 \columnwidth]{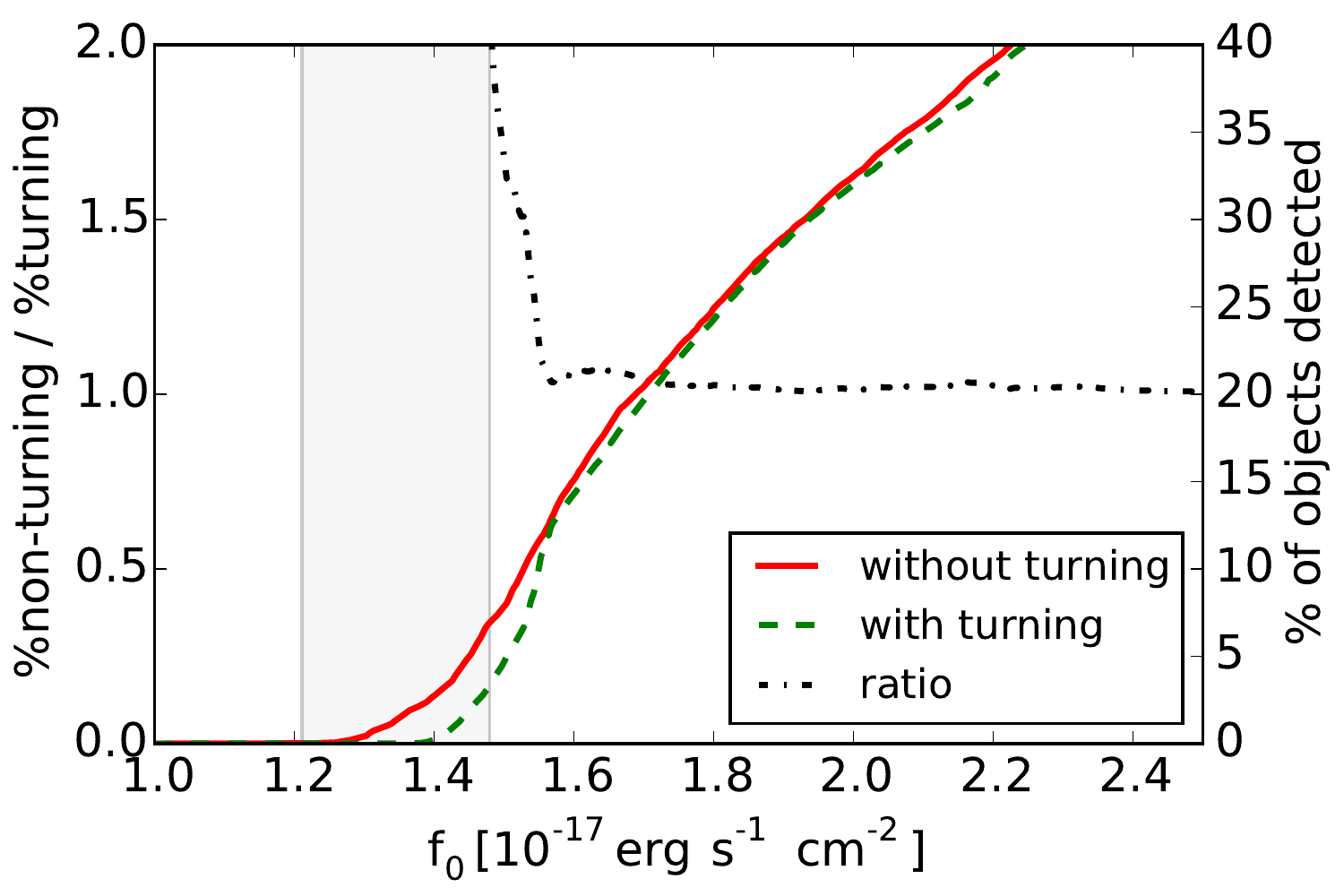}
\caption{Simulated detection completeness as a function of line flux
  both for the standard \ultvis{} observing pattern (without turning;
  solid) and a possible modification (with turning; dashed). The
  dotted-dashed line gives the ratio between the two completeness
  curves. Fluxes for which twice as many objects are expected in the
  standard pattern as in the modified pattern are indicated by the
  shaded area.  No objects will be detected left of this area in
  either of the two observing patterns. The completeness curves were
  calculated for point sources with a spectrum consisting of a single
  infinite $EW$ emission line.  The underlying detection criterion was
  a color-significance of the NB118 excess in the stack including data
  from the 16 filters jointly at least at the 5$\sigma$ level (cf.
  sec. \ref{sec:nb118:ultravista:turning} for more details).  }
\label{fig_turning_nonturning}
\end{figure}

%%%%%%%%%%%

As discussed in sec. \ref{sec_mno}, the best suited pairs have
$\Delta\,mag \mbox{--} \lambda_0$ curves with average slopes in
$\Delta\,mag$ of $0.10\,\mathrm{mag}\;\mathrm{nm}^{-1}$ and are
monotonous over the relevant wavelength range. Unfortunately, the
standard \ultvis{} observing pattern leads to only three cases where
the same position on the sky is observed with such ideal filter
combinations, namely the filter combinations 9 \& 10, 14 \& 15, and 15
\& 16.  One reason for this is that the filter arrangement within the
VISTA camera was chosen to maximize the obtainable depth and hence the
filters are as similar as possible within each column
\citep{Nilsson:2007:106,Milvang-Jensen:2013:94}.  Therefore, the
overlapping filters are in most cases more similar than the overall
spread between the 16 NB118 filters suggests
(cf. Fig. \ref{fig_all_filter}).

In order to make use of the full potential of the proposed TPV method, the
observing pattern should be modified to increase the number of cases where
repeated observations of the same field use dissimilar filters. One easy to
implement strategy is to turn the telescope by 180\degr{} for half of the
observation time. With this strategy, every observed position is covered by at
least two filters, whereas in the standard observing pattern, about 75\% of all
positions are covered with only one filter.  Furthermore, in the other patches,
where already in the standard observing pattern two filters contribute, there
would be data from four filters.  This increases the number of combinations,
which are different enough to allow for a good parameter estimation. Out of the
44 patches more than half (24) have an average slope of the $\Delta mag
\mbox{--} \lambda_0$ curve larger than
$0.10\,\mathrm{mag}\;\mathrm{nm}^{-1}$.\footnote{The average slope was
calculated as further described in Appendix
  \ref{sec:appendix:quantitativeall}. For those patches, where four
  filters contribute, we calculated an average slope from the three
  largest slopes for all six possible combinations.}

While the turning pattern would be an enormous step forward for the
method, it comes at a prize.  The specific positioning of similar
filters is the logical step in order to maximize the reachable depth,
being most crucial for one of the main science goals of the VISTA
NB118 observations, the search for $z=8.8$ Ly$\alpha$ emitters. We
evaluated the impact of this loss in depth with the simulation
described in Appendix \ref{sec:ultravista:simu}. This simulation takes
account for the difference in filter profiles and background in the
individual filters.  In short, we randomly assigned positions on the
sky and central-wavelengths to $3\times10^5$ objects, simulated the
observations, and determined the fraction of input objects which would
be detected as a function of wavelength.  Different from the selection
criteria in sec.  \ref{sec:selcriteria}, we required here a $5\sigma$
color-significance in the complete stack instead of the two-filter
criteria, and considered the complete field. To test effects of the
finite number of simulated objects, we also tested the variance
between 10 subsets of $3\times10^4$ objects, and found negligible
impact on the results.

In Fig. \ref{fig_turning_nonturning}, the fraction of detected objects
as function of the flux is shown both for the standard observing
pattern and the turning modified one.  There is clearly a
non-negligible fraction of objects that will be missed at the very
faint end when using the turning pattern. Therefore, deciding against
or in favor of the turning pattern is a difficult trade-off.

%##############################
\section{Summary and Conclusions}
%##############################

In this paper, we have carefully demonstrated the usefulness of the TPV method
to derive redshifts and line fluxes from wavelength dependent throughput
differences between NB filters with slightly differing, yet overlapping
passbands, and corresponding broadband filters.  While it is possible to
specifically design NB filters for this method, suitable filters and data taken
with such filters already exist, e.g. the \ultvis{} survey taken with ESO's
VISTA/VIRCAM. For our analysis, we focused on the \ha{} line in the narrowband
NB118 filters of that survey. About 1/4 of the \ultvis{} field is covered by at
least two different NB118 filters. We used simulations to assess the expected
accuracy of our method given the current exposure time of the survey. We found
that for the most suitable filter pair, the simulations predict that it is
possible to measure wavelengths with random and systematic errors as low as
$1\;\mathrm{nm}$ for a line with $f_0 =
10\times10^{-17}\;\mathrm{erg}\;\mathrm{s}^{-1}\;\mathrm{cm}^{-2}$ independent
of $EW$.  The wavelength estimation results were also shown not to be strongly
affected by assumptions on the line width and the required \fion{N}{ii}$\lambda
6583$/H$\alpha$ ratio.

The accuracy in redshift compares favorable to photometric redshifts.
A wavelength error of $1\,\mathrm{nm}$ corresponds to a
$\sigma \frac{\Delta z}{1+z} = 0.001$.  By comparison, highest quality
photometric redshifts in the COSMOS field can at best reach a
resolution of about $\sigma \frac{\Delta z}{1+z} = 0.01$
\citep[e.g.][]{Ilbert:2013:55}.

In addition to the simulations, we also applied the method to the actual
\ultvis{} DR2 data. A comparison of wavelengths estimated with our method and
those obtained from the spectroscopic zCOSMOS-bright 20k catalog
\citep{Lilly:2009:218} shows an excellent agreement with a measured scatter of
$\sigma \frac{\Delta z}{1+z} = 0.0016$. This value is similar to the $\sigma
\frac{\Delta z}{1+z} = 0.002$ found by \citet{Hayashi:2014ec} in their
observations of a galaxy-cluster with two optical NB filters in Suprime-Cam on
the Subaru Telescope. Independent predictions for the \ha{} line flux, both
based on spectroscopic \fion{O}{ii} and H$\beta$ fluxes and photometric data,
also confirm that the proposed method works very well and is a significant
improvement compared to the results from a generic line flux estimation. This
improvement is shown again in Fig. \ref{fig:alt_ratios}.

We therefore conclude that the TPV method is a powerful tool to derive redshift
and flux estimates from NB surveys that employ multiple versions of similar
filters. One of the advantages of the method is that it exploits information
that is routinely collected with some instruments.  The improvement over the
standard analysis is as much as an order of magnitude reduction in the errors.

%##############################
\begin{acknowledgements}
  We thank the anonymous referee for very constructive comments. The Dark
  Cosmology Centre is funded by the Danish National Research Foundation. JZ
  acknowledges support from the ERC Consolidator Grant funding scheme (project
  ConTExt, grant number 648179). BMJ and JPUF acknowledge support from the
  ERC-StG grant EGGS-278202.  Part of this research was funded by an ESO DGDF
  grant to WF and PM.  Based on data products from observations made with ESO
  Telescopes at the La Silla Paranal Observatory under ESO programme ID
  179.A-2005 and on data products produced by TERAPIX and the Cambridge
  Astronomy Survey Unit on behalf of the \ultvis{} consortium. The zCOSMOS
  observations are based on observations made with ESO Telescopes at the La
  Silla or Paranal Observatories under programme ID 175.A-0839. This research
  made use of several community-developed Python packages: Astropy \citep{astropy}, Matplotlib \citep{matplotlib}, SciPy and NumPy \citep{numpy}.

\end{acknowledgements}
%##############################

\bibliographystyle{aa}
\bibliography{final}

%##############################
\begin{appendix}
\section{The NB118 filter curves}
\label{app_nb118}

%%%%%%%%%%%
\begin{figure}
\centering
%\resizebox{1.0\hsize}{!}{\includegraphics{images/all_filtercurves_ndcvalues_preafter.pdf}}
\resizebox{0.9\hsize}{!}{\includegraphics{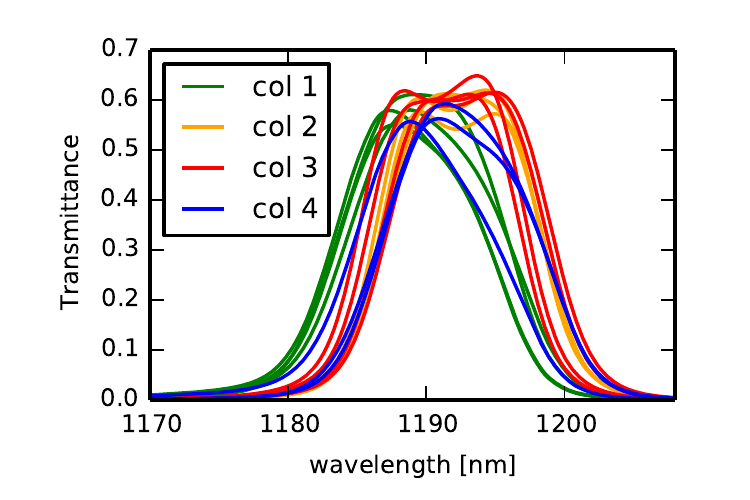}}
\caption{Passbands of 16 NB118 filters after convergent beam
  transformation, with QE and mirror reflectivities applied, and
  artificially shifted by $3.5\mathrm{nm}$ towards the red, as
  motivated by the results in \citet{Milvang-Jensen:2013:94}. Filters
  in the the four different columns of the observing pattern are
  marked in four different colors. Columns numbers are counted from
  right to left in Fig. \ref{fig:nb118:uvista:filters}.}
\label{fig_all_filter}
\end{figure}
%%%%%%%%%%%

\begin{figure}
  \centering
  \resizebox{0.9\hsize}{!}{\includegraphics{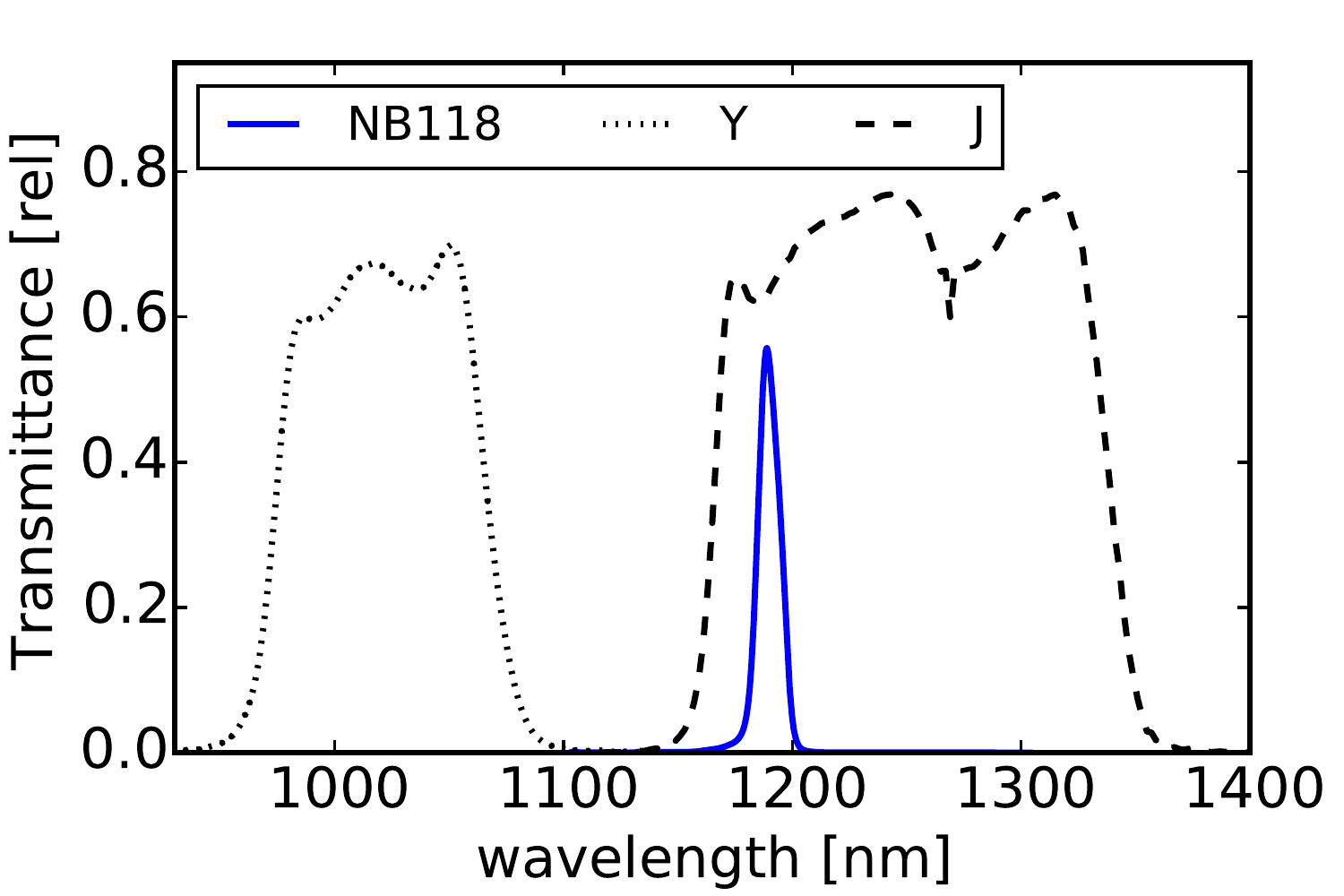}}
  \caption{Passbands of one of the NB118 filters (filter 15) in
  comparison to the VISTA/VIRCAM Y and J passbands. }
  \label{fig:nb118_y_j}
\end{figure}

Throughout this paper our results were based on the set of VISTA NB118 filters. In Fig. \ref{fig_all_filter} we show all 16 filter curves used in this study and in Fig. \ref{fig:nb118_y_j} one of the filters (filter 15) is shown in comparison to the Y and J filters. 

So far, we neglected the uncertainties in the measured filter curves.  Unfortunately, there are several sources of potentially significant errors in the available filter curves. We therefore summarize in this appendix the origin of the assumed filter curves and our current understanding of their accuracy.

The VIRCAM filter curves are based on laboratory scans carried out at
room temperature in the normal incidence collimated beam.  These
measurements have been supplied by the filter manufacturer, NDC
\footnote{NDC Infrared Engineering; http://www.ndcinfrared.com}.

However, as the NB118 filters are multilayer dielectric interference
filters, the transmittance curves depend both on the temperature and
the angle of incidence of the beam.  \noindent Qualitatively, both a
cooling and the change from the collimated beam to larger incidence
angles lead to a shift of the passband towards shorter wavelengths
\citep{Morelli:1991}.  \noindent This is relevant, as in VIRCAM the
filters are located in a fast convergent beam
\citep[e.g.~][]{Atad-Ettedgui:2003:95} at cryogenic temperatures.  The
convergent beam can be understood as formed by rays coming from
different incidence angles, each of which sees a filter curve
corresponding to its incidence angle.

One way to approximate the actual filter curves within the cryogenic
convergent beam is to do an entirely theoretical conversion from the
collimated beam measurement.  Assuming a temperature dependence of
$0.0186\,\mathrm{nm}\,\mathrm{K}^{-1}$ (NDC), the difference of 205K
between room temperature and filter temperature in VIRCAM
($\sim90\,\mathrm{K}$; private communication ESO), equals a blueward
shift of 4.26nm.  We made the assumption that the shape of the
transmittance curve is preserved under the temperature shift. The
justification of this assumption was confirmed by a re-measurement of
a witness sample done by NDC in 2013.

The transformation between collimated and convergent beam was based on
the assumption that it is possible to approximate the filter curve in
a collimated beam with non-normal incidence angle $\theta$ from that
for normal incidence by:

% See email correspondence with J. Emerson
\begin{equation}
T_\theta\,(\lambda)  \approx T_0\left(\frac{1}{\cos(\frac{\theta}{n_{eff}})}\lambda\right)
\label{eq_transf}
\end{equation}

As the range of relevant incidence angles in the VISTA beam extents up
to $\sim20\deg$, eq. \ref{eq_transf} must be a good approximation for
a large range of angles.  According to \citet{Morelli:1991} a
conservation of the general filter curve shape is a good approximation
up to angles of $30\deg$. In the literature exist a few examples,
where measurements of similar NIR filters have been published for
different incidence angles.  For some of them the results seem
approximately consistent with eq. \ref{eq_transf}
\citep[e.g.][]{Ghinassi:2002:1157}, while there is for others a
stronger discrepancy \citep[e.g][]{Vanzi:1998:177}. Again based on a
witness-sample, NDC provided us with measurements of the material for
incidence angles up to $12\;\mathrm{deg}$. The shape was indeed
approximately conserved.

VISTA's beam can be characterized by the radiant intensity as a
function of the angle of incidence, $\epsilon(\phi,\theta)$.  Here,
$\phi$ and $\theta$ are the two dimensional polar-coordinates
characterizing the latter.  Neglecting all effects of wave optics,
$\epsilon(\phi,\theta)$ can be described to first order by an annulus
with constant value.  The annulus's inner radius $\rho_{in} $ and
outer radius $\rho_{out}$ have values of $3.85\deg$ and $8.75\deg$,
respectively \citep{Nilsson:2007:71}.  This annulus is shifted
corresponding to the object's position in the field of view in the
$\phi$,$\theta$ plane.  The angle of incidence of the annulus's center
is related to the object's position on the sky by
$1.208^\circ \times \frac{\vec{d}_\mathrm{per}}{100}$
\citep{Findlay:2012}.  The position $\vec{d}_\mathrm{per}$ is stated
here in percentages of detectors.

With this input, the effective filter curve for the complete beam can be
calculated by \citep[e.g.][]{Lissberger:1970:197}:

\begin{equation}
T(\lambda)  =  \frac{\int_{\Omega} \epsilon(\phi,\theta)T_\theta(\lambda)d\Omega}{\int_{\Omega} \epsilon(\phi,\theta)d\Omega}
\label{eq_beam}
\end{equation}

Consequently, we can finally estimate the shape of the filter curves
in VIRCAM's convergent beam by combining equations \ref{eq_transf} and
\ref{eq_beam}.

\begin{equation}
T(\lambda) = \frac{\int_{annulus}T_0\left(\frac{1}{\cos \frac{\sqrt{(\theta_x + \theta_x^c)^2 + (\theta_y + \theta_y^c)^2}}{n_eff}}\lambda \right) d\theta_x d\theta_y}{\int_{annulus}d\theta_x d\theta_y}
\label{eq_beam_final}
\end{equation}

We tested our script doing the actual convergent beam conversion based
on a 1\% top-hat filter and comparing the results to those presented
in \cite{Bland-Hawthorn:2001:611}.

Filter curves for the different steps in the conversion were shown in
Fig. 4 of
\citet{Milvang-Jensen:2013:94}. \citet{Milvang-Jensen:2013:94} found
an unexpected shift of the filter curves towards the red by about
$3.5\,\mathrm{nm}$. While we are still investigating possible physical
reasons for the shift, we assume in this work the predicted convergent
beam curves shifted by $3.5\,\mathrm{nm}$ towards the red to account
for this finding. The fact that the presented TPV method works well
for the actual data under this assumption, further indicates that
these filter curves are a reasonable assumption.

%##############################
\section{Quantitative assessment of throughput variation for all possible NB118 pairs }
\label{sec:appendix:quantitativeall}
%##############################

The suitability of a filter combination for the TPV can be assessed
through $\Delta mag$-$\lambda_0$ curves, as discussed in
sec. \ref{sec_mno}.  We characterized the $\Delta mag$-$\lambda_0$
curves for all 120 possible VIRCAM NB118 combinations by following
three quantities:

\begin{itemize}
\item Difference between maximum and minimum $\Delta mag$
\item Percentage of the wavelength range, where the $\Delta mag$ values are unique
\item Average absolute slope ($\vert \mathrm{d} \Delta mag / \mathrm{d} \lambda_0 \vert$) and its standard deviation
\end{itemize}

For the calculation of these values a relevant wavelength interval
needs to be chosen.  We assumed the wavelength range, where the
transmittance of the combined effective filter is not below 2/30 of
its maximum value. This threshold is a reasonable number, as it
approximately corresponds to the ratio between the FWHMs of the NB118
filters and J. Consequently, an emission line causes a stronger excess
in the NB118 filters than in J within the included range.

The resulting values for all 120 combinations are listed in Table
\ref{tab:nb118:appendix:charact_alldeltabmag}.  For the example
filter-combination 14 \& 15, as shown in Fig.
\ref{fig:nb118:fig_sol_deltab}, the range in $\Delta mag$ values is 3.12 with a
uniqueness of 100\%. The average absolute slope is
$0.14\,\mathrm{mag}\;\mathrm{nm}^{-1}$. The assumed wavelength interval, as
defined above, is indicated in the left panel of Fig.
\ref{fig:nb118:fig_sol_deltab}. Assuming e.g. 5$\sigma$ detections in each of
the two filters, corresponding to an error of $\delta\Delta\mathrm{mag} \sim
0.3$ for the magnitude difference, this would allow on average for a very good
wavelength resolution of about $2\,\mathrm{nm}$.

%%%%%%%%%%%
\begin{sidewaystable*}

\addtolength{\tabcolsep}{-5pt}
   \centering
   \tiny
\vspace{100ex}

\begin{tabular}{c||c|c|c|c|c|c|c|c|c|c|c|c|c|c|c|c||}
  &   1 &  2 &  3 &  4 &  5 &  6 &  7 &  8 &  9 & 10 & 11 & 12 & 13 & 14 & 15 & 16 \\
\hline
\hline
 1 &  &  0.07$\pm$0.05 \cellcolor[gray]{0.85}  &  0.06$\pm$0.06 &  0.04$\pm$0.03 &  0.15$\pm$0.09 &  0.12$\pm$0.09 &  0.13$\pm$0.10 &  0.13$\pm$0.07 &  0.16$\pm$0.09 &  0.08$\pm$0.07 &  0.09$\pm$0.08 &  0.14$\pm$0.08 &  0.13$\pm$0.06 &  0.14$\pm$0.08 &  0.03$\pm$0.01 &  0.16$\pm$0.10 \\
\hline
 2 &  77./ 1.39 \cellcolor[gray]{0.85}  &  &  0.01$\pm$0.00 \cellcolor[gray]{0.85}  &  0.05$\pm$0.03 &  0.21$\pm$0.12 &  0.18$\pm$0.11 &  0.19$\pm$0.13 &  0.20$\pm$0.10 &  0.21$\pm$0.12 &  0.12$\pm$0.08 &  0.14$\pm$0.09 &  0.20$\pm$0.11 &  0.19$\pm$0.09 &  0.19$\pm$0.11 &  0.07$\pm$0.05 &  0.22$\pm$0.13 \\
\hline
 3 &  71./ 1.30 &  40./ 0.20 \cellcolor[gray]{0.85}  &  &  0.05$\pm$0.03 \cellcolor[gray]{0.85}  &  0.21$\pm$0.12 &  0.18$\pm$0.12 &  0.19$\pm$0.13 &  0.19$\pm$0.10 &  0.21$\pm$0.12 &  0.11$\pm$0.08 &  0.14$\pm$0.08 &  0.20$\pm$0.11 &  0.19$\pm$0.09 &  0.19$\pm$0.11 &  0.07$\pm$0.05 &  0.22$\pm$0.13 \\
\hline
 4 &  17./ 0.57 &  95./ 1.05 &  72./ 0.92 \cellcolor[gray]{0.85}  &  &  0.18$\pm$0.11 &  0.15$\pm$0.11 &  0.17$\pm$0.12 &  0.16$\pm$0.09 &  0.19$\pm$0.11 &  0.09$\pm$0.08 &  0.11$\pm$0.09 &  0.17$\pm$0.10 &  0.15$\pm$0.09 &  0.16$\pm$0.10 &  0.06$\pm$0.04 &  0.19$\pm$0.12 \\
\hline
 5 &  90./ 3.36 &  83./ 4.71 &  86./ 4.64 &  82./ 3.86 &  &  0.04$\pm$0.03 \cellcolor[gray]{0.85}  &  0.04$\pm$0.03 &  0.03$\pm$0.02 &  0.03$\pm$0.02 &  0.12$\pm$0.08 &  0.09$\pm$0.06 &  0.01$\pm$0.01 &  0.06$\pm$0.04 &  0.04$\pm$0.03 &  0.16$\pm$0.07 &  0.08$\pm$0.06 \\
\hline
 6 &  82./ 2.52 &  76./ 3.88 &  80./ 3.81 &  63./ 3.05 &  48./ 0.69 \cellcolor[gray]{0.85}  &  &  0.02$\pm$0.01 \cellcolor[gray]{0.85}  &  0.04$\pm$0.02 &  0.05$\pm$0.04 &  0.09$\pm$0.06 &  0.06$\pm$0.04 &  0.03$\pm$0.03 &  0.08$\pm$0.05 &  0.07$\pm$0.04 &  0.12$\pm$0.08 &  0.10$\pm$0.08 \\
\hline
 7 &  79./ 2.74 &  73./ 4.10 &  77./ 4.04 &  59./ 3.27 &  29./ 0.43 &  28./ 0.26 \cellcolor[gray]{0.85}  &  &  0.04$\pm$0.02 \cellcolor[gray]{0.85}  &  0.05$\pm$0.04 &  0.11$\pm$0.08 &  0.08$\pm$0.05 &  0.03$\pm$0.02 &  0.08$\pm$0.05 &  0.07$\pm$0.04 &  0.13$\pm$0.09 &  0.10$\pm$0.07 \\
\hline
 8 &  82./ 2.79 &  78./ 4.16 &  81./ 4.08 &  77./ 3.32 &  68./ 0.56 &   2./ 0.34 &   0./ 0.25 \cellcolor[gray]{0.85}  &  &  0.04$\pm$0.03 &  0.10$\pm$0.08 &  0.07$\pm$0.06 &  0.02$\pm$0.01 &  0.05$\pm$0.04 &  0.04$\pm$0.04 &  0.13$\pm$0.06 &  0.08$\pm$0.08 \\
\hline
 9 &  86./ 3.44 &  75./ 4.76 &  79./ 4.72 &  77./ 3.95 &  37./ 0.42 &  50./ 0.98 &  48./ 0.80 &  81./ 0.85 &  &  0.13$\pm$0.10 \cellcolor[gray]{0.85}  &  0.10$\pm$0.08 &  0.03$\pm$0.02 &  0.05$\pm$0.04 &  0.04$\pm$0.02 &  0.16$\pm$0.07 &  0.06$\pm$0.05 \\
\hline
10 &  27./ 1.30 &  83./ 2.38 &  92./ 2.28 &  43./ 1.56 &  81./ 2.37 &  55./ 1.68 &  51./ 1.93 &  55./ 1.85 &  65./ 2.60 \cellcolor[gray]{0.85}  &  &  0.04$\pm$0.02 \cellcolor[gray]{0.85}  &  0.11$\pm$0.08 &  0.10$\pm$0.09 &  0.11$\pm$0.10 &  0.08$\pm$0.05 &  0.15$\pm$0.12 \\
\hline
11 &  53./ 1.74 &  78./ 2.97 &  87./ 2.89 &  71./ 2.16 &  68./ 1.69 &  45./ 1.00 &  43./ 1.25 &  42./ 1.18 &  58./ 1.93 &  78./ 0.68 \cellcolor[gray]{0.85}  &  &  0.08$\pm$0.06 \cellcolor[gray]{0.85}  &  0.09$\pm$0.08 &  0.09$\pm$0.08 &  0.09$\pm$0.06 &  0.13$\pm$0.11 \\
\hline
12 &  87./ 3.09 &  81./ 4.45 &  84./ 4.38 &  79./ 3.61 &  37./ 0.20 &  30./ 0.50 &  20./ 0.35 &  46./ 0.36 &  79./ 0.53 &  69./ 2.14 &  65./ 1.48 \cellcolor[gray]{0.85}  &  &  0.05$\pm$0.03 &  0.04$\pm$0.02 &  0.15$\pm$0.06 &  0.08$\pm$0.06 \\
\hline
13 &  99./ 2.96 &  93./ 4.29 &  95./ 4.20 &  90./ 3.35 &  22./ 0.81 &   5./ 0.58 &   3./ 0.61 &   1./ 0.46 &  42./ 0.87 &  44./ 1.88 &  24./ 1.31 &  18./ 0.61 &  &  0.02$\pm$0.02 \cellcolor[gray]{0.85}  &  0.14$\pm$0.05 &  0.06$\pm$0.04 \\
\hline
14 &  93./ 3.04 &  83./ 4.38 &  86./ 4.31 &  83./ 3.53 &  16./ 0.59 &  17./ 0.64 &  13./ 0.55 &  16./ 0.54 &  52./ 0.62 &  53./ 2.05 &  35./ 1.42 &   9./ 0.39 &  28./ 0.31 \cellcolor[gray]{0.85}  &  &  0.14$\pm$0.06 \cellcolor[gray]{0.85}  &  0.05$\pm$0.04 \\
\hline
15 &   6./ 0.25 &  57./ 1.37 &  48./ 1.29 &   8./ 0.56 & 100./ 3.41 &  91./ 2.55 &  88./ 2.77 &  93./ 2.82 &  92./ 3.49 &  41./ 1.25 &  66./ 1.69 & 100./ 3.14 &  99./ 3.00 & 100./ 3.12 \cellcolor[gray]{0.85}  &  &  0.17$\pm$0.09 \cellcolor[gray]{0.85}  \\
\hline
16 &  91./ 3.70 &  84./ 4.99 &  86./ 4.95 &  84./ 4.17 &  12./ 0.79 &  15./ 1.24 &  14./ 1.08 &  19./ 1.01 &   3./ 0.55 &  50./ 2.76 &  39./ 2.10 &  12./ 0.80 &  44./ 1.01 &  32./ 0.71 &  98./ 3.77 \cellcolor[gray]{0.85}  &  \\
\hline
\end{tabular}
\caption{
\label{tab:nb118:appendix:charact_alldeltabmag}
Stated are the quantities as introduced in
sec. \ref{sec:appendix:quantitativeall} to characterize the
suitability of a given filter pair for the TPV. The table includes
these values for all possible 120 NB118 filter-combinations. Both the
part of the well-defined wavelength range, where the $\Delta mag$
values are unique (in percentage), and the span between maximum and
minimum $\Delta mag$ are stated. In the upper right triangle, the
average absolute slope
($\vert \mathrm{d} \Delta mag / \mathrm{d} \mathrm{\lambda_0} \vert$
[$\mathrm{mag}/\mathrm{nm}$]) and its standard deviation are listed.
The grey-shaded combinations are those NB118 combinations, for which
observations will become available within the default \ultvis{}
survey.  }

\end{sidewaystable*}

%%%%%%%%%%%

%##############################
\section{Expected number of \ha{} emitters in regions of filter overlap}
\label{sec:ultravista:expnumber}
%##############################

%###############
\subsection{Simulation}
%###############
\label{sec:ultravista:simu}
We estimated the number of \ha{} emitters expected to be found in the
field covered by two differing filters as a function of line flux both
for the DR2 and the finalized \ultvis{}.  For this purpose we created
300000 simulated objects with a continuum flat in $f_{\nu}$, a fixed
\fion{N}{ii} ratio of either $w_{6583}=0$ or $w_{6583}=0.3$, a random
\ha{} central wavelength between $1167\mbox{--}1209\,\mathrm{nm}$, and
assigned to each of these objects the same fixed input line
luminosity, $\mathcal{L}_{0}[in]$. The \ha{} $EW_0$ were drawn from a
log-normal distribution with a
$\langle \log_{10}(\mathrm{EW}_{0}/\mathrm{nm})\rangle = 0.35$ and
$\sigma[\log_{10}(\mathrm{EW}/\mathrm{nm})] = 0.4$. These values were
taken from the best fit distribution obtained by \citet{Ly:2011:109}
based on NEWFIRM narrowband observations at a similar wavelength as
the \ultvis{} NB118 filters.

After assigning to these 300000 objects random positions within a
$1.4x1.4\, \deg^2$ field, we determined for each of the jitter
positions in each of the three pawprints in the \ultvis{} NB118
observing pattern, whether an object is observed and if yes, in which
filter.  For computational reasons, we only did 28 random jitters per pawprint
drawn from a 2''x 2'' box.

In each of the individual simulated pointings we determined for each
object falling within the boundaries corresponding to a filter the
synthetic $\mathrm{f}_\nu$ (eq.  \ref{eq:fnu}) and the corresponding
error on the $\mathrm{f}_\nu$, which we separated into $\delta_{o}f$
and $\delta_{b}f$ for object and background, respectively.\footnote{As
  the considered observations are background-limited, we could in
  principle ignore $\delta_{o}f$. We kept it for generality of our
  simulator.} The calculations were based on the same ZPs, gains, and
detector-depended sky-counts, and observation times for the DR2 and
final \ultvis{}, as described in sec. \ref{sec_test}.  Finally, we
combined for each of the individual pawprints the signal from the
different jitter positions by weighting with $\frac{1}{\delta_{b}^2f}$
and propagated the errors on the noise. This simulated observing and
stacking strategy resembles that for the actual \ultvis{}
observations.

%###############
\subsection{Method}
%###############

As the measured source flux density scales for fixed $EW_{obs}$
linearly with line luminosity both for the NB and the BB filter, we
can based on $f_{\nu;NB118/J}$, $\delta_o f_{NB118/J}$,
$\delta_b f_{NB118/J}$ obtained for the input line luminosity,
$\mathcal{L}_\mathrm{0;in}$, directly determine for a chosen
color-significance $\kappa$ the required $\mathcal{L}_\mathrm{0,req}$
to fulfill the inequality \ref{eq:colsig}. This is to solve a
quadratic equation
$a_{quad}\;\alpha_{col}^2 + b_{quad}\;\alpha_{col} + c_{quad} = 0 $ in
the common way, where $a_{quad}$, $b_{quad}$, and $c_{quad}$ are given
as:\footnote{For the strongly background limited \ultvis{}
  observations, the inclusion of the source noise is in principle not
  necessary and is only included for generality.}

\begin{eqnarray}
	a_{quad} = & {\left(f_{NB118} - f_{J}\right)}^2 \\
	b_{quad} = & -1\;\kappa^2(\delta^2_o f_{NB118} + \delta^2_o f_{J})\\
	c_{quad} = & -1\;\kappa^2(\delta^2_b f_{NB118} + \delta^2_b f_{J})\\
\end{eqnarray}

$\alpha_{\mathrm{col}}$ is the ratio between $\mathcal{L}_\mathrm{0,req}$ and
$\mathcal{L}_\mathrm{0;in}$. Similarly, we can determine the factor
$\alpha_{detsig}$, required to fulfill eq. \ref{eq:detsig}.  Having determined
the factors $\alpha_{col}$ and $\alpha_{detsig}$ for each of the two
contributing filters, we find the minimum factor $\alpha$, which fulfills the
color-significance combined with the color-cut in one of the two filters and
the detection significance in the other filter.  If any of the criteria is not
fulfilled at any flux, we set $\alpha$ to $\infty$. We also applied the same
region mask as used for the actual data (cf. eq. \ref{eq:mask}).

Based on the $\alpha$'s obtained for each simulated object, we directly
determined the fraction of input objects being detected as a function of line
luminosity.  Eventually, multiplying this luminosity completeness function with
luminosity functions from the literature and the volume covered by the random
box allowed for an estimate of the number of objects expected to be detected as
a function of line luminosity. This takes fully account for the filter curve
shapes and the different background brightnesses in the individual filters.

We used the \citet{Schechter:1976:297} parameterizations of the three LFs
stated by \citet{Ly:2011:109}, including their own and the two $z=0.84$ LFs of
\cite{Villar:2008:169} and \citet{Sobral:2009:75}. The LFs stated in
\citet{Ly:2011:109} are reddening and completeness corrected.  As we need for
the purpose of our simulation non-reddening corrected LFs, we convert their
Schechter LFs to reddened LFs, by inverting the same SFR depended correction as
used in
\cite{Ly:2011:109}, which is based on
\citet{Hopkins:2001:L31}.\footnote{There are small differences in the two
  versions, as mentioned in \cite{Ly:2007:738}.} Assuming the underlying direct
  proportionality between SFR and H$\alpha$ luminosity
\citep{Kennicutt:1998:189}, the relation between intrinsic,
$L_{\mathrm{H}\alpha;\mathrm{int}}$, and observed \ha{} luminosity,
$L_{\mathrm{H},\alpha;\mathrm{int}}$, can be written as
\citep{Ly:2011:109}:

\begin{equation} L_{\mathrm{H}\alpha;\mathrm{obs}} =
		L_{\mathrm{H}\alpha;\mathrm{int}}\times{\left(\frac{0.797\log(SFR_{int}[H\alpha])
	+ 3.786} {2.86}\right)}^{-2.360}
\end{equation}

\subsection{Results}

\begin{figure}
  \centering
  \resizebox{1.0\hsize}{!}{\includegraphics[]{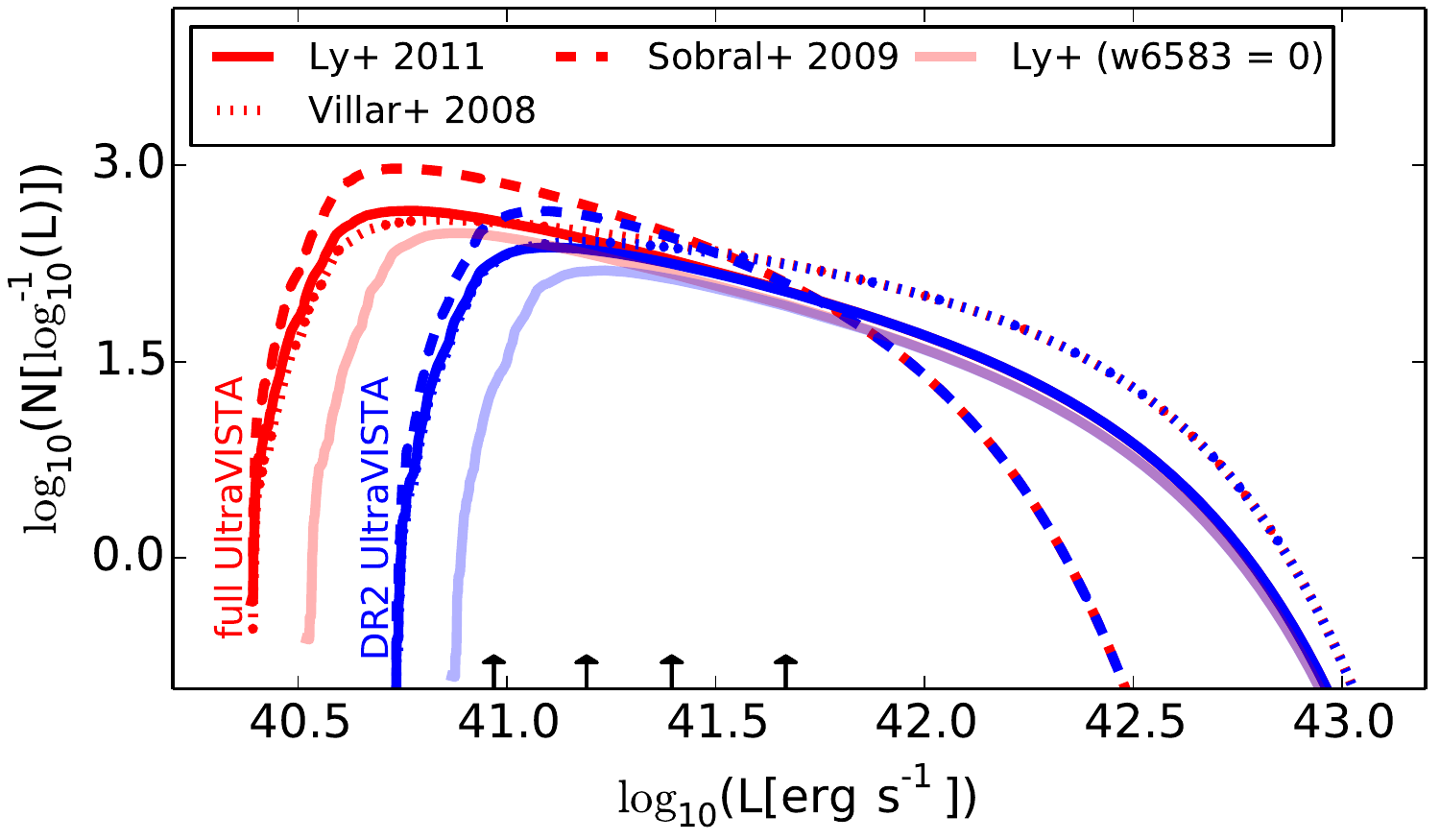}}
  \caption{\label{fig:nb118:expected} Expected number of \ha{} emitters in
    the part of the \ultvis{} NB118 field with data from two different NB118
    filters is shown per unit logarithmic interval as a function of line
    luminosity. Predicted curves are included both for the \ultvis{} DR2 (blue)
    and the final \ultvis{} data (red) based on three different $z=0.8$ \ha{}
    LFs from the literature. Results are in all three cases available for the
    assumption of $w_{6583}=0.3$ and in the case of the
    \citet{Ly:2011:109} LF also for $w_{6583}=0$.  Small arrows
    indicate luminosities corresponding to line fluxes of $3.0$,
    $5.0$, $10.0$, and
    $20.0\times10^{-17}\,\mathrm{erg}\,\mathrm{s}^{-1}\,\mathrm{cm}^{-2}$. The
    integrated numbers for objects brighter than the respective marked
    fluxes are stated in Table \ref{tab:nb118:expected}. }
\end{figure}

The number of galaxies expected to be selected per unit logarithmic interval by
the criteria stated in eq. \ref{eq:pos}--\ref{eq:mask} are shown in Fig.
\ref{fig:nb118:expected} for the three different LFs. For the
\citet{Ly:2011:109} LF, the result is shown in addition to assuming
$w_{6583}=0.3$ also for the assumption of $w_{6583} = 0$. For orientation,
luminosities corresponding to fluxes of approximately $3.0$, $5.0$, $10.0$, and
$20.0 \times 10^{-17}\,\mathrm{erg}\,\mathrm{s}^{-1}\,\mathrm{cm}^{-2}$ are
marked with small arrows.\footnote{Approximately, as the actual ratio between
line flux and line luminosity depends on the luminosity distance, which
slightly varies over the considered wavelength range.}

Integrated numbers within the 12 patches with contribution of two different
filters are stated down to the detection limit and to the four reference fluxes
in Table \ref{tab:nb118:expected}.  Already in the DR2, we would expect to have
on the order of 200 \ha{} emitters in the patches of overlapping filters.  The
total \ultvis{} field is expected to have about three times the numbers stated
in Table
\ref{tab:nb118:expected} and the number in the final \ultvis{} data
will almost double the number compared to the DR2.

%%%%%%%%%%%
%%%%%%%%%%%

%%%%%%%%%%%
\begin{table}
  \caption{\label{tab:nb118:expected} Integrated numbers of \ha{} emitters expected in the \ultvis{} survey above different flux thresholds. The numbers are for those parts with coverage in two different NB118 filters only. Values outside the brackets are for $w_{6583}=0.3$ and inside brackets for $w_{6583}=0$. For more details see caption of Fig. \ref{fig:nb118:expected}.}

\begin{tabular}{cccccc}
    \hline
    \hline

	& \multicolumn{5}{c}{flux
		[$10^{-17}\,\mathrm{erg}\;\mathrm{s}^{-1}\;\mathrm{cm}^{-2}$]} \\
		LF  & all    & $> 3.0$ & $> 5.0$  & $>10.0$ & $> 20.0$ \\

\hline
\multicolumn{6}{c}{\ultvis{} DR2} \\
\hline
LY11 & 184 (113) & 169 (113) & 119 (90) & 77 (60) & 39 (30) \\
SO09 & 250 (165) & 237 (164) & 184 (140) & 132 (102) & 76 (59) \\
VI08 & 275 (156) & 246 (155) & 150 (113) & 79 (61) & 25 (19) \\
\hline
\multicolumn{6}{c}{Full \ultvis} \\
\hline
LY11 & 373 (235) & 198 (154) & 126 (98) & 79 (62) & 39 (30) \\
SO09 & 413 (280) & 268 (209) & 192 (150) & 135 (106) & 76 (60) \\
VI08 & 666 (400) & 299 (233) & 161 (126) & 81 (64) & 25 (19) \\

\end{tabular}\end{table}
%%%%%%%%%%%

%##############################
\section{Expected line S/N in NB and BB filters}
%##############################
\label{sec:nb118:app:sn}

%%%%%%%%%%%
\begin{figure}
\centering
\resizebox{0.8\hsize}{!}{\includegraphics{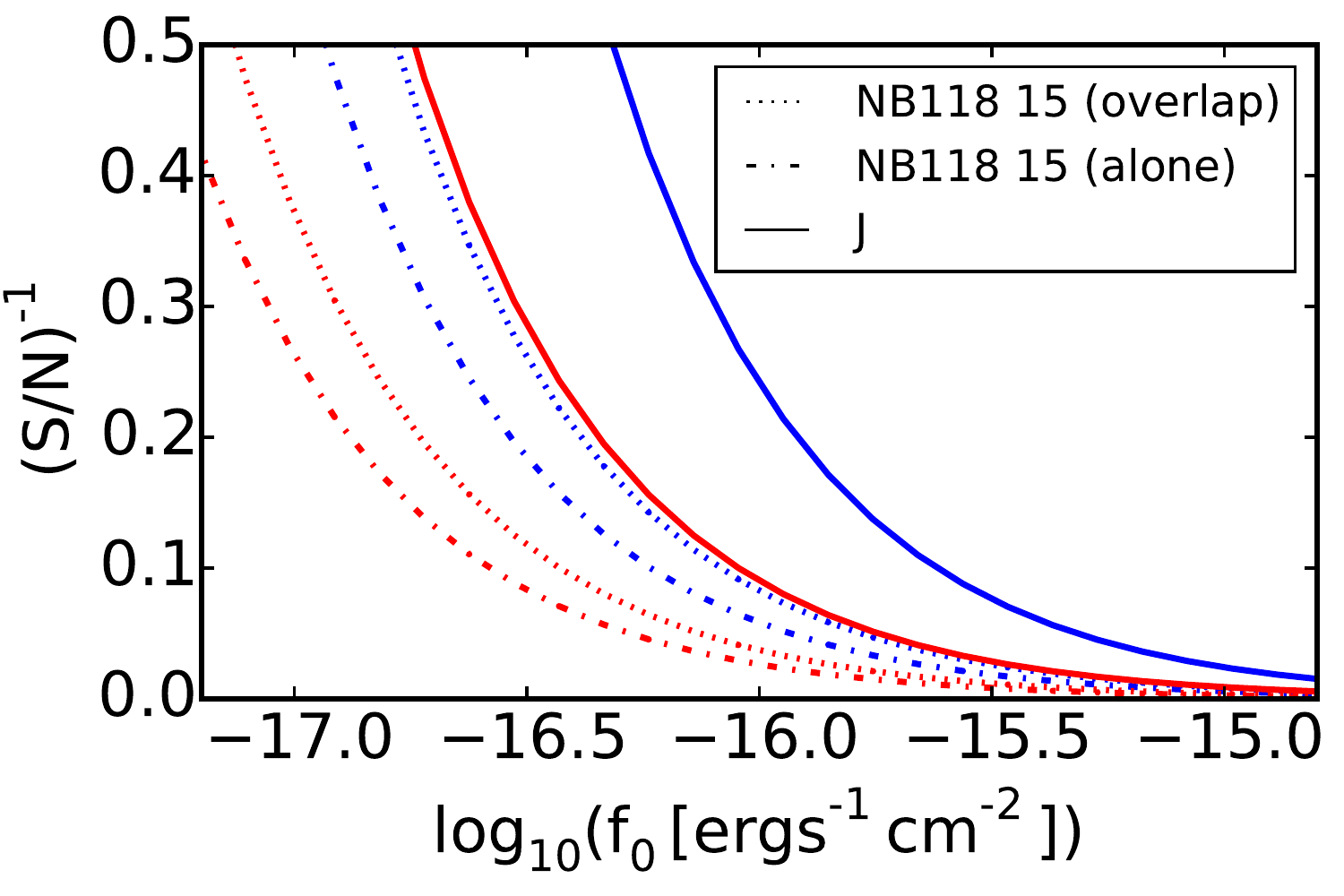}}
\caption{\label{fig:inverseSN} Inverse of the signal-to-noise as a
  function of line flux for an infinite $EW$ line in J and at peak
  transmittance of NB118 filter 15. Curves are shown both for exposure
  times as in the \ultvis{} DR2 (blue) and for the final \ultvis{}
  survey (red).  Moreover, for the NB118 filter results are included
  for the typical per pixel integration time and for half this value,
  with the latter being relevant for the regions of overlapping
  filters (cf. sec. \ref{sec:nb118:observingpattern:ultravista}).}
\end{figure}
%%%%%%%%%%%

Fig. \ref{fig:inverseSN} shows the inverse of the $S/N$ that an
emission line point source with infinite $EW$ would reach for a given
line flux in J and NB118, respectively, both for the \ultvis{} DR2 and
the expected full \ultvis{}. The calculation was based on
eq. \ref{eq_snrfinal} and the simulation inputs described in
sec. \ref{sec:nb118:mockobs}.  Assuming that the continuum flux
density could be estimated without uncertainty, these would be the S/N
values for the line alone, independent of the EW.

NB118 results are shown for the peak of filter 15, which has a typical
sky-brightness, and are plotted both for the typical per-pixel
integration time and half its value. The latter would be applicable
when sharing the available time equally between two contributing
filters.

Even so the \ultvis{} BB data is extremely deep, the line S/N in the
NB118 is a factor 4.1 or 3.4 higher than that in J, for the DR2 and
the expected final survey data respectively. On the other hand, at low
transmittances of the NB filters, the line S/N in J becomes equivalent
or even higher than that in the NB filter.

\section{Full SED fitting results}

The full results from the SED fitting for the sample of NB excess
objects with observations available in either of the NB118 pairs
9 \& 10, 14 \& 15, or 15 \& 16, as described in
sec. \ref{sec:estimation_sed}, are listed in Table
\ref{tab:nb118:table_results2}.  Additionally, the table includes the
$EW_\mathrm{obs}$ estimated through the TPV.

\begin{table*}
\caption{This table extends Table \ref{tab:nb118:table_results1}. Here mainly the properties of the best fit SEDs are listed. In addition, $EW_{obs}$ from the TPV is included. \label{tab:nb118:table_results2}}
\centering

\begin{tabular}{crrccccccccc}
& & &  \multicolumn{8}{|c|}{SED Fitting} & TPV   \\
\hline
ID\tablefootmark{a} & RA (J200) & DEC (J2000) & Mass\tablefootmark{b} & $E{\scriptscriptstyle B-V}$ 		 \tablefootmark{c} & $Age$ \tablefootmark{d} & $\tau$ \tablefootmark{e} & $SFR$\tablefootmark{f} & $\overline{SFR}$ \tablefootmark{g} & $Z$ \tablefootmark{h} & $f_\mathrm{cov}$ \tablefootmark{i} & $EW_\mathrm{obs}$ \tablefootmark{j} \\
\hline
\hline
7 & +10:01:54.356 & +01:53:18.36 & $10.3_{-0.1}^{+0.1}$ & $0.14_{-0.06}^{+0.03}$ & $9.1_{-0.2}^{+0.3}$ & $9.9_{-0.6}^{+0.6}$ & $13.3_{-5.1}^{+3.6}$ & $13.3_{-5.1}^{+3.9}$ & $1.00_{-0.06}^{+0.65}$ & $0.7_{-0.0}^{+0.3}$ & $21.4_{-1.2}^{+1.2}$ \\
14 & +10:01:57.962 & +01:53:57.58 & $10.3_{-0.1}^{+0.0}$ & $0.24_{-0.05}^{+0.13}$ & $9.3_{-0.5}^{+0.1}$ & $9.1_{-0.6}^{+0.7}$ & $4.0_{-1.4}^{+6.4}$ & $4.2_{-1.3}^{+7.8}$ & $0.20_{-0.00}^{+0.58}$ & $0.7_{-0.0}^{+0.3}$ & $3.9_{-1.1}^{+1.3}$ \\
33 & +10:01:45.444 & +01:55:22.99 & $10.5_{-0.0}^{+0.0}$ & $0.20_{-0.03}^{+0.07}$ & $9.1_{-0.1}^{+0.3}$ & $8.9_{-0.1}^{+1.2}$ & $9.3_{-1.5}^{+7.3}$ & $9.9_{-1.3}^{+7.7}$ & $0.40_{-0.20}^{+0.15}$ & $0.7_{-0.0}^{+0.3}$ & $11.6_{-1.1}^{+1.1}$ \\
94 & +10:01:37.653 & +02:10:33.39 & $10.4_{--0.0}^{+0.2}$ & $0.22_{-0.04}^{+0.10}$ & $9.2_{-0.1}^{+0.3}$ & $8.5_{-0.1}^{+0.3}$ & $0.5_{-0.1}^{+0.7}$ & $0.6_{-0.1}^{+0.7}$ & $0.40_{-0.20}^{+0.56}$ & $1.0_{-0.3}^{+0.0}$ & $14.9_{-1.4}^{+1.4}$ \\
96 & +10:02:12.744 & +02:10:47.98 & $9.9_{--0.0}^{+0.2}$ & $0.34_{-0.09}^{+0.00}$ & $8.3_{-0.0}^{+0.8}$ & $7.9_{-0.0}^{+2.2}$ & $9.4_{-2.8}^{+5.4}$ & $18.6_{-11.1}^{+0.0}$ & $0.40_{-0.20}^{+0.29}$ & $1.0_{-0.3}^{+0.0}$ & $12.8_{-1.2}^{+1.2}$ \\
97 & +10:02:16.988 & +02:10:55.96 & $10.0_{-0.2}^{+0.1}$ & $0.30_{-0.06}^{+0.04}$ & $8.8_{-0.4}^{+0.2}$ & $9.9_{-1.2}^{+0.5}$ & $13.7_{-5.4}^{+5.1}$ & $13.8_{-5.2}^{+7.0}$ & $0.40_{-0.19}^{+0.40}$ & $0.7_{-0.0}^{+0.3}$ & $14.5_{-1.1}^{+1.1}$ \\
99 & +10:02:15.195 & +02:10:54.36 & $10.8_{-0.1}^{+0.0}$ & $0.28_{-0.10}^{+0.04}$ & $9.3_{-0.3}^{+0.1}$ & $8.9_{-0.4}^{+0.0}$ & $6.3_{-3.2}^{+1.8}$ & $6.7_{-3.4}^{+2.7}$ & $0.20_{-0.00}^{+0.76}$ & $1.0_{-0.3}^{+0.0}$ & $4.5_{-0.5}^{+0.7}$ \\
104 & +10:02:15.978 & +02:11:18.89 & $9.0_{-0.0}^{+0.2}$ & $0.02_{-0.02}^{+0.05}$ & $8.6_{-0.1}^{+0.4}$ & $8.3_{-0.0}^{+1.8}$ & $0.7_{-0.0}^{+0.7}$ & $0.9_{-0.1}^{+0.6}$ & $0.20_{-0.00}^{+0.36}$ & $1.0_{-0.3}^{+0.0}$ & $19.0_{-2.4}^{+3.5}$ \\
105 & +10:02:07.660 & +02:11:20.09 & $9.6_{-0.3}^{+0.0}$ & $0.28_{-0.01}^{+0.01}$ & $7.6_{-0.3}^{+0.1}$ & $7.9_{-0.0}^{+1.9}$ & $82.0_{-9.2}^{+36.5}$ & $106.5_{-17.9}^{+13.7}$ & $1.00_{-0.00}^{+1.50}$ & $0.7_{-0.0}^{+0.1}$ & $28.1_{-1.2}^{+1.2}$ \\
111 & +10:02:16.354 & +02:12:00.30 & $9.6_{-0.0}^{+0.2}$ & $0.24_{-0.14}^{+0.01}$ & $8.7_{-0.0}^{+0.7}$ & $8.5_{-0.0}^{+1.7}$ & $3.0_{-1.8}^{+0.8}$ & $3.5_{-2.3}^{+0.6}$ & $0.40_{-0.20}^{+0.18}$ & $1.0_{-0.3}^{+0.0}$ & $14.7_{-3.2}^{+2.8}$ \\
113 & +10:01:41.651 & +02:12:02.75 & $9.6_{-0.1}^{+0.0}$ & $0.08_{-0.00}^{+0.10}$ & $8.8_{-0.3}^{+0.2}$ & $8.5_{-0.2}^{+1.7}$ & $2.1_{-0.0}^{+3.0}$ & $2.5_{-0.0}^{+3.2}$ & $1.00_{-0.80}^{+0.00}$ & $1.0_{-0.3}^{+0.0}$ & $13.7_{-2.1}^{+2.0}$ \\
114 & +10:02:09.637 & +02:12:02.47 & $10.1_{-0.1}^{+0.0}$ & $0.18_{-0.01}^{+0.08}$ & $9.0_{-0.3}^{+0.2}$ & $8.9_{-0.2}^{+1.5}$ & $5.9_{-0.3}^{+5.9}$ & $6.3_{-0.3}^{+6.2}$ & $0.40_{-0.20}^{+0.28}$ & $1.0_{-0.3}^{+0.0}$ & $12.0_{-0.8}^{+0.9}$ \\
117 & +10:02:17.543 & +02:12:12.54 & $11.1_{--0.0}^{+0.1}$ & $0.26_{-0.03}^{+0.09}$ & $9.1_{-0.0}^{+0.5}$ & $8.7_{-0.0}^{+1.3}$ & $21.3_{-4.2}^{+21.7}$ & $23.5_{-4.7}^{+20.0}$ & $1.00_{-0.80}^{+0.43}$ & $1.0_{-0.3}^{+0.0}$ & $5.1_{-0.3}^{+0.3}$ \\
121 & +10:02:03.952 & +02:12:39.54 & $9.7_{-0.1}^{+0.1}$ & $0.26_{-0.05}^{+0.04}$ & $8.6_{-0.3}^{+0.3}$ & $9.5_{-0.8}^{+0.9}$ & $12.2_{-4.3}^{+5.5}$ & $12.4_{-4.1}^{+7.1}$ & $0.40_{-0.13}^{+0.53}$ & $1.0_{-0.3}^{+0.0}$ & $18.3_{-1.4}^{+1.4}$ \\
122 & +10:02:01.838 & +02:12:38.81 & $10.4_{--0.0}^{+0.1}$ & $0.52_{-0.06}^{+0.04}$ & $8.4_{-0.0}^{+0.7}$ & $7.9_{-0.0}^{+2.2}$ & $15.4_{-0.0}^{+19.3}$ & $30.5_{-11.4}^{+9.2}$ & $1.00_{-0.80}^{+0.07}$ & $1.0_{-0.3}^{+0.0}$ & $8.9_{-1.1}^{+1.0}$ \\
124 & +10:02:16.714 & +02:12:55.00 &  &  &  &  &  &  &  &  & $15.3_{-2.2}^{+4.2}$ \\
125 & +10:02:18.112 & +02:13:02.53 & $10.7_{-0.1}^{+0.0}$ & $0.24_{-0.07}^{+0.06}$ & $9.3_{-0.2}^{+0.0}$ & $8.7_{-0.3}^{+0.0}$ & $2.1_{-1.1}^{+0.8}$ & $2.3_{-1.2}^{+1.1}$ & $0.20_{-0.00}^{+0.53}$ & $1.0_{-0.3}^{+0.0}$ & $5.7_{-0.8}^{+0.9}$ \\
126 & +10:01:47.095 & +02:13:26.64 & $9.5_{-0.1}^{+0.0}$ & $0.06_{-0.01}^{+0.08}$ & $9.1_{-0.4}^{+0.1}$ & $9.1_{-0.6}^{+1.0}$ & $1.3_{-0.3}^{+1.3}$ & $1.4_{-0.2}^{+1.5}$ & $0.20_{-0.00}^{+0.29}$ & $1.0_{-0.3}^{+0.0}$ & $19.5_{-3.1}^{+3.2}$ \\
128 & +10:01:46.900 & +02:13:30.85 & $9.1_{-0.1}^{+0.1}$ & $0.12_{-0.06}^{+0.03}$ & $8.0_{-0.0}^{+0.5}$ & $7.9_{-0.0}^{+2.5}$ & $6.1_{-1.7}^{+3.7}$ & $12.1_{-7.4}^{+0.0}$ & $0.40_{-0.15}^{+0.65}$ & $1.0_{-0.3}^{+0.0}$ & $30.0_{-2.3}^{+2.4}$ \\
131 & +10:02:06.263 & +02:13:40.26 & $8.8_{-0.1}^{+0.2}$ & $0.16_{-0.07}^{+0.06}$ & $8.2_{-0.0}^{+0.7}$ & $7.9_{-0.0}^{+2.5}$ & $1.3_{-0.3}^{+1.9}$ & $2.6_{-1.5}^{+0.9}$ & $0.40_{-0.18}^{+0.61}$ & $1.0_{-0.3}^{+0.0}$ & $20.6_{-4.1}^{+6.7}$ \\
135 & +10:01:52.737 & +02:13:53.75 & $10.4_{-0.0}^{+0.1}$ & $0.46_{-0.06}^{+0.04}$ & $8.7_{-0.1}^{+0.4}$ & $8.7_{-0.2}^{+1.6}$ & $27.2_{-9.8}^{+12.2}$ & $30.0_{-10.8}^{+12.0}$ & $0.20_{-0.00}^{+0.33}$ & $1.0_{-0.3}^{+0.0}$ & $6.5_{-0.7}^{+0.8}$ \\
138 & +10:02:17.661 & +02:14:02.29 & $9.8_{-0.1}^{+0.0}$ & $0.10_{-0.03}^{+0.06}$ & $9.0_{-0.3}^{+0.2}$ & $9.1_{-0.3}^{+1.3}$ & $4.1_{-0.8}^{+2.5}$ & $4.2_{-0.8}^{+2.9}$ & $0.40_{-0.20}^{+0.24}$ & $1.0_{-0.1}^{+0.0}$ & $24.1_{-0.8}^{+1.0}$ \\
147 & +10:01:46.836 & +02:14:24.08 & $9.5_{-0.1}^{+0.0}$ & $0.10_{-0.01}^{+0.13}$ & $9.1_{-0.5}^{+0.2}$ & $8.9_{-0.5}^{+1.2}$ & $1.0_{-0.1}^{+1.9}$ & $1.0_{-0.0}^{+2.1}$ & $0.20_{-0.00}^{+0.37}$ & $1.0_{-0.3}^{+0.0}$ & $13.9_{-0.9}^{+2.7}$ \\
150 & +10:02:12.970 & +02:14:28.42 & $9.6_{-0.0}^{+0.1}$ & $0.20_{-0.02}^{+0.05}$ & $8.4_{-0.0}^{+0.5}$ & $7.9_{-0.0}^{+1.2}$ & $2.3_{-0.0}^{+3.8}$ & $4.6_{-0.6}^{+3.3}$ & $0.40_{-0.20}^{+0.35}$ & $1.0_{-0.3}^{+0.0}$ & $10.8_{-1.5}^{+1.4}$ \\
153 & +10:02:18.594 & +02:14:41.36 & $9.8_{-0.1}^{+0.0}$ & $0.10_{-0.02}^{+0.07}$ & $8.9_{-0.3}^{+0.2}$ & $8.7_{-0.3}^{+1.4}$ & $3.0_{-0.4}^{+2.6}$ & $3.4_{-0.3}^{+2.8}$ & $0.40_{-0.20}^{+0.37}$ & $1.0_{-0.3}^{+0.0}$ & $8.9_{-1.4}^{+1.4}$ \\
161 & +10:01:54.279 & +02:15:26.80 & $10.3_{-0.3}^{+-0.0}$ & $0.12_{-0.00}^{+0.15}$ & $9.8_{-0.7}^{+0.0}$ & $9.7_{-0.7}^{+0.6}$ & $1.5_{-0.0}^{+3.4}$ & $1.5_{-0.0}^{+3.5}$ & $0.20_{-0.00}^{+0.23}$ & $1.0_{-0.3}^{+0.0}$ & $11.1_{-1.9}^{+1.7}$ \\
164 & +10:02:04.380 & +02:15:30.63 & $9.4_{--0.0}^{+0.2}$ & $0.16_{-0.10}^{+0.00}$ & $8.4_{-0.0}^{+0.8}$ & $8.1_{-0.0}^{+2.3}$ & $3.5_{-1.4}^{+0.8}$ & $5.3_{-3.0}^{+0.0}$ & $0.40_{-0.20}^{+0.31}$ & $0.7_{-0.0}^{+0.3}$ & $16.2_{-2.4}^{+2.4}$ \\
167 & +10:01:53.203 & +02:15:49.45 & $9.8_{-0.1}^{+0.0}$ & $0.16_{-0.04}^{+0.03}$ & $8.8_{-0.3}^{+0.1}$ & $10.7_{-1.8}^{+0.0}$ & $9.4_{-2.5}^{+2.2}$ & $9.4_{-2.2}^{+2.9}$ & $0.40_{-0.16}^{+0.31}$ & $1.0_{-0.1}^{+0.0}$ & $28.5_{-0.4}^{+0.5}$ \\
170 & +10:02:15.102 & +02:15:59.41 & $9.4_{-0.0}^{+0.2}$ & $0.14_{-0.09}^{+0.01}$ & $8.5_{-0.0}^{+0.6}$ & $8.1_{-0.0}^{+1.5}$ & $1.9_{-0.7}^{+0.9}$ & $2.9_{-1.5}^{+0.3}$ & $0.40_{-0.20}^{+0.31}$ & $0.7_{-0.0}^{+0.3}$ & $10.9_{-1.2}^{+1.6}$ \\
172 & +10:00:46.944 & +02:26:10.89 & $9.8_{-0.2}^{+-0.0}$ & $0.00_{-0.00}^{+0.13}$ & $9.4_{-0.5}^{+0.0}$ & $9.1_{-0.5}^{+0.3}$ & $0.7_{-0.0}^{+1.3}$ & $0.7_{-0.0}^{+1.5}$ & $0.20_{-0.00}^{+0.34}$ & $1.0_{-0.3}^{+0.0}$ & $15.1_{-2.4}^{+2.4}$ \\
186 & +10:00:12.440 & +02:27:46.31 & $10.2_{-0.1}^{+0.0}$ & $0.16_{-0.00}^{+0.14}$ & $9.2_{-0.4}^{+0.1}$ & $8.9_{-0.4}^{+0.7}$ & $3.3_{-0.1}^{+6.6}$ & $3.5_{-0.0}^{+7.7}$ & $0.20_{-0.00}^{+0.34}$ & $1.0_{-0.3}^{+0.0}$ & $14.4_{-1.1}^{+1.1}$ \\
204 & +10:00:41.641 & +02:29:02.47 & $11.1_{-0.0}^{+0.0}$ & $0.40_{-0.05}^{+0.08}$ & $9.2_{-0.2}^{+0.0}$ & $8.5_{-0.3}^{+0.1}$ & $2.7_{-1.7}^{+1.4}$ & $3.2_{-1.8}^{+1.8}$ & $0.20_{-0.00}^{+0.25}$ & $1.0_{-0.3}^{+0.0}$ & $7.6_{-1.4}^{+1.3}$ \\
205 & +10:00:41.331 & +02:29:04.57 & $10.6_{--0.0}^{+0.1}$ & $0.16_{-0.02}^{+0.09}$ & $9.2_{-0.2}^{+0.1}$ & $8.7_{-0.2}^{+0.1}$ & $3.7_{-0.2}^{+3.5}$ & $4.1_{-0.5}^{+4.5}$ & $0.20_{-0.00}^{+0.20}$ & $0.7_{-0.0}^{+0.3}$ & $7.7_{-1.3}^{+1.1}$ \\
226 & +10:00:36.526 & +02:31:07.13 & $10.5_{-0.0}^{+0.1}$ & $0.20_{-0.02}^{+0.09}$ & $9.2_{-0.2}^{+0.3}$ & $8.9_{-0.2}^{+0.7}$ & $5.8_{-0.8}^{+7.4}$ & $6.1_{-0.8}^{+7.9}$ & $0.20_{-0.00}^{+0.26}$ & $0.7_{-0.0}^{+0.3}$ & $5.1_{-0.6}^{+0.7}$ \\
235 & +10:00:44.244 & +02:32:18.36 & $9.3_{-0.3}^{+0.0}$ & $0.10_{-0.01}^{+0.16}$ & $9.2_{-0.9}^{+0.0}$ & $9.9_{-1.1}^{+0.6}$ & $1.1_{-0.1}^{+3.0}$ & $1.1_{-0.1}^{+3.3}$ & $0.40_{-0.20}^{+0.48}$ & $1.0_{-0.3}^{+0.0}$ & $35.6_{-8.2}^{+6.4}$ \\
\end{tabular}
\tablefoot{
    \tablefoottext{a}{NBES}
    \tablefoottext{b}{$\log_{10}(M[\mathrm{M}_\odot]$; mass in stars at time of observation }
    \tablefoottext{c}{Assuming \citet{Calzetti:2000:682} extinction law}
	\tablefoottext{d}{$\log_{10}(Age[\mathrm{yr}])$}
	\tablefoottext{e}{$\log_{10}(\tau[\mathrm{yr}])$}
	\tablefoottext{f}{$[\mathrm{M}_\odot\;\mathrm{yr}^{-1}]$; instantaneous SFR at time of observation}
	\tablefoottext{g}{$[\mathrm{M}_\odot\;\mathrm{yr}^{-1}]$; SFR averaged over $100\;\mathrm{Myr}$ before time of observation}
	\tablefoottext{h}{metallicity in $[\mathrm{Z}_\odot]$}
	\tablefoottext{i}{Covering fraction of the gas; related to the escape of ionizing radiation through $f_\mathrm{esc;ion} = 1-f_\mathrm{cov}$}
	\tablefoottext{j}{$[\mathrm{nm}]$}
 }
\end{table*}

\section{Details about the selection of NB excess objects}
 \label{app:selcriteria}
\subsection{Color correction for J band magnitudes}

The NB118 filter is at the blue end of the J passband (Fig.
\ref{fig:nb118_y_j}).  Consequently, an estimate of the continuum at the
wavelength of the NB118 filter needs to include more information than $J$
alone, as it is necessary to correct for the galaxies' intrinsic colors.

Therefore, we estimated the continuum magnitude at the wavelength of the NB118
filter, $J_\mathrm{corr}$, through a linear combination of \emph{Y} and
\emph{J}. While this approach is identical to \citet{Milvang-Jensen:2013:94},
we adjusted the exact linear combination for two reasons: The zeropoints for
the broadband data have been adjusted between the UltraVISTA DR1
\citep{McCracken:2012:156}, which was used by them, and the DR2\footnotemark[9] used by us.
Further, we applied in this work corrections between the Vega magnitude system
and the AB system, which differ slightly from those used for both UltraVISTA
data releases and the work of \citet{Milvang-Jensen:2013:94}. In the following,
we justify the chosen color correction.

Under the assumption that SEDs are power laws over the wavelength range covered
by the Y and J filters, the appropriate combination can be determined based on
the filters' central wavelengths. This results in:

\begin{equation}
J_\mathrm{corr} = J + 0.25\,(Y-J) \label{eq:colcorr}
\end{equation}
\noindent
This corresponds to eq. \ref{eq:colcorr:flux4}, when using flux densities
instead of magnitudes. 

\begin{figure}
\centering
\includegraphics[width=1\columnwidth]{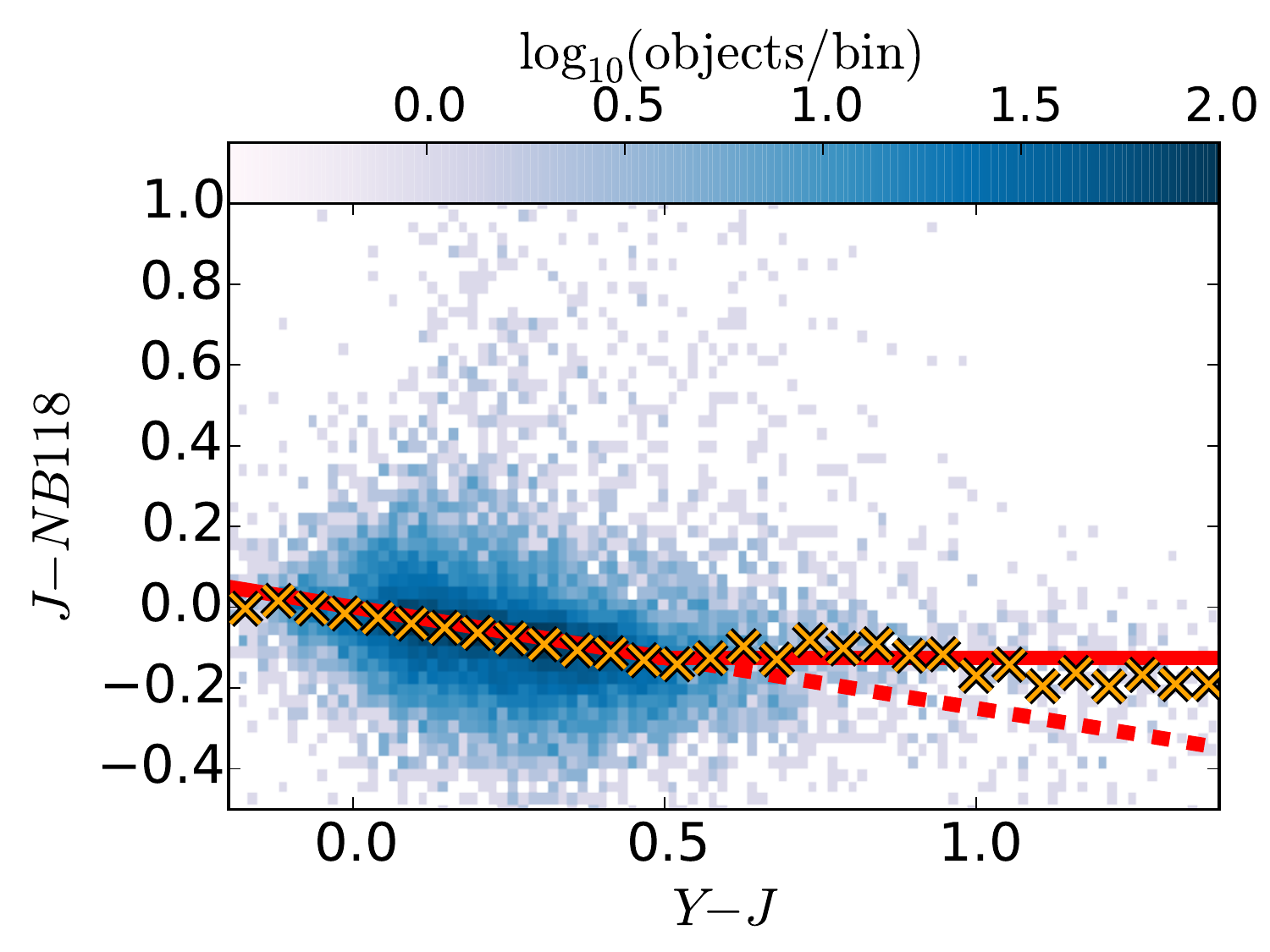}
	\caption{\label{fig:colcorr} $J-NB118$ versus $Y-J$ for all objects with a
	$5\,\sigma$ NB118 detection in the stack and at least a $2\sigma$ detection
	in the two broadband filters. Orange crosses give the median of $J-NB118$
	for all objects within equidistant $Y-J$ bins. The solid red line is the
	assumed color correction (cf. eq. \ref{eq:colcorr} and
	\ref{eq:colcorr2}), meaning that we defined objects on this line to have a
	$J_\mathrm{corr}-NB118$ of zero. The $J-NB118 = -0.25 (Y-J)$ line (eq.
	\ref{eq:colcorr}) is also shown beyond $Y-J>0.5$, but there as dotted
	line.}
\end{figure}

The validity of eq. \ref{eq:colcorr} can be verified empirically. Due to the
simplicity of a one-color correction, this can be easily visualized. In Fig.
\ref{fig:colcorr} we show the 2d histogram of the number of objects with $J -
NB118$ as a function of $Y - J$. All sources from the NB118 detected catalog
which have a NB118 detection above $5\,\sigma$, and at least $3\sigma$
detections in $Y$ and $J$, are included.
It is clear that the locus of the objects follows the line very well for
$Y-J<0.5$. This confirms empirically that the relation is justified.

The number of objects is small for $Y-J\gtrsim0.5$. Nevertheless, it can be
concluded that the line with a slope of -0.25 is not a good representation of
their typical colors. The main reason is that this part of the color space is
mainly populated by passive galaxies at $z\sim2$. Their red colors are not
caused by dust but by the Balmer/$4000\AA$ break located at the interface of Y
and J. While there might be some identifiable trend in $J-NB118$ as a function
of $Y-J$ also beyond $Y-J>0.5$, the number of objects is small, and we decided
to follow also in this part of the color space \citet{Milvang-Jensen:2013:94}.
They used for the reddest objects a constant correction to $J$. We have
determined this constant correction as minus the median $J-NB118$ color of all
objects with $Y-J>0.5$ and derived a value of
0.126. In order to have a continuous transition from \ref{eq:colcorr} into the
constant part, we use:

\begin{equation}
	J_\mathrm{corr} = J + 0.125 \textnormal{ for } Y-J > 0.5 \label{eq:colcorr2}
\end{equation}
\noindent
Expressed as flux densities, this is equivalent to eq. \ref{eq:colcorr:flux3}.

Formally, we set in absence of a Y detection ($<2\sigma$) $J_\mathrm{corr}$
simply to $J$. However, this is not really relevant, as we do not have
any NB excess objects without Y detection in the NB118 catalog created
with the conservative \sext{} parameters (cf. sec.
\ref{sec:reqnbexcess_detthresh}) used for this work.

\subsection{Narrowband excess and detection thresholds}
\label{sec:reqnbexcess_detthresh}

% TODO this needs to be written nicer
When selecting emission line galaxies from NB data, thresholds for the
magnitude excess (cf. eq. \ref{eq:col_crit}) and its significance need to be
set (cf. eq. \ref{eq:colsig}). Good selection criteria provide a compromise
between the inclusion of low $EW$ emitters and a small contamination from
objects without emission line in the NB filter.

\citet{Milvang-Jensen:2013:94} concluded based on an analysis of stellar
population models that $J_\mathrm{corr} - NB118$, which is for power law SEDs
expected to be close to zero, does also for realistic stellar SEDs not exceed
0.2 by much. The largest deviations from zero are expected at redshifts where
the $4000\mathrm{\AA}$/Balmer break and strong absorption lines fall into the
wavelength range of the Y and J filters, especially at population ages where
these features are pronounced.\footnote{Even larger deviations are
theoretically possible for galaxies with $z\sim8$, where the Lyman
break would be in this
range.} A selection threshold of $J_\mathrm{corr} - NB118 = 0.2$ can be
considered as a conservative choice to identify emission line galaxies.
$J_\mathrm{corr} - NB118 > 0.2$ corresponds to an $EW_\mathrm{obs}$ of
$27.4\,\AA$.\footnote{Averaged over all 16 filters, assuming
$250\,\mathrm{km}\,\mathrm{s}^{-1}$ emission lines at the wavelengths
corresponding to the peak of the respective filters and assuming a continuum
flat in $f_\lambda$.}

\begin{figure}
	\centering
	\includegraphics[width=0.95\columnwidth]{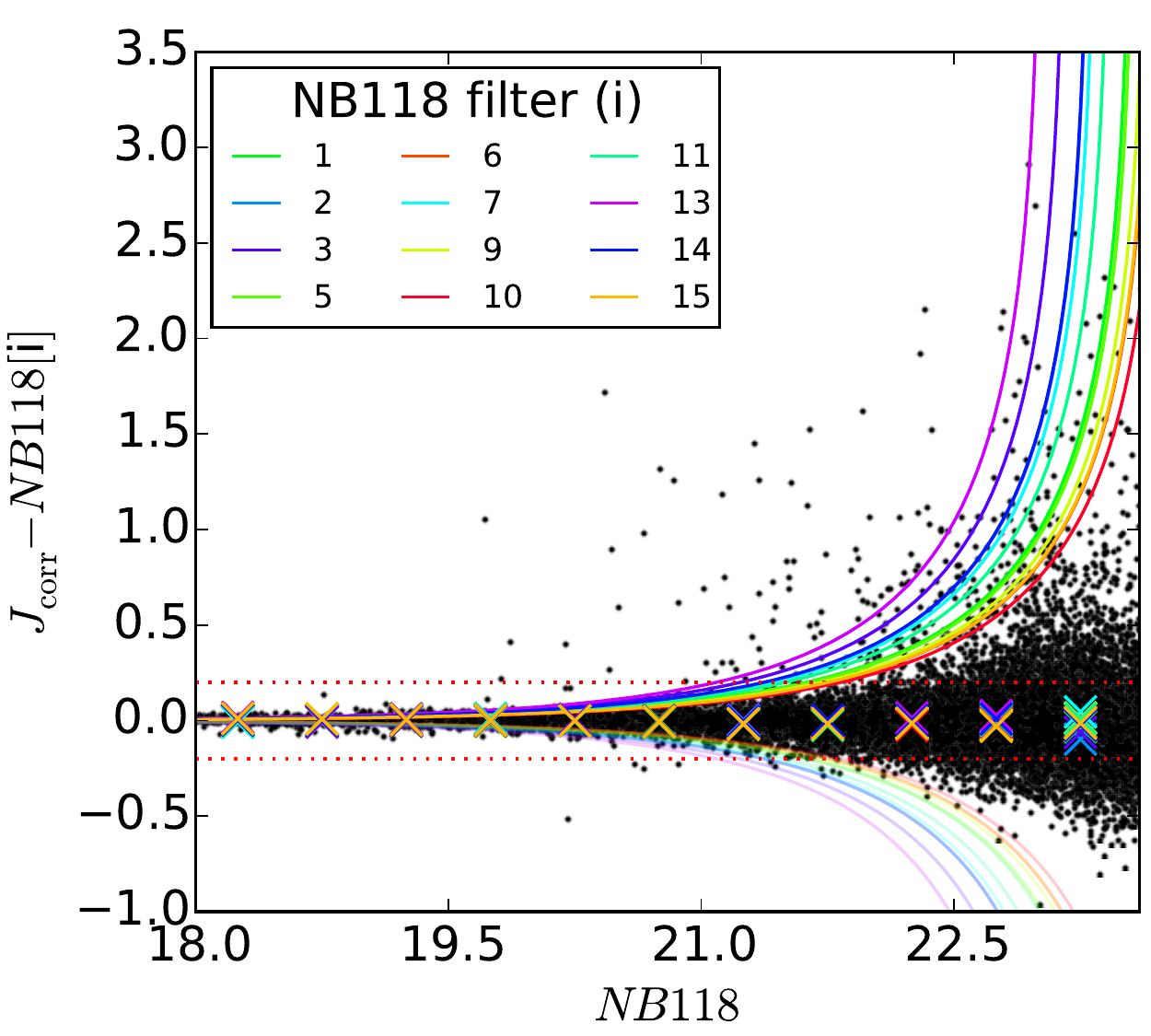}

	\caption{\label{fig:color_excess} Measured NB excess as a function of the
	NB118 magnitude for all objects in the main NB118 detected catalog. Data is
	only included for the regions of overlapping filters and in each of the
	overlapping regions only for one of the two filters. The red-dotted
	horizontal line indicates the minimum required NB118 excess for objects
	which we classify as NB excess objects. In addition, the curves which
	indicate a $4\sigma$ positive $J_\mathrm{corr}-NB118[i]$ are shown for each
	of the 12 filters The crosses show the median of the $J_\mathrm
	{corr}-NB118$ in $NB118$ bins with a width of $0.5\,\mathrm{mag}$. Finally,
	all mentioned lines are also plotted mirrored at
	$J_\mathrm{corr}-NB118=0$.}

\end{figure}

The required $J_\mathrm{corr} - NB118$ color combined with a $4\,\sigma$
significant color excess (eq. \ref{eq:colsig}) is expected to result in a
nearly pure sample of NB excess objects. We demonstrate this in Fig.
\ref{fig:color_excess}, where data for one filter from each of the 12 regions
with overlapping filters is included. The plot shows 12 different lines for the
color-excess criterion (eq. \ref{eq:colsig}). The reason for this is that the
depth in the 12 relevant filters strongly differs.  In addition to the relevant
selection curves, we also show these curves mirrored at $J_\mathrm{corr} -
NB118= 0$. This allows, at least to some extent, to judge the contamination
fraction due to statistical noise. It is as expected very low.

It is noteworthy that these conservative selection criteria will miss \fion{O}
{ii} emitters at $z=2.2$ (cf. also \citealt{Milvang-Jensen:2013:94}). However,
this is no problem for the present work, as we are here not interested in
\fion{O}{ii} emitters. Refined selection criteria for these objects
will be discussed in a forthcoming work.

In addition to the criteria described above, we decided to use a relatively
high detection and analysis threshold of $2\,\sigma$ in a least four
neighboring pixels for \sext{}. This detection threshold is at the limit of
affecting the completeness within our selection criteria: A lower detection
threshold would slightly increase the number of objects in the NBES. E.g. at a
very low threshold of $0.9\,\sigma$, the NBES sample would include 13 more
objects (252 vs 239). Nevertheless, we decided to use the $2\sigma$ catalog, as
we aimed in this work to include only clean and well centered detections.

\end{appendix}

\end{document}